\documentclass[onecolumn,showpacs,preprintnumbers,amsmath,amssymb,
superscriptaddress,prd,aps,nofootinbib,tightenlines,12pt]{revtex4-1}

\usepackage{graphicx}
\usepackage{bm}
\usepackage{epsfig}
\usepackage{verbatim}
\usepackage{upgreek}
\usepackage{bbm}

\usepackage{tikz}
\usetikzlibrary{backgrounds}

\newcommand{\msbar}{$\overline{\text{MS}}$}
\newcommand{\msbarformel}{\overline{\text{MS}}}
\newcommand{\lp}{\left(}
\newcommand{\rp}{\right)}
\newcommand{\lbk}{\left \lbrack}
\newcommand{\rbk}{\right \rbrack}
\newcommand{\lbc}{\left \lbrace}
\newcommand{\rbc}{\right \rbrace}
\newcommand{\la}{\left.}
\newcommand{\ra}{\right.}

\DeclareMathOperator{\Tr}{tr}
\DeclareMathOperator{\Diag}{Diag}
\DeclareMathOperator{\BlockDiag}{BlockDiag}

\renewcommand{\mathbbm}[1]{\rm 1\!\!1}

\allowdisplaybreaks

\begin{document}


\preprint{SFB/CPP-12-61, TTP12-30}
\title{Renormalization constants and beta functions for the gauge couplings of
  the Standard Model to three-loop order}

\author{Luminita N. Mihaila, Jens Salomon, Matthias Steinhauser}
\affiliation{Institut f\"ur Theoretische Teilchenphysik, Karlsruhe
  Institute of Technology (KIT), D-76128 Karlsruhe, Germany}


\begin{abstract}
  We compute the beta functions for the three gauge couplings of the Standard
  Model in the minimal subtraction scheme to three loops. We take into account
  contributions from all sectors of the Standard Model.  The calculation is
  performed using both Lorenz gauge in the unbroken phase of the Standard
  Model and background field gauge in the spontaneously broken phase.
  Furthermore, we describe in detail the treatment of $\gamma_5$ and present
  the automated setup which we use for the calculation of the Feynman
  diagrams. It starts with the generation of the Feynman rules and leads to
  the bare result for the Green's function of a given process.
\end{abstract}


\maketitle


\section{\label{sec::introduction}Introduction}

Renormalization group functions are fundamental quantities of each quantum
field theory. They provide insights in the energy dependence of cross
sections, hints to phase transitions and can provide evidence to the energy
range in which a particular theory is valid. The renormalization group
functions of the gauge couplings in the Standard Model (SM) are of particular
importance in the context of Grand Unified Theories allowing
the extrapolation
of low-energy precision data to high energies, not accessible by collider
experiments.

Important milestones for the calculation of the gauge coupling beta functions
in the Standard Model are the following computations:
\begin{itemize}
\item The one-loop beta functions in gauge theories along with the discovery
  of asymptotic freedom have been presented in
  Refs.~\cite{Gross:1973id,Politzer:1973fx}.
\item The corresponding two-loop corrections 
  \begin{itemize}
  \item in gauge theories without fermions~\cite{Jones:1974mm,Tarasov:1976ef},
  \item in gauge theories with fermions neglecting Yukawa
    couplings~\cite{Caswell:1974gg,Egorian:1978zx,Jones:1981we},
  \item with corrections involving Yukawa
    couplings~\cite{Fischler:1981is},
  \end{itemize}
  are also available.
\item The two-loop gauge coupling beta functions in an arbitrary quantum field
  theory have been considered in Ref.~\cite{Machacek:1983tz,Jack:1984vj}.
\item The contribution of the scalar self-interaction at three-loop
  order has been computed in~\cite{Curtright:1979mg,Jones:1980fx}.
\item The gauge coupling beta function in quantum chromodynamics (QCD) 
  to three loops is known from Ref.~\cite{Tarasov:1980au,Larin:1993tp}.
\item The three-loop corrections to the gauge coupling beta function 
  involving two strong and one top quark Yukawa coupling
  have been computed in Ref.~\cite{Steinhauser:1998cm}.
\item The three-loop corrections for a general quantum field theory based on a
  single gauge group have been computed in~\cite{Pickering:2001aq}.
\item The four-loop corrections in QCD 
  are known from Refs.~\cite{vanRitbergen:1997va,Czakon:2004bu}.
\end{itemize}
Two-loop corrections to the renormalization group functions for the Yukawa and
Higgs boson self-couplings in the Standard Model are also
known~\cite{Fischler:1982du,Machacek:1983fi,Machacek:1984zw,Jack:1984vj,Ford:1992pn,Luo:2002ey}.
Recently, the dominant three-loop corrections to the renormalization group
functions of the top quark Yukawa and the Higgs boson self-coupling have been
computed in Ref.~\cite{Chetyrkin:2012rz}. In that calculation the gauge
couplings and all the Yukawa couplings except the one of the top quark have
been set to zero.

In this paper we present details to the three-loop calculation of the gauge
coupling renormalization constants and the corresponding beta functions in the
SM taking into account all sectors. The results have already been presented in
Ref.~\cite{Mihaila:2012fm}.

The remainder of the paper is organized as follows: In the next
Section we
introduce our notation and describe in detail how we proceed to obtain the
beta functions of the gauge couplings.  In particular, we describe the
calculation in Lorenz gauge and background field gauge
(BFG), discuss our setup for an automated calculation, and explain our
treatment of $\gamma_5$.  The analytical results for the beta functions are
presented in Section~\ref{sec::results}.  In contrast to
Ref.~\cite{Mihaila:2012fm} we show the results including all
Yukawa couplings.  Section~\ref{sec::checks} is
devoted to a description of the checks which have been performed to verify our
result. A discussion of the numerical impact of the newly obtained corrections
is given in Section~\ref{sec::num}. We conclude in Section~\ref{sec::concl}.
Explicit results for the renormalization constants are relegated to
Appendix~\ref{app::renconst} and Appendix~\ref{app::renconst2}.
In Appendix~\ref{app::betaQED}, we present
three-loop results for the beta functions of the QED coupling constant and the
weak mixing angle. Furthermore, we present in Appendix~\ref{app::compare}
translation rules which are useful in order to compare parts of our findings
with the results of Ref.~\cite{Pickering:2001aq}.


\section{The calculation of the beta functions}

In this paper we present the beta functions for the three gauge couplings of
the SM up to three loops in the modified minimal subtraction (\msbar)
renormalization scheme. In the corresponding calculation we take into account
contributions involving the three gauge couplings of the SM, the top, the
bottom, and the tau Yukawa couplings and the Higgs self-coupling.  
We derive the beta functions for a general SM Yukawa
sector from the calculation involving the aforementioned seven couplings. We
postpone the discussion of this issue to the next Section. In the following,
we give details on the computation at the three-loop order.

We define the beta functions as
\begin{equation}
  \mu^2\frac{d}{d\mu^2}\frac{\alpha_i}{\pi}=\beta_i(\{\alpha_j\},\epsilon)\,,
\label{eq::beta_fc}
\end{equation}
where $\epsilon=(4-d)/2$ is the regulator of Dimensional Regularization with
$d$ being the space-time dimension used for the evaluation of the momentum
integrals.  The dependence of the couplings $\alpha_i$ on the renormalization
scale is suppressed in the above equation.

The three gauge couplings $\alpha_1$, $\alpha_2$ and $\alpha_3$ used in this
paper are related to the quantities usually used in the SM by
the following all-order relations
\begin{align}
  \alpha_1 &= \frac{5}{3}\frac{\alpha_{\rm QED}}{\cos^2\theta_W}\,,\notag\\
  \alpha_2 &= \frac{\alpha_{\rm QED}}{\sin^2\theta_W}\,,\notag\\
  \alpha_3 &= \alpha_s\,,
  \label{eq::alpha_123}
\end{align}
where all quantities are defined in the $\overline{\rm MS}$ scheme.
$\alpha_{\rm QED}$ is the fine structure constant, $\theta_W$ the weak
mixing angle and $\alpha_s$ the strong coupling. We adopt the SU(5)
normalization which leads to the factor 5/3 in the equation for $\alpha_1$.
Eq.~(\ref{eq::alpha_123}) serves as a definition for $\alpha_{\rm QED}$
and $\theta_W$.

To lowest order, the Yukawa couplings are given by
\begin{equation}\label{eq::def Yukawa}
  \alpha_x = \frac{\alpha_{\rm QED} m_x^2}{2 \sin^2\theta_W M_W^2}
  {\quad \mbox{with}\quad} x=t,b,\tau\,,
\end{equation}
where $m_x$ and $M_W$ are the fermion and W boson mass, respectively.
In the numerical analysis below also one-loop corrections to
Eq.~(\ref{eq::def Yukawa}) are taken into account~\cite{Pierce:1996zz}.

We denote the Higgs boson self-coupling by $\hat{\lambda}$, where
the Lagrange density contains the following term
\begin{eqnarray}
  {\cal L}_{\rm SM} &=& \ldots - (4\pi\hat{\lambda})(H^{\dagger}H)^2 + \ldots
  \,,
\end{eqnarray}
describing the quartic Higgs boson self-interaction.

The beta functions are obtained by calculating the renormalization constants
relating bare and renormalized couplings via
\begin{eqnarray}
  \alpha_i^{\text{bare}} &=&
  \mu^{2\epsilon}Z_{\alpha_i}(\{\alpha_j\},\epsilon)\alpha_i
  \,.
  \label{eq::alpha_bare}
\end{eqnarray}
Taking into account that $\alpha_i^{\text{bare}}$ does not depend on
$\mu$, 
Eqs.~(\ref{eq::beta_fc})  and (\ref{eq::alpha_bare}) lead to
\begin{eqnarray}
  \label{eq::renconst_beta}
  \beta_i &=& 
  -\left[\epsilon\frac{\alpha_i}{\pi}
    +\frac{\alpha_i}{Z_{\alpha_i}}
    \sum_{{j=1},{j \neq i}}^7
    \frac{\partial Z_{\alpha_i}}{\partial \alpha_j}\beta_j\right]
  \left(1+\frac{\alpha_i}{Z_{\alpha_i}}
    \frac{\partial Z_{\alpha_i}}{\partial \alpha_i}\right)^{-1}
  \,,
\end{eqnarray}
where $i=1,2$ or $3$. We furthermore set $\alpha_4=\alpha_t$,
$\alpha_5=\alpha_b$, $\alpha_6=\alpha_{\tau}$ and $\alpha_7=\hat{\lambda}$.

The first term in the first factor of Eq.~(\ref{eq::renconst_beta}) originates
from the term $\mu^{2\epsilon}$ in Eq.~(\ref{eq::alpha_bare}) and vanishes in
four space-time dimensions. 
The second term in the first factor
contains the beta functions of the remaining six couplings of the SM. Note
that (for the gauge couplings) the one-loop term of $Z_{\alpha_i}$ only
contains $\alpha_i$, whereas at two loops all couplings are present, except
$\hat{\lambda}$. The latter appears for the first time at
three-loop level. As a
consequence, it is necessary to know $\beta_j$ for $j=4,5,6$ to
one-loop order and only the $\epsilon$-dependent term for
  $\beta_7$, namely $\beta_7 = - \epsilon \alpha_7/\pi$.
From the second term in the first factor and the
second factor of Eq.~(\ref{eq::renconst_beta}) one can read
off that three-loop corrections to $Z_{\alpha_i}$ are required 
for the computation of $\beta_i$ to the same loop order.  

We have followed two distinct paths to obtain our results for the three-loop
renormalization constants, which we discuss in the following Subsections, where
we discuss their features and differences.


\subsection{\label{sub::lorenz}Lorenz gauge in the unbroken phase of the SM}

The first method used for the calculation of the renormalization constants is
based on Feynman rules derived for the SM in the unbroken phase in a general
Lorenz gauge with three independent gauge parameters corresponding to the
three simple gauge groups. All building blocks of our calculation are
evaluated for general gauge parameters in order to have a strong check
of the
final results for the beta functions which have to be gauge parameter
independent.  It is possible to use the unbroken phase of the SM since the
gauge beta functions
in the \msbar{} scheme are independent of all mass parameters and thus the
spontaneous symmetry breaking does not affect the final result.
  Note that this choice is advantageous for the calculation because in
the unbroken phase much less different
types of vertices have to be considered as compared to
the phase in which the spontaneous symmetry breaking is present.

In principle each vertex containing the coupling
$g_i=\sqrt{4\pi\alpha_i}$ can be used
to determine the corresponding renormalization constant 
via
\begin{eqnarray}
  Z_{\alpha_i} &=& \frac{(Z_{\text{vrtx}})^2}{\prod_k Z_{k,{\text{wf}}}}\,, 
  \label{eq::Zalpha}
\end{eqnarray}
where $Z_{\text{vrtx}}$ stands for the renormalization constant of the
vertex and $Z_{k,{\text{wf}}}$ for the wave function renormalization constant;
$k$ runs over all external particles. 

We have computed $Z_{\alpha_3}$ using both the ghost-gluon and the three-gluon
vertex. $Z_{\alpha_2}$ has been evaluated with the help of the
ghost-$W_3$, the $W_1 W_2 W_3$ and the $\phi^+ \phi^- W_3$ vertices where
$\phi^{\pm}$ is the charged component of the Higgs doublet corresponding to
the charged Goldstone boson in the broken phase and $W_1$, $W_2$ and $W_3$ are
the W boson components.  As to $Z_{\alpha_1}$, a Ward identity
guarantees that there is a cancellation
between the vertex and some of the wave function
renormalization constants yielding 
\begin{eqnarray}
  Z_{\alpha_1} &=& \frac{1}{Z_B},
\end{eqnarray}
where $Z_B$ is the wave function renormalization constant 
for the  gauge boson of the $U(1)$ subgroup of the  SM in 
the unbroken phase.

\begin{figure}
  \begin{center}
    \includegraphics[width=.85\textwidth]{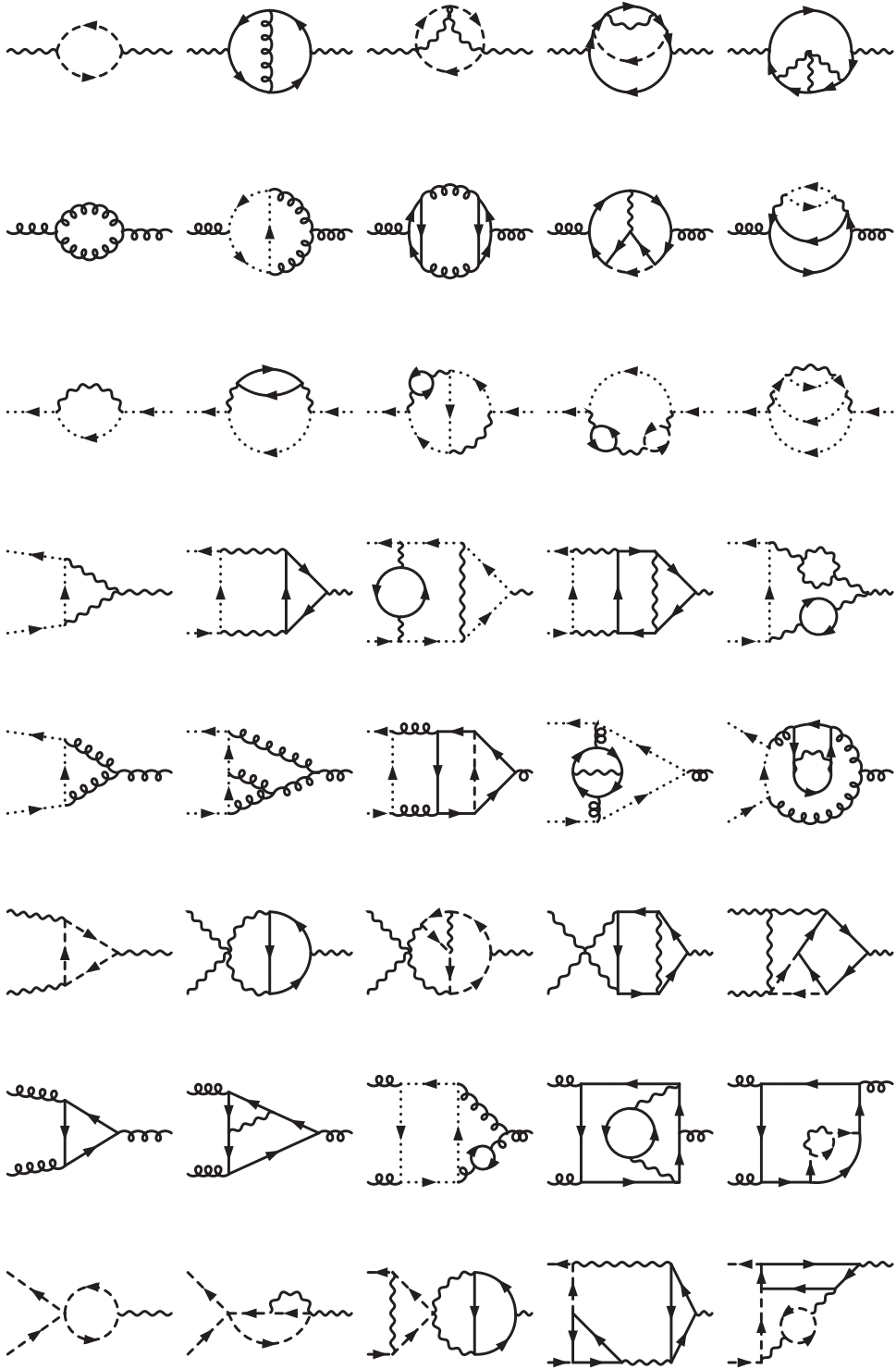}
    \caption{\label{fig::diags}Sample Feynman diagrams contributing to 
      the Green's functions which have been used for our calculation of the 
      gauge coupling renormalization constants.
      Solid, dashed, dotted, curly and wavy lines correspond to fermions,
      Higgs bosons, ghosts, gluons and electroweak gauge bosons,
      respectively.} 
  \end{center}
\end{figure}

In Fig.~\ref{fig::diags} we show several
one-, two- and three-loop sample diagrams contributing to the considered two-
and three-point functions.

We have not used vertices involving fermions as external particles as
they may lead to problems in connection with the
treatment of $\gamma_5$ in $d\not=4$ dimensions. The vertices
selected by us are safe in this 
respect.  A detailed discussion of our prescription for
$\gamma_5$ is given below.

In order to compute the individual renormalization constants entering
Eq.~(\ref{eq::Zalpha}) we proceed as outlined, e.g., in
Ref.~\cite{Steinhauser:1998cm}. The underlying formula can be written in the
form
\begin{eqnarray}
  Z_{\Gamma} &=& 1 - K_\epsilon\left( Z_{\Gamma} \Gamma \right)
  \,,
  \label{eq::Z}
\end{eqnarray}
where $\Gamma$ represents the two- or three-point function corresponding to
the renormalization constant $Z_{\Gamma}$ and the operator $K_\epsilon$ extracts the
pole part of its argument.
From the structure of Eq.~(\ref{eq::Z}) it is clear that $Z_{\Gamma}$ is computed
order-by-order in perturbation theory in a recursive way. It
is understood that the bare parameters 
entering $\Gamma$  on 
the right-hand side are replaced by the renormalized ones
before applying $K_\epsilon$. The corresponding counterterms are only needed to
lower loop orders than the one which is requested 
for $\Gamma$.
In our approach the three-loop calculation of $Z_{\alpha_i}$ requires 
--- besides the result for $Z_{\alpha_i}$ to two loops --- the one-loop
renormalization constants for the other two gauge and the Yukawa
couplings. Furthermore we have to renormalize the gauge parameters; the
corresponding renormalization constants are given by the wave function 
renormalization constants of the corresponding gauge
bosons which we anyway have 
to evaluate in the course of our calculation.


\subsection{\label{sub::background}Background field gauge in the spontaneously
  broken phase}

The second method that we used in order to get an independent result for the
renormalization constants of the gauge couplings is a calculation in
the BFG~\cite{Abbott:1980hw,Denner:1994xt}.  The
basic idea of the BFG is the splitting of all gauge fields
in a ``quantum'' and a ``classical''
part where in practical calculations the latter only occurs as external
particle.

The BFG has the advantage that Ward identities guarantee
that renormalization constants for gauge couplings can be obtained from the
exclusive knowledge of the corresponding wave function renormalization
constants.\footnote{In Lorenz gauge, this only works for $U(1)$ gauge groups,
  cf. Subsection~\ref{sub::lorenz}.}  
Thus we have the following formula

\begin{eqnarray}
  Z_{\alpha_i} &=& \frac{1}{Z_{A_i,\text{wf}}}
  \,,
  \label{eq::Z_BFG}
\end{eqnarray}
where $A$ denotes the gauge boson corresponding to the gauge coupling
$\alpha_i$. 

In contrast to the calculation using Lorenz gauge, we performed the
calculation in the BFG in the spontaneously broken phase of the
SM.\footnote{A
  {\tt FeynArts} model file (see Subsection~\ref{sub::setup}) for BFG in the
  unbroken phase has not been at our disposal. Furthermore, {\tt FeynRules}
  (see Subsection~\ref{sub::setup}) can not produce Feynman rules in BFG.}  As
discussed in the last Subsection, such a calculation is more involved than a
calculation in the unbroken phase since more vertices are present. On the
other hand, it constitutes an additional check of our result, that allows us
not only to compare the BFG and the Lorenz gauge but also to switch from the
broken to the unbroken phase of the SM.

Since the calculation has been performed in the broken phase we have computed
the transverse part of the two-point functions of the (background) photon,
$Z$ boson, photon-$Z$ mixing, $W$ boson and gluon which we denote by
$\Pi_\gamma$, $\Pi_Z$, $\Pi_{\gamma Z}$, $\Pi_W$, $\Pi_g$, respectively.
Sample Feynman diagrams up to three loops can be found in the first two
lines of~Fig.~\ref{fig::diags}. 

$\Pi_W$ and $\Pi_g$ can be used in analogy to Subsection~\ref{sub::lorenz} in
order to obtain the corresponding renormalization constants which leads in
combination with Eq.~(\ref{eq::Z_BFG}) to the renormalization constants for
$\alpha_2$ and $\alpha_s$. We found complete agreement with the calculation
performed in Lorenz gauge.

As far as the self energies involving photon and $Z$ boson are concerned, we
consider at the bare level appropriate linear combinations in order to obtain
the $B$ and $W$ boson self-energy contributions.  To be precise, we have
\begin{eqnarray}
  \Pi_B &=& \cos^2\theta_W^{\rm bare} \Pi_\gamma 
  + 2 \cos\theta_W^{\rm bare}\sin\theta_W^{\rm bare} \Pi_{\gamma Z}
  + \sin^2\theta_W^{\rm bare} \Pi_Z
  \,, \nonumber\\
  \Pi_W &=& \sin^2\theta_W^{\rm bare} \Pi_\gamma 
  - 2 \cos\theta_W^{\rm bare}\sin\theta_W^{\rm bare} \Pi_{\gamma Z}
  + \cos^2\theta_W^{\rm bare} \Pi_Z
  \,.
\end{eqnarray}
The second linear combination can immediately be compared with the bare result
obtained from the charged $W$ boson self energy and complete agreement up to
the three-loop order has been found. This constitutes a strong consistency
check on the implementation of the BFG Feynman rules.  $\Pi_B$ is used
together with Eq.~(\ref{eq::Z_BFG}) in order to obtain the renormalization
constant for $\alpha_1$. Again, complete agreement with the calculation
described in the previous Subsection has been found.

In our BFG calculation we want to adopt Landau gauge in order to avoid the
renormalization of the gauge parameters $\xi_i$. However, it
is not possible to choose Landau gauge from the very beginning since some
Feynman rules for vertices involving a background gauge boson contain terms
proportional to $1/\xi_i$ where $\xi_i=0$ corresponds to Landau gauge.  To
circumvent this problem we evaluate the bare integrals for arbitrary gauge
parameters. In the final result all inverse powers of $\xi_i$ cancel and thus
the limit $\xi_i=0$ can be taken at the bare level.


\subsection{\label{sub::setup}Automated Calculation}

Higher order calculations in the SM taking into account all
contributions are quite
involved. Apart from the complicated loop integrals, there are many different
Feynman rules and plenty of Feynman diagrams which have to be considered.
In our calculation we have used a setup which to a large extend avoids manual
interventions in order to
keep the error-proneness to a minimum. 

As far as the loop integrals are concerned we exploit the fact that the beta
function in the \msbar{} scheme is independent of the external momenta and the
 particle masses. Thus, we can choose a convenient kinematical
configuration which leads to simple loop integrals as long as the infra-red
structure is not modified. In our case we set all particle masses to zero and
only keep one external momentum different from zero.  We have
checked that no infra-red divergences are introduced as we will discuss in
detail in Section~\ref{sec::checks}.  In this way all loop-integrals are
mapped to massless two-point functions that up to three loops can be computed
with the help of the package \verb|MINCER|~\cite{Larin:1991fz}.

\begin{figure}[t]
  \includegraphics[width=.85\textwidth]{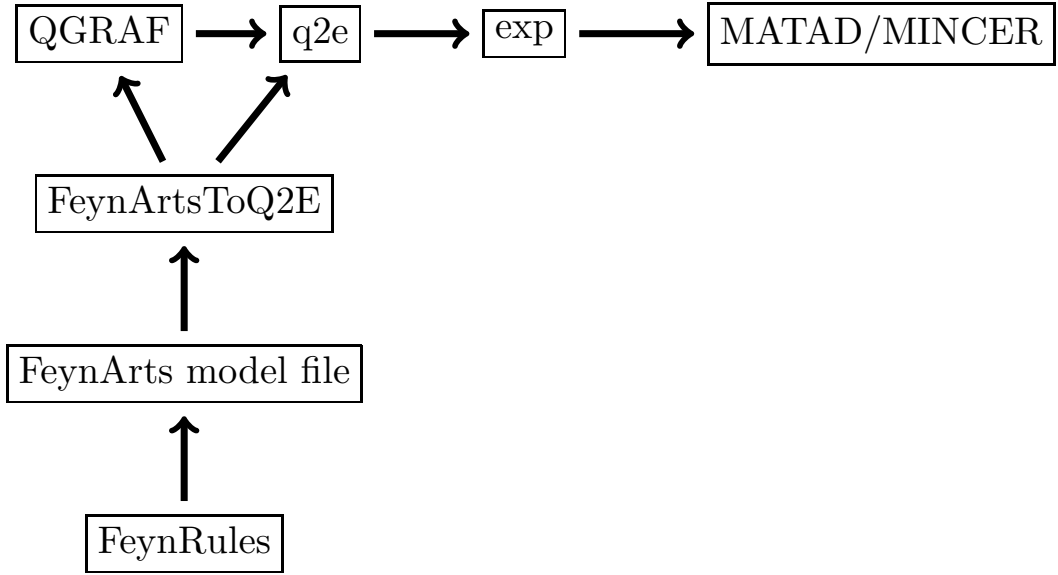}
  \caption{Overview of our automated setup. 
      Calling up the programs in the
    uppermost line determines and evaluates a given process in a given
    model. The vertical workflow leads to the implementation of a new model in
    the setup. The programs are discussed in more detail in the
    text.}\label{fig::setup}
\end{figure}

As core of our setup we use a well-tested chain of programs that work
hand-in-hand: \verb|QGRAF|~\cite{Nogueira:1991ex} generates all contributing
Feynman diagrams. The output is passed via
\verb|q2e|~\cite{Harlander:1997zb,Seidensticker:1999bb}, which transforms
Feynman diagrams into Feynman amplitudes, to
\verb|exp|~\cite{Harlander:1997zb,Seidensticker:1999bb} that generates
\verb|FORM|~\cite{Vermaseren:2000nd} code.  The latter is processed by
\verb|MINCER|~\cite{Larin:1991fz} and/or
\verb|MATAD|~\cite{Steinhauser:2000ry} that compute the Feynman integrals and
output the $\epsilon$ expansion of the result.  The parallelization of the
latter part is straightforward as the evaluation of each Feynman diagram
corresponds to an independent calculation. We have also parallelized the part
performed by \verb|q2e| and \verb|exp| which is essential for our calculation
since it may happen that a few times $10^5$ diagrams contribute at three-loop
level to a single Green's function.  The described workflow is illustrated on
the top of Fig.~\ref{fig::setup}.

In order to perform the calculation described in this paper we have extended
the above setup by the vertical program chain in Fig.~\ref{fig::setup}. The
core of the new part is the program \verb|FeynArtsToQ2E| which translates
\verb|FeynArts|~\cite{Hahn:2000kx} model files into model files processable by
\verb|QGRAF| and \verb|q2e|.  In this way we can exploit the well-tested input
files of \verb|FeynArts| in our effective and flexible setup based on
\verb|QGRAF|, \verb|q2e|, \verb|exp| and \verb|MINCER|.  This avoids the
coding of the Feynman rules by hand which for the SM would require a dedicated
debugging process.

For the part of our calculation based on the BFG we have used the
\verb|FeynArts| model files which come together with version 3.5.
However, for Lorenz gauge in the unbroken phase there
is no publicly available \verb|FeynArts| model. For this reason we have
used
the package
\verb|FeynRules|~\cite{Christensen:2008py} in order to generate such a file
which is also indicated in Fig.~\ref{fig::setup}.

Let us mention that \verb|FeynArtsToQ2E| is  not restricted to the SM but can
process all model files available for \verb|FeynArts|.


\subsection{\label{sub::gamma_5}Treatment of $\gamma_5$}

An important issue in multi-loop calculations is the definition of $\gamma_5$
away from $d=4$ dimensions. A first possibility is the naive regularization
that requires that $\gamma_5$ anti-commutes with all other
$\gamma$-matrices. This approach has the advantage that its
implementation is very simple.  However, it can lead to
wrong results, especially for Feynman diagrams involving several
fermion loops. For example, the naive regularization of $\gamma_5$ leads to
the problematic result (see, e.g., Ref.~\cite{Collins:1984xc})
\begin{equation}
  \text{tr}(\gamma^{\mu}\gamma^{\nu}\gamma^{\rho}\gamma^{\sigma}\gamma_5)=0
  \qquad
  (d \neq 4)\,.
\label{eq::naive}
\end{equation}
The limit of this expression for $d\rightarrow 4$ does  not agree with
its value  in the physical case,  when the regularization is
turned off
\begin{equation}\label{tr d=4}
  \text{tr}(\gamma^{\mu}\gamma^{\nu}\gamma^{\rho}\gamma^{\sigma}\gamma_5)
  = - 4\text{i}{\varepsilon}^{\mu\nu\rho\sigma}\ (d = 4)\,.
\end{equation} 
Here the totally anti-symmetric Levi-Civita tensor is defined by
${\varepsilon}^{0123}=1$.

It is therefore reassuring that one can show explicitly that in the
computation via the ghost-ghost-gauge boson vertex\footnote{We restrict
  the discussion in this Subsection to the ghost-ghost-gauge boson vertex. For
  all other vertices without external fermions the reasoning is in full
  analogy.}  all contributions stemming from this kind of traces vanish.  To
prove this, we notice that this kind of traces can only lead to non-vanishing
contributions if there are at least two of them in a diagram. Only in this
case the ${\varepsilon}$-tensors originating from Eq.~(\ref{tr d=4}) can be
contracted, providing Lorentz structures that may contribute to the
renormalization constants. We observe that the fermion loops can only
yield problematic non-vanishing contributions if at least \emph{three} lines
are attached to them.
Otherwise, there are too few external momenta and too
few open Lorentz indices available. A general three-loop diagram with at least
two closed fermion loops has the following form
\begin{center}
  \includegraphics[width=4cm]{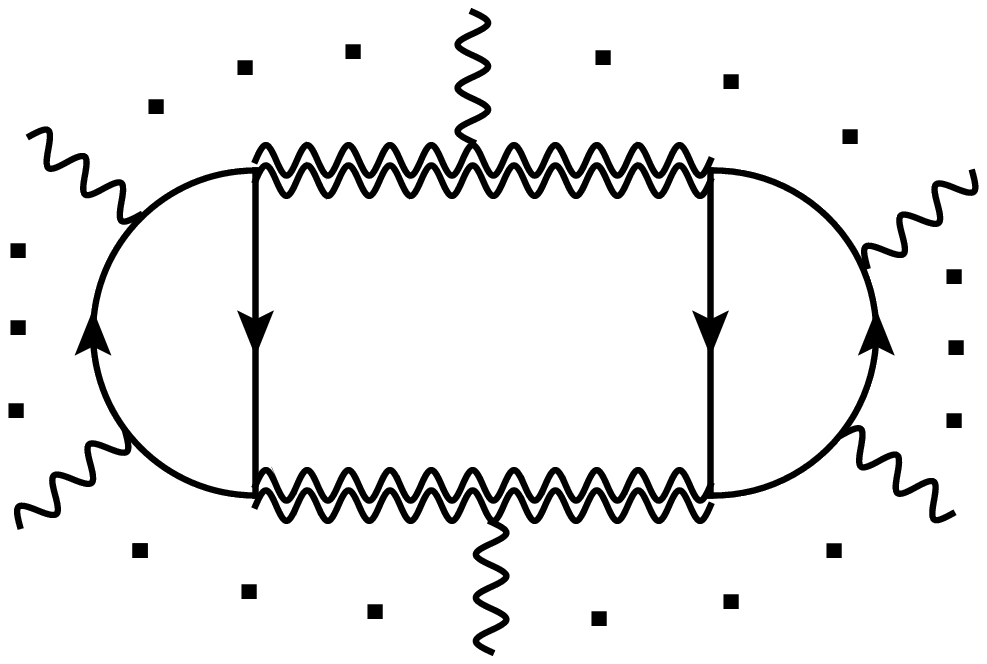}.
\end{center}
Here the solid lines represent fermions while the double lines can be
either scalar bosons or gauge bosons.

One can easily show that one cannot attach ghosts as external particles in the
above diagram, since there are no vertices involving ghosts and fermions. 
Thus, the ghost self-energy and the ghost-ghost-gauge boson vertex
do not contain diagrams of this type. Therefore, we need to consider only
diagrams with two external gauge bosons for the following discussion.  So the
diagrams which still have to be discussed have the structure
\begin{center}
  \includegraphics[width=6cm]{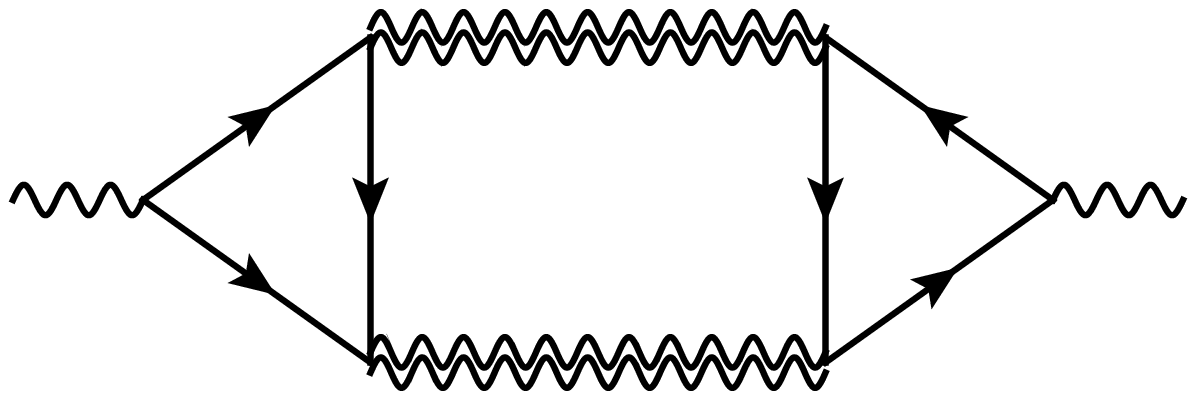}.
\end{center}
There are three ways to replace the double lines:
\begin{center}
  \mbox{
    \includegraphics[width=12em]{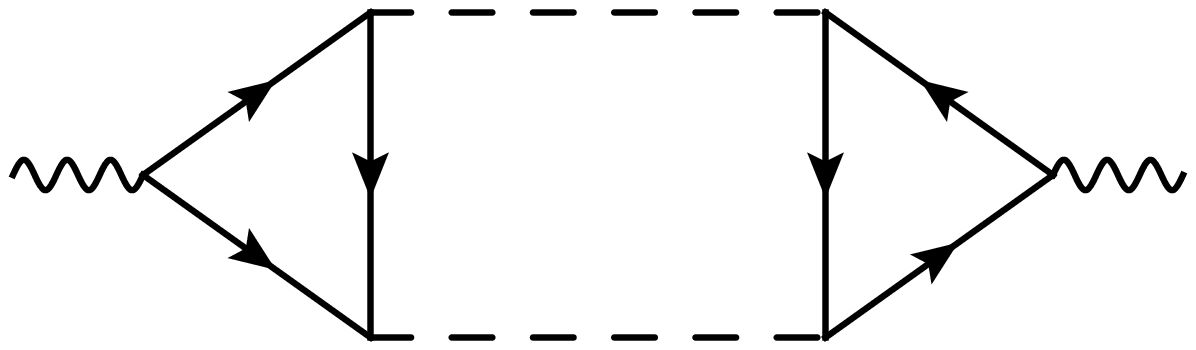}
    \quad
    \includegraphics[width=12em]{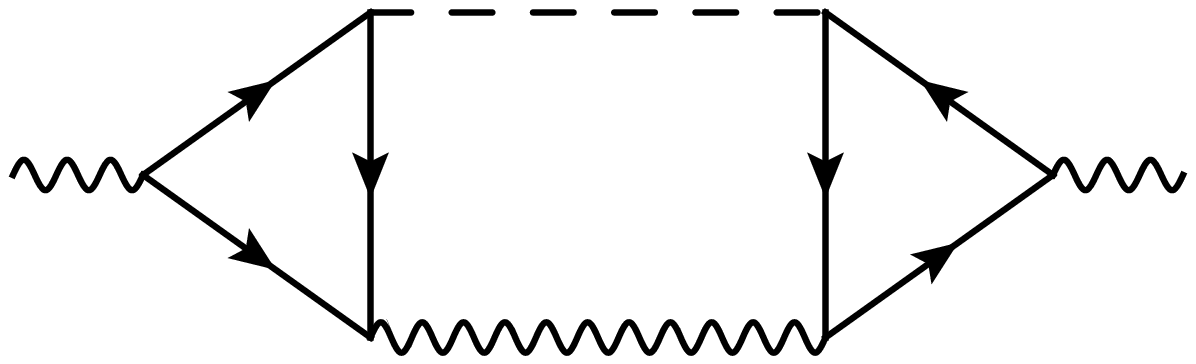}
    \quad 
    \includegraphics[width=12em]{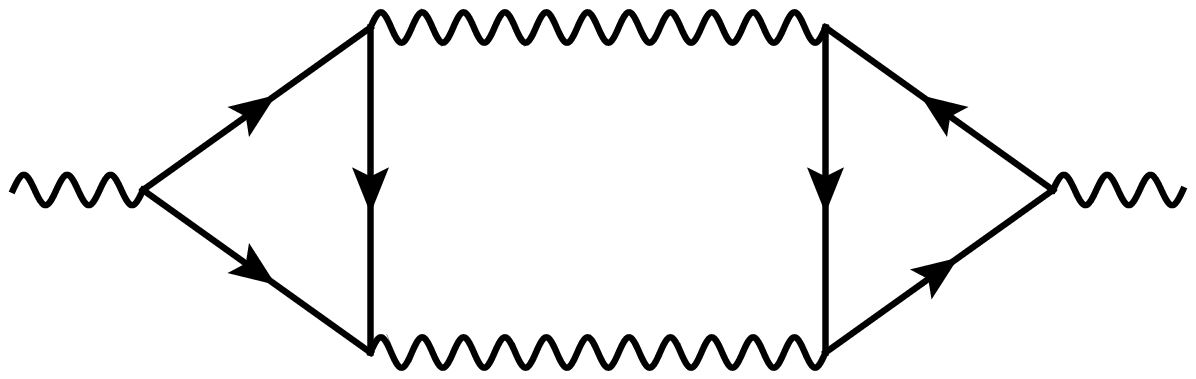}
  .}
\end{center}
The first diagram type cannot yield problematic contributions for the same
reason as mentioned above.  The second diagram type vanishes as the fermion
traces involve exactly five $\gamma$-matrices, not counting $\gamma_5$
matrices. (Remember that we deal with diagrams in which all the propagators
are massless.)  Finally, the third diagram type involves fermion loops with
three external gauge bosons. Such diagrams can indeed contain contributions
originating from traces of $\gamma_5$ and an even number of other
$\gamma$-matrices. However, the sum over all possible fermion species that can
circulate in the loops, including also the diagrams in which the fermions
circle in opposite directions, vanishes.\footnote{The proof of this statement
  can be based only on considerations about group theoretic invariants. For
  details see Chapter~20 of Ref.~\cite{Peskin:1995ev}.} This is of course a
consequence of the cancellation of the Adler-Bell-Jackiw
anomaly~\cite{Adler:1969gk,Bell:1969ts} within the SM, as required by
gauge invariance.
Therefore, we are allowed to calculate the Feynman diagrams contributing to
$Z_{\alpha_i}$ 
using a naive regularization prescription for $\gamma_5$, in which the
diagrams containing triangle anomalies are set to zero from the very
beginning, according to Eq.~(\ref{eq::naive}) .

As an additional check of the calculation we implemented also a ``semi-naive''
regularization prescription for $\gamma_5$. Explicitly, we evaluate the
expression 
$\text{tr}(\gamma^{\mu}\gamma^{\nu}\gamma^{\rho}\gamma^{\sigma}\gamma_5)$ by
applying the \emph{formal} replacement
\begin{equation}
  \text{tr}(\gamma^{\mu}\gamma^{\nu}\gamma^{\rho}\gamma^{\sigma}\gamma_5)=
  -4\text{i}\tilde{\varepsilon}^{\mu\nu\rho\sigma}+
  {\cal O}(\epsilon)\,.
\label{eq::seminaive}
\end{equation} 
The tensor $\tilde{\varepsilon}^{\mu\nu\rho\sigma}$ has some similarities with
the four-dimensional Levi-Civita tensor: $(i)$ it is completely antisymmetric in
all indices; $(ii)$ when contracted with a second one of its kind one obtains the
following result
\begin{equation}
  \tilde{\varepsilon}^{\mu\nu\rho\sigma}
\tilde{\varepsilon}_{\mu^{\prime}\nu^{\prime}\rho^{\prime}\sigma^{\prime}}=
 g^{[\mu\phantom{]}}_{[\mu'\phantom{]}}
  g^{\phantom{[}\nu\phantom{]}}_{\phantom{[}\nu'\phantom{]}}
  g^{\phantom{[}\rho\phantom{]}}_{\phantom{[}\rho'\phantom{]}}
  g^{\phantom{[}\sigma\,]}_{\phantom{[}\sigma']}\,, 
\label{eq::leci}
\end{equation} 
where the square brackets denote complete anti-symmetrization.  When taking the limit
$d\to 4$, $\tilde{\varepsilon}^{\mu\nu\rho\sigma}$ converts into the
four-dimensional Levi-Civita tensor and Eqs.~(\ref{eq::seminaive})
and~(\ref{eq::leci}) ensure that it provides the correct four-dimensional 
result. 

At this point a comment on Eq.~(\ref{eq::seminaive}) is in order. It is
straightforward to see that the combination of this equation and the cyclic
property of traces leads to an ambiguity of order ${\cal
  O}(\epsilon)$. Therefore, we made sure that the terms that need to be
treated in this way generate at most simple poles in $\epsilon$,
which are thus unambiguous, and the above
procedure can be applied directly without introducing additional finite
counterterms.

Let us stress again that we find the same result for the renormalization
constants $Z_{\alpha_i}$
both from the ghost--ghost--gauge boson vertex and
by using other vertices and both by applying the ``naive'' as well as the
``semi-naive'' scheme. These findings strongly support the above reasoning.


\subsection{Comparison of the methods}

This Subsection is devoted to a brief comparison of the calculation via the
Lorenz gauge and the one involving the BFG. As has been mentioned
before, in the BFG it is sufficient to consider only the gauge
boson propagators. This is advantageous as the use of
Lorenz gauge also requires the
evaluation of  three- (or four-) point functions and in
most cases it also demands
the consideration of additional two-point functions apart from
the gauge boson ones. Disadvantages of the
BFG are the increased  number of
vertices and the more involved structure of the vertices
containing a background field.

\begin{table}[t]
  \begin{center}
    \begin{tabular}[t]{c||r|r|r|r}
      \multicolumn{5}{c}{Lorenz gauge} \\
      \hline
      \# loops & 1 & 2 & 3 & 4\\
      \hline
      $BB$                    &   14 & 410 &  45\,926 & 7\,111\,021\\
      $W_3W_3$                &   17 & 502 &  55\,063 & 8\,438\,172\\
      $gg$                    &    9 & 188 &  17\,611 & 2\,455\,714\\
      $c_g \bar{c}_g$         &    1 &  12 &      521 &     46\,390\\
      $c_{W_3} \bar{c}_{W_3}$ &    2 &  42 &   2\,480 &    251\,200\\
      $\phi^+ \phi^-$         &   10 & 429 &  46\,418 & 6\,918\,256\\
      \hline
      $BBB$                       &   34 & 2\,172 & 358\,716 & 73\,709\,886\\
      $W_1W_2W_3$                 &   34 & 2\,216 & 382\,767 & 79\,674\,008\\
      $ggg$                       &   21 &    946 & 118\,086 & 20\,216\,024\\
      $c_g \bar{c}_g g$           &    2 &     66 &   4\,240 &     460\,389\\
      $c_{W_1} \bar{c}_{W_2} W_3$ &    2 &    107 &  10\,577 &  1\,517\,631\\
      $\phi^+\phi^- W_3$          &   24 & 2\,353 & 387\,338 & 77\,292\,771\\
      \hline
    \end{tabular}
    \hspace*{3em}
    \begin{tabular}[t]{c||r|r|r|r}
      \multicolumn{5}{c}{BFG} \\
      \hline
      \# loops & 1 & 2 & 3 & 4\\
      \hline
      $\gamma^B \gamma^B$ &   13 &  416 &  61\,968 & 13\,683\,693\\
      $\gamma^B Z^B$      &   13 &  604 & 100\,952 & 23\,640\,897\\
      $Z^B Z^B$           &   20 & 1064 & 183\,465 & 44\,049\,196\\
      $W^{+B} W^{-B}$     &   18 & 1438 & 252\,162 & 42\,423\,978\\
      $g^B g^B$           &   10 &  186 &  17\,494 &  2\,775\,946 \\
      \hline
    \end{tabular}
  \end{center}
  \caption[]{\label{tab::num_diag}The number of Feynman diagrams contributing to
    the Green's functions evaluated in this work.  Left table: two- and three-point
    functions computed in Lorenz gauge; right table: two-point functions
    computed in BFG. The superscript ``B'' denotes background fields.
    The first column
    indicates the external legs of the Green's function, the other columns
    show the number of diagrams at the individual loop orders.
    Note that the $BBB$ vertex is computed in order to have a cross check as
    we will explain in Section~\ref{sec::checks}.}
\end{table}

In Tab.~\ref{tab::num_diag} 
we list the number of diagrams for each Green's
function contributing to the \mbox{one-,} two- and three-loop order. The number of
diagrams computed in this work is obtained from the sum of the numbers in
these columns. For comparison we also provide the corresponding number of
diagrams which 
contribute to the four-loop order. 

It is tempting to compare the number of contributing Feynman diagrams in
Lorenz gauge and in BFG which is, however, not straightforward since we use
the former in the broken and the latter in the unbroken phase. Nevertheless,
one observes that in the case of $\beta_3$ the
number of diagrams  entering the BFG calculation is roughly the same
as in
case the gluon-ghost vertex is used in Lorenz gauge, even up to
four-loop order. All other vertices lead to
significantly more diagrams.  In the case of $\beta_2$ there are about three
to four times more diagrams to be considered in the BFG as compared to Lorenz
gauge.  Whereas at three-loop order the difference between approximately
70\,000 and 250\,000 diagrams is probably not substantial it is striking at
four-loop order where the number of Feynman diagrams goes from about
10\,000\,000 ($W_3W_3$ , $c_{W_3} \bar{c}_{W_3}$ and $c_{W_1} \bar{c}_{W_2}
W_3$ Green's function) to 42\,000\,000 ($W^{+B} W^{-B}$ Green's function) when
switching from Lorenz gauge to BFG. Thus, starting from four loops it is
probably less attractive to use the BFG.

Let us add that the precise number of Feynman diagrams depends on the detailed
setup as, e.g., on the implementation of the four-particle vertices. Because
of the colour structure we split in our calculation the four-gluon vertex into
two cubic vertices by introducing non-propagating auxiliary particles, whereas
all other four-particle vertices are left untouched.

Let us finally mention that the CPU time for the evaluation of an individual
diagram ranges from less than a second to few minutes. For general gauge
parameters it may take up to the order of an hour.  Thus the use of about 100
cores leads to a wall-clock time which ranges from a few hours for a
calculation in Feynman gauge up to about one day for general gauge parameters.
For the preparation of the {\tt FORM} files using \verb|QGRAF|, \verb|q2e| and
\verb|exp| also a few hours of CPU time are needed which is because of the
large amount of Feynman diagrams. The use of about 100 cores leads to a
wall-clock time of a few minutes.



\section{\label{sec::results}Analytical Results}

In this Section we present our analytical results for the beta functions. As
mentioned before, we are able to present the results involving all
contributions of the SM Yukawa sector.

The SM Yukawa interactions are described by (see, e.g., Chapter~11 of
Ref.~\cite{Nakamura:2010zzi})
\begin{equation}
  {\cal L}_{\rm Yukawa} = - \bar{Q}_i^LY^U_{ij} \epsilon H^{\star} u_j^R -
  \bar{Q}_i^LY^D_{ij} H d_j^R
 - \bar{L}_i^LY^L_{ij} H l_j^R+ \text{h.c.}\,,
\end{equation}
where $Y^{U,D,L}$ are complex $3\times 3$ matrices, $i,j$ are generation
labels, $H$ denotes the Higgs field and $\epsilon$ is the $2 \times 2$
antisymmetric tensor. $Q^L,L^L$ are the left-handed quark and lepton
doublets, and $u^R,d^R,l^R$ are the right-handed up- and down-type
quark and lepton singlets, respectively. The physical mass-eigenstates
are obtained by diagonalizing $Y^{U,D,L}$ by six unitary matrices
$V_{L,R}^{U,D,L}$ as follows
\begin{equation}
 \tilde{Y}^{f}_{\rm diag}=V_L^f Y^f V_R^{f\dagger}\,, \quad f=U,D,L\,.
 \label{eq::yuk1}
\end{equation}
As a result the charged-current $W^{\pm}$ couples to the physical quark
states 
with couplings parametrized by the Cabibbo-Kobayashi-Maskawa (CKM) matrix
$V_{CKM}\equiv V_L^UV_L^{D\dagger}$.
We furthermore introduce the notation
\begin{equation}
  \hat{T} = \frac{1}{4\pi} Y^U {Y^U}^{\dagger},\ \hat{B} =
  \frac{1}{4\pi}  Y^D {Y^D}^{\dagger},\ \hat{L} = \frac{1}{4\pi}  Y^L
  {Y^L}^{\dagger}. 
 \label{eq::yuk2}
\end{equation}

In order to
reconstruct the results for a general Yukawa sector, we have multiplied each
Feynman diagram by a factor $(n_h)^m$, where $m$ denotes the number of fermion
loops involving Yukawa couplings.  After analyzing the structure of the
diagrams that can arise, we could establish the following set of replacements
that have to be performed in order to take into account a generalized Yukawa
sector
\begin{align}\label{eq::nh_to_tr}
  n_h \alpha_{t} &\rightarrow \text{tr}\hat{T},&
  n_h \alpha_{b} &\rightarrow \text{tr}\hat{B},\notag \\
  n_h \alpha_{\tau} &\rightarrow \text{tr}\hat{L},&
  n_h \alpha_{t}^2 &\rightarrow \text{tr}\hat{T}^2,\notag \\
  n_h \alpha_{b}^2 &\rightarrow \text{tr}\hat{B}^2,&
  n_h \alpha_t \alpha_b &\rightarrow \text{tr}\hat{T}\hat{B},\notag \\
  n_h \alpha_{\tau}^2 &\rightarrow \text{tr}\hat{L}^2,&
  n_h^2 \alpha_t^2 &\rightarrow (\text{tr}\hat{T})^2,\notag \\
  n_h^2 \alpha_b^2 &\rightarrow (\text{tr}\hat{B})^2,&
  n_h^2 \alpha_{\tau}^2 &\rightarrow (\text{tr}\hat{L})^2,\notag \\
  n_h^2 \alpha_t \alpha_b &\rightarrow \text{tr}\hat{T}\text{tr}\hat{B},&
  n_h^2 \alpha_t \alpha_{\tau} &\rightarrow \text{tr}\hat{T}\text{tr}\hat{L},\notag \\
  n_h^2 \alpha_b \alpha_{\tau} &\rightarrow \text{tr}\hat{B}\text{tr}\hat{L}.
\end{align}
Of course, only traces over products of Yukawa matrices can occur because
they arise from closed fermion loops.  Using Eqs.~(\ref{eq::yuk1})
and~(\ref{eq::yuk2}), it is straightforward to see that in
Eq.~(\ref{eq::nh_to_tr}) only traces of diagonal matrices have to be taken
except for $\text{tr}\hat{T}\hat{B}$ which is given by
\begin{equation}
  \text{tr}\hat{T}\hat{B} = \text{tr}\left[\left(\begin{matrix}
        \alpha_u & 0 & 0 \\
        0        & \alpha_c & 0 \\
        0 & 0 & \alpha_t
      \end{matrix}
    \right) V_{\text{CKM}}
    \left(\begin{matrix}
        \alpha_d & 0 & 0 \\
        0        & \alpha_s & 0 \\
        0 & 0 & \alpha_b
      \end{matrix}
    \right) V_{\text{CKM}}^{\dagger}\right].
  \label{eq:ckm}
\end{equation}

The addition of a fourth generation of fermions to the SM particle content can
be also easily accounted for by this general notation. In this case, the
Yukawa matrices become $4\times 4$ dimensional.  If we assume that the fourth
generation is just a repetition of the existing generation pattern but much
heavier and if we neglect all SM Yukawa interactions, then the explicit form of
Yukawa matrices reads
\begin{equation}
  \hat{F_4} = \left(\begin{matrix}
      0_{3\times 3} & 0  \\
      0        & \alpha_F 
    \end{matrix}
  \right)\,,\quad \mbox{with} \quad F=T, B, L\,.
\end{equation}
Here $T$ and $B$ stand for the up- and down-type heavy quarks, and $L$ for
the heavy charged leptons, while
$\alpha_F$ denotes the  corresponding Yukawa couplings
as defined in Eq.~(\ref{eq::def Yukawa}).
Since in our calculation no Yukawa couplings for neutrinos have been
introduced we cannot
incorporate heavy neutrinos. This would require a dedicated calculation which,
however, does not pose any principle problem.

We are now in the position to present our
results for the beta functions of the gauge couplings. They are
given by
\begin{align}
\beta_1 &=
      \frac{\alpha_1^2}{\lp4\pi\rp^2} \bigg\{ \frac{2}{5} + \frac{16 n_G}{3} \bigg\} \notag \\
 &  + \frac{\alpha_1^2}{\lp4\pi\rp^3} \bigg\{ \frac{18 \alpha_1}{25} + \frac{18 \alpha_2}{5} - \frac{34 \text{tr}\hat{T}}{5} - 2 \text{tr}\hat{B} - 6 \text{tr}\hat{L} + n_G \bigg[ \frac{76 \alpha_1}{15} + \frac{12 \alpha_2}{5} + \frac{176 \alpha_3}{15} \bigg] \bigg\} \notag \\
 &  + \frac{\alpha_1^2}{\lp4\pi\rp^4} \bigg\{ \frac{489 \alpha_1^2}{2000} + \frac{783 \alpha_1 \alpha_2}{200} + \frac{3401 \alpha_2^2}{80} + \frac{54 \alpha_1 \hat{\lambda}}{25} + \frac{18 \alpha_2 \hat{\lambda}}{5} - \frac{36 \hat{\lambda}^2}{5} - \frac{2827 \alpha_1 \text{tr}\hat{T}}{200} \notag \\
 &  - \frac{471 \alpha_2 \text{tr}\hat{T}}{8} - \frac{116 \alpha_3 \text{tr}\hat{T}}{5} - \frac{1267 \alpha_1 \text{tr}\hat{B}}{200} - \frac{1311 \alpha_2 \text{tr}\hat{B}}{40} - \frac{68 \alpha_3 \text{tr}\hat{B}}{5} - \frac{2529 \alpha_1 \text{tr}\hat{L}}{200} \notag \\
 &  - \frac{1629 \alpha_2 \text{tr}\hat{L}}{40} + \frac{183 \text{tr}\hat{B}^2}{20} + \frac{51 (\text{tr}\hat{B})^2}{10} + \frac{157 \text{tr}\hat{B}\text{tr}\hat{L}}{5} + \frac{261 \text{tr}\hat{L}^2}{20} + \frac{99 (\text{tr}\hat{L})^2}{10} \notag \\
 &  + \frac{3 \text{tr}\hat{T}\hat{B}}{2} + \frac{339 \text{tr}\hat{T}^2}{20} + \frac{177 \text{tr}\hat{T}\text{tr}\hat{B}}{5} + \frac{199 \text{tr}\hat{T}\text{tr}\hat{L}}{5} + \frac{303 (\text{tr}\hat{T})^2}{10} \notag \\
 &  + n_G \bigg[ - \frac{232 \alpha_1^2}{75} - \frac{7 \alpha_1 \alpha_2}{25} + \frac{166 \alpha_2^2}{15} - \frac{548 \alpha_1 \alpha_3}{225} - \frac{4 \alpha_2 \alpha_3}{5} + \frac{1100 \alpha_3^2}{9} \bigg] \notag \\
 &  + n_G^2 \bigg[ - \frac{836 \alpha_1^2}{135} - \frac{44 \alpha_2^2}{15} - \frac{1936 \alpha_3^2}{135} \bigg] \bigg\},
\end{align}
\begin{align}
\beta_2 &=
      \frac{\alpha_2^2}{\lp4\pi\rp^2} \bigg\{ - \frac{86}{3} + \frac{16 n_G}{3} \bigg\} \notag \\
 &  + \frac{\alpha_2^2}{\lp4\pi\rp^3} \bigg\{ \frac{6 \alpha_1}{5} - \frac{518 \alpha_2}{3} - 6 \text{tr}\hat{T} - 6 \text{tr}\hat{B} - 2 \text{tr}\hat{L} + n_G \bigg[ \frac{4 \alpha_1}{5} + \frac{196 \alpha_2}{3} + 16 \alpha_3 \bigg] \bigg\} \notag \\
 &  + \frac{\alpha_2^2}{\lp4\pi\rp^4} \bigg\{ \frac{163 \alpha_1^2}{400} + \frac{561 \alpha_1 \alpha_2}{40} - \frac{667111 \alpha_2^2}{432} + \frac{6 \alpha_1 \hat{\lambda}}{5} + 6 \alpha_2 \hat{\lambda} - 12 \hat{\lambda}^2 - \frac{593 \alpha_1 \text{tr}\hat{T}}{40} \notag \\
 &  - \frac{729 \alpha_2 \text{tr}\hat{T}}{8} - 28 \alpha_3 \text{tr}\hat{T} - \frac{533 \alpha_1 \text{tr}\hat{B}}{40} - \frac{729 \alpha_2 \text{tr}\hat{B}}{8} - 28 \alpha_3 \text{tr}\hat{B} - \frac{51 \alpha_1 \text{tr}\hat{L}}{8} \notag \\
 &  - \frac{243 \alpha_2 \text{tr}\hat{L}}{8} + \frac{57 \text{tr}\hat{B}^2}{4} + \frac{45 (\text{tr}\hat{B})^2}{2} + 15 \text{tr}\hat{B}\text{tr}\hat{L} + \frac{19 \text{tr}\hat{L}^2}{4} + \frac{5 (\text{tr}\hat{L})^2}{2} + \frac{27 \text{tr}\hat{T}\hat{B}}{2} \notag \\
 &  + \frac{57 \text{tr}\hat{T}^2}{4} + 45 \text{tr}\hat{T}\text{tr}\hat{B} + 15 \text{tr}\hat{T}\text{tr}\hat{L} + \frac{45 (\text{tr}\hat{T})^2}{2} \notag \\ 
 &  + n_G \bigg[ - \frac{28 \alpha_1^2}{15} + \frac{13 \alpha_1 \alpha_2}{5} + \frac{25648 \alpha_2^2}{27} - \frac{4 \alpha_1 \alpha_3}{15} + 52 \alpha_2 \alpha_3 + \frac{500 \alpha_3^2}{3} \bigg] \notag \\
 &  + n_G^2 \bigg[ - \frac{44 \alpha_1^2}{45} - \frac{1660 \alpha_2^2}{27} - \frac{176 \alpha_3^2}{9} \bigg]\bigg\}
\end{align}
and
\begin{align}
\beta_3 &=
      \frac{\alpha_3^2}{\lp4\pi\rp^2} \bigg\{ - 44 + \frac{16 n_G}{3} \bigg\} \notag \\
 &  + \frac{\alpha_3^2}{\lp4\pi\rp^3} \bigg\{ - 408 \alpha_3 - 8 \text{tr}\hat{T} - 8 \text{tr}\hat{B} + n_G \bigg[ \frac{22 \alpha_1}{15} + 6 \alpha_2 + \frac{304 \alpha_3}{3} \bigg] \bigg\} \notag \\
 &  + \frac{\alpha_3^2}{\lp4\pi\rp^4} \bigg\{ - 5714 \alpha_3^2 - \frac{101 \alpha_1 \text{tr}\hat{T}}{10} - \frac{93 \alpha_2 \text{tr}\hat{T}}{2} - 160 \alpha_3 \text{tr}\hat{T} - \frac{89 \alpha_1 \text{tr}\hat{B}}{10} - \frac{93 \alpha_2 \text{tr}\hat{B}}{2} \notag \\
 &  - 160 \alpha_3 \text{tr}\hat{B} + 18 \text{tr}\hat{B}^2 + 42 (\text{tr}\hat{B})^2 + 14 \text{tr}\hat{B}\text{tr}\hat{L} - 12 \text{tr}\hat{T}\hat{B} + 18 \text{tr}\hat{T}^2 + 84 \text{tr}\hat{T}\text{tr}\hat{B} \notag \\ 
 &  + 14 \text{tr}\hat{T}\text{tr}\hat{L} + 42 (\text{tr}\hat{T})^2 \notag \\ 
 &  + n_G \bigg[ - \frac{13 \alpha_1^2}{30} - \frac{\alpha_1 \alpha_2}{10} + \frac{241 \alpha_2^2}{6} + \frac{308 \alpha_1 \alpha_3}{45} + 28 \alpha_2 \alpha_3 + \frac{20132 \alpha_3^2}{9} \bigg] \notag \\
 &  + n_G^2 \bigg[ - \frac{242 \alpha_1^2}{135} - \frac{22 \alpha_2^2}{3} - \frac{2600 \alpha_3^2}{27} \bigg] \bigg\}.
\end{align}
In the above formulas $n_G$ denotes the number of fermion generations. It is
obtained by labeling the closed quark and lepton loops present in the
diagrams.

To obtain the results for the three-loop gauge beta functions, one also needs
the one-loop beta functions of the Yukawa couplings,
cf. Eq.~(\ref{eq::renconst_beta}). They can be found in the literature, of
course. Nevertheless we decided to re-calculate them as an additional
check of our setup. For completeness, we present the analytical two-loop
expressions which read
\begin{align}
\beta_{\alpha_{{t}}} =&\ 
   - \epsilon\frac{\alpha_{{t}}}{\pi}
   +  \frac{\alpha_{{t}}}{\left(4\pi\right)^2}
     \lbc - 6 \alpha_{{b}}
     + 6 \alpha_{{t}}
     + 4 \text{tr}\hat{L} + 12 \text{tr}\hat{B} + 12 \text{tr}\hat{T}
     - \frac{17}{5}\alpha_1
     - 9 \alpha_2
     - 32 \alpha_3
     \rbc 
    \notag \\ &
+ \frac{\alpha_t}{\left(4\pi\right)^3} 
     \lbc
 \frac{9 \alpha_1^2}{50} 
     - \frac{9 \alpha_1 \alpha_2}{5} 
     - 35 \alpha_2^2 
     + \frac{76 \alpha_1 \alpha_3}{15} 
     + 36 \alpha_2 \alpha_3
     - \frac{1616 \alpha_3^2}{3} 
     + n_G  \lbk \frac{116 \alpha_1^2}{45} \ra \ra \notag \\ &
 \la  + 4 \alpha_2^2  + \frac{320 \alpha_3^2}{9}  \rbk 
     + 24 \hat{\lambda}^2 
     + \frac{393 \alpha_1 \alpha_t}{20} 
     + \frac{225 \alpha_2 \alpha_t}{4} 
     + 144 \alpha_3 \alpha_t
     - 48 \hat{\lambda} \alpha_t
     - 48 \alpha_t^2 
     \notag \\ &
 +\frac{7 \alpha_1 \alpha_b}{20} 
     + \frac{99 \alpha_2 \alpha_b}{4} 
     + 16 \alpha_3 \alpha_b
     - 11 \alpha_t \alpha_b
     - \alpha_b^2 
     + \frac{15 \alpha_1 \alpha_{\tau}}{2} 
     + \frac{15 \alpha_2 \alpha_{\tau}}{2} 
     - 9 \alpha_t \alpha_{\tau} \notag \\ &
     + 5 \alpha_b \alpha_{\tau}
     - 9 \alpha_{\tau}^2 
     \bigg\}
,
\label{eq::betaalphat}
\end{align}
\begin{align}
\beta_{\alpha_{{b}}} =&\ 
   - \epsilon\frac{\alpha_{{b}}}{\pi}
   +  \frac{\alpha_{{b}}}{\left(4\pi\right)^2}
     \lbc 6 \alpha_{{b}}
     - 6 \alpha_{{t}}
     + 4 \text{tr}\hat{L} + 12 \text{tr}\hat{B} + 12 \text{tr}\hat{T}
     - \alpha_1
     - 9 \alpha_2
     - 32 \alpha_3
     \rbc 
    \notag \\ &
   + \frac{\alpha_b}{\left(4\pi\right)^3} 
     \lbc \frac{ - 29 \alpha_1^2}{50} \ra
     - \frac{27 \alpha_1 \alpha_2}{5} 
     - 35 \alpha_2^2 
     + \frac{124 \alpha_1 \alpha_3}{15} 
     + 36 \alpha_2 \alpha_3
     - \frac{1616 \alpha_3^2}{3} \notag \\ &
+ n_G  \lbk \frac{ - 4 \alpha_1^2}{45}  + 4 \alpha_2^2  
+  \frac{320 \alpha_3^2}{9}  \rbk 
     + 24 \hat{\lambda}^2 
     + \frac{91 \alpha_1 \alpha_t}{20} 
     + \frac{99 \alpha_2 \alpha_t}{4} 
     + 16 \alpha_3 \alpha_t
     - \alpha_t^2 
     + \frac{237 \alpha_1 \alpha_b}{20} 
      \notag \\ &
 +\frac{225 \alpha_2 \alpha_b}{4} 
     + 144 \alpha_3 \alpha_b
     - 48 \hat{\lambda} \alpha_b
     - 11 \alpha_t \alpha_b
     - 48 \alpha_b^2 
     + \frac{15 \alpha_1 \alpha_{\tau}}{2} 
     + \frac{15 \alpha_2 \alpha_{\tau}}{2} 
     + 5 \alpha_t \alpha_{\tau}
      \notag \\ &
 - 9 \alpha_b \alpha_{\tau}
     - 9 \alpha_{\tau}^2 
     \bigg\}
\,,
\label{eq::betaalphab}
\end{align}
and
\begin{align}
\beta_{\alpha_{\tau}} =&\ 
   - \epsilon\frac{\alpha_{\tau}}{\pi}
   +  \frac{\alpha_{\tau}}{\left(4\pi\right)^2}
     \lbc 6 \alpha_{\tau}
     + 4 \text{tr}\hat{L} + 12 \text{tr}\hat{B} + 12 \text{tr}\hat{T}
     - 9 \alpha_1
     - 9 \alpha_2
     \rbc 
    \notag \\ &
   + \frac{\alpha_{\tau}}{\left(4\pi\right)^3}
     \lbc \frac{51 \alpha_1^2}{50} \ra
     + \frac{27 \alpha_1 \alpha_2}{5} 
     - 35 \alpha_2^2 
     + n_G  \lbk \frac{44 \alpha_1^2}{5}  + 4 \alpha_2^2  \rbk 
     + 24 \hat{\lambda}^2 
     + \frac{17 \alpha_1 \alpha_t}{2} 
      \notag \\ &
     + \frac{45 \alpha_2 \alpha_t}{2} 
     + 80 \alpha_3 \alpha_t
     - 27 \alpha_t^2 
     + \frac{5 \alpha_1 \alpha_b}{2} 
     + \frac{45 \alpha_2 \alpha_b}{2} 
     + 80 \alpha_3 \alpha_b
     + 6 \alpha_t \alpha_b
     - 27 \alpha_b^2 
     \notag \\ &
     + \frac{537 \alpha_1 \alpha_{\tau}}{20} 
 +\la \frac{165 \alpha_2 \alpha_{\tau}}{4} 
     - 48 \hat{\lambda} \alpha_{\tau}
     - 27 \alpha_t \alpha_{\tau}
     - 27 \alpha_b \alpha_{\tau}
     - 12 \alpha_{\tau}^2 
     \rbc
\,.
\end{align}
The one-loop results have been expressed in terms of Yukawa matrices since
these expressions enter the three-loop beta functions. At two-loop order,
however, we refrain from reconstructing the general expression which would
require an extension of the rules given in Eq.~(\ref{eq::nh_to_tr}).

Our independent calculation of the two-loop Yukawa beta functions is
also interesting as there has been a discrepancy between~\cite{Machacek:1983fi}
and~\cite{Luo:2002ey} concerning the absence of terms proportional to
$\alpha_b\alpha_t\hat{\lambda}$ in Eqs.~(\ref{eq::betaalphat})
and~(\ref{eq::betaalphab}).
We were able to confirm the results
of Ref.~\cite{Luo:2002ey}.

In Appendices~\ref{app::renconst} and~\ref{app::renconst2} we provide the
results for the renormalization 
constants which lead to the beta functions discussed in this Section.


\section{\label{sec::checks}Checks}

We have successfully performed a number of consistency checks and compared our
results with those
already available in the literature.  We describe these
checks in detail in this Section.

The consistency checks show that all computed renormalization constants are
local (i.e. there are no $\ln\mu$ terms in the final
expression), that the renormalization constants of the gauge couplings are
gauge parameter independent and that the beta functions are finite. We also
find that the beta functions calculated by considering different vertices in
Lorenz gauge agree among themselves and with the results of the computation in
BFG.

In order to test that the program \verb|FeynArtsToQ2E| correctly translates
\verb|FeynArts| model files into model files for \verb|QGRAF|/\verb|q2e|, we
reproduced the beta function for the Higgs self-coupling to one-loop order and
the beta functions for the top and bottom quark, and the tau lepton Yukawa
couplings to two-loop order (cf. previous Section). We have even considered
quantities within the Minimal Supersymmetric Standard Model, like the relation
between the squark masses within one generation, which is quite involved in
case electroweak interactions are kept non-zero. Furthermore, we find that the
divergent loop corrections to the $BBB$ vertex vanish in the Lorenz
gauge, as expected 
since for this vertex no renormalization is required. 
We performed the latter check up to three-loop order.

Another check consists in verifying
that in the vertex diagrams no infra-red
divergences are introduced although one external momentum is set to zero.  We
do not have to consider two-point functions since they are infra-red safe.
One can avoid infra-red divergences by introducing a common mass for the
internal particles. Afterwards, the resulting integrals are evaluated in the
limit $q^2\gg m^2$ where $q$ is the non-vanishing external momentum of the
vertex diagrams.  This is conveniently done by applying the rules of
asymptotic expansion~\cite{Smirnov:2002pj} which are encoded in the program
\verb|exp|. The setup described in
Section~\ref{sub::setup} is particularly useful for this test
since \verb|exp| takes over
the task of generating {\tt FORM} code for all relevant sub-diagrams which can
be up to 35 for some of the diagrams;
this  makes the calculation significantly
more complex. In our case the asymptotic expansion either leads to massless
two-point functions or massive vacuum integrals or a combination of both. The
former are computed with the help of the package
\verb|MINCER|, for the latter
the  package
\verb|MATAD| is used.  As a result one obtains a
series in $m^2/q^2$ where the coefficients contain numbers and $\ln(m^2/q^2)$
terms. For our purpose it is sufficient to restrict ourselves to the term
$(m^2/q^2)^0$ and check that no logarithms appear in the final result.  With this
method we have explicitly checked that the $W_1 W_2 W_3$ and three-gluon
vertices are free from infra-red divergences.  Since the results for the
  gauge coupling renormalization constants agree with the
ones obtained from the other vertices also the latter are infra-red safe.

Let us finally comment on the comparison of our findings with the literature.
We have successfully compared our results for the two-loop gauge beta
functions with~\cite{Machacek:1983tz} and the two-loop Yukawa beta functions
with~\cite{Luo:2002ey}.  Even partial results for the three-loop gauge beta
functions were available in the literature~\cite{Pickering:2001aq}.  That
paper comprises all hitherto known three-loop corrections in a general quantum
field theory based on a single gauge group, however, the presentation of the
results relies on a quite intricate notation. Its specification to the SM is
straightforward, however, a bit tedious.  For convenience, the translation
rules needed to convert the notation of~\cite{Pickering:2001aq} to ours is
given in Appendix~\ref{app::compare}.

After converting the notation of Ref.~\cite{Pickering:2001aq} to ours we find
complete agreement, taking into account the following modifications in
Eq.~(33) of~\cite{Pickering:2001aq}:\footnote{We want to thank the authors
  of~\cite{Pickering:2001aq} for pointing out the issue of symmetrization and
  for assistance in deriving the translation rules.}  The terms
\begin{eqnarray}
  &&-\frac{7 g^2 \Tr \left(Y^a \bar{Y}^b\right)
    \Tr\left(\bar{Y}^b Y^a C(R)\right)}{12r}\nonumber\\
  &&+\frac{g^2 \Tr \left(Y^a \bar{Y}^b\right)
    \Tr\left(Y^b \bar{Y}^c\right)C(S)_{ca}}{12r}
\end{eqnarray}
have to be ``symmetrized'', so that they read
\begin{eqnarray}
  &&\frac{1}{2} \left(
    -\frac{7 g^2 \Tr \left(Y^a \bar{Y}^b\right)
      \Tr\left(\bar{Y}^b Y^a C(R)\right)}{12r}
    -\frac{7 g^2 \Tr \left(Y^a \bar{Y}^b\right)
      \Tr\left(Y^b \bar{Y}^a C(R)\right)}{12r}
  \right) \nonumber\\
  &+&\frac{1}{2} \left(
    +\frac{g^2 \Tr \left(Y^a \bar{Y}^b\right)
      \Tr\left(Y^b \bar{Y}^c\right)C(S)_{ca}}{12r}
    +\frac{g^2 \Tr \left(Y^a \bar{Y}^b\right)
      \Tr\left(\bar{Y}^b Y^c\right)C(S)_{ca}}{12r}
  \right)\,.
  \label{eq::trans}
\end{eqnarray}
Furthermore, one has to correct the obvious misprint
\begin{equation}
  +\frac{5 g^4 \Tr \left(C(R)^2 \bar{Y}^a Y^b\right)}{12r}\ 
  \rightarrow\ 
  +\frac{5 g^4 \Tr \left(C(R)^2 \bar{Y}^a Y^a\right)}{12r}\,.
\end{equation}


\section{\label{sec::num}Numerical Analysis}

In this Section we discuss the numerical effect of the new contributions to
the gauge beta functions.  We solve the corresponding renormalization group
equations of the gauge couplings numerically and take into account the 
contributions from the Yukawa couplings and the Higgs self-coupling to
two-loops. As boundary conditions we choose
\begin{align}\label{eq::values}
  \alpha_1^{\msbarformel}\left(M_{{Z}}\right) &= 0.0169225 \pm 0.0000039\,,\notag\\
  \alpha_2^{\msbarformel}\left(M_{{Z}}\right) &= 0.033735 \pm 0.000020\,,\notag\\
  \alpha_3^{\msbarformel}\left(M_{{Z}}\right) &= 0.1173 \pm 0.00069\,,\notag\\
  \alpha_{{t}}^{\msbarformel}\left(M_{{Z}}\right) &= 0.07514\,,\notag\\
  \alpha_{{b}}^{\msbarformel}\left(M_{{Z}}\right) &= 0.00002064\,,\notag\\
  \alpha_{\tau}^{\msbarformel}\left(M_{{Z}}\right) &=
  8.077\cdot10^{-6}\,,
  \notag\\
  4\pi\hat{\lambda} &= 0.13\,,
\end{align}
where the first six entries correspond to experimentally determined values
while the value for the Higgs coupling is determined assuming a Higgs boson
with mass 125 GeV:
\begin{equation}
  4\pi\hat{\lambda} = 
  \frac{m_{{H}}^2}{2 v^2} \approx 
  \frac{125^2\,\textrm{GeV}^2}
  {2\cdot\left(\sqrt{2}\cdot174\right)^2\,\textrm{GeV}^2} 
  \approx 0.13\,.
  \label{eq::lamhat}
\end{equation}
Since $\hat{\lambda}$ only occurs at three loops the tree-level
relation~(\ref{eq::lamhat}) is sufficient for our purpose.
Note that the values in Eq.~(\ref{eq::values}) are given in
the full SM, the top quark being not integrated out. For a description how
these values are determined from the knowledge of their directly measured
counterparts~\cite{Nakamura:2010zzi}, we refer
to~\cite{Martens:2010nm,Nakamura:2010zzi}.

It is noteworthy that for all three gauge couplings the sum of
all three-loop terms
involving at least one of the couplings $\alpha_b$,
$\alpha_\tau$ or $\lambda$ leads to corrections which are less than $0.1\%$
of the difference between the two- and three-loop prediction.

\begin{figure}
  \begin{center}
  \includegraphics[width=0.8\linewidth]{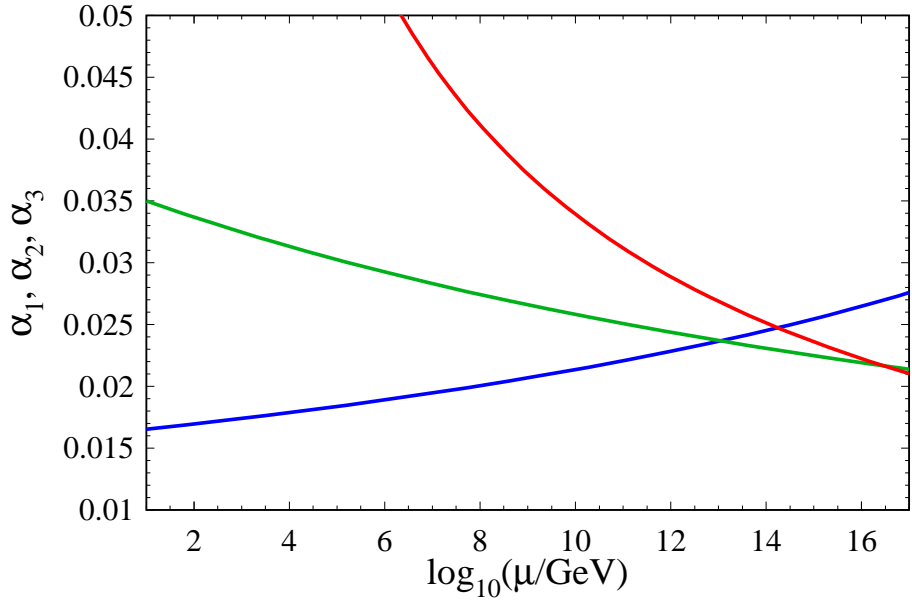}
  \caption{The running of the gauge couplings at three loops. The curve with
    the smallest initial value corresponds to $\alpha_1$, the middle curve to
    $\alpha_2$, and the curve with the highest initial value to
    $\alpha_3$.}\label{fig::run}
  \end{center}
\end{figure}

In Fig.~\ref{fig::run} the running of the couplings $\alpha_1$, $\alpha_2$ and
$\alpha_3$, is shown from $\mu=M_Z$ up to high energies. At this scale no
difference between one, two and three loops is visible, all curves lie on top
of each other.

\begin{figure}
  \begin{center}
  \includegraphics[width=0.8\linewidth]{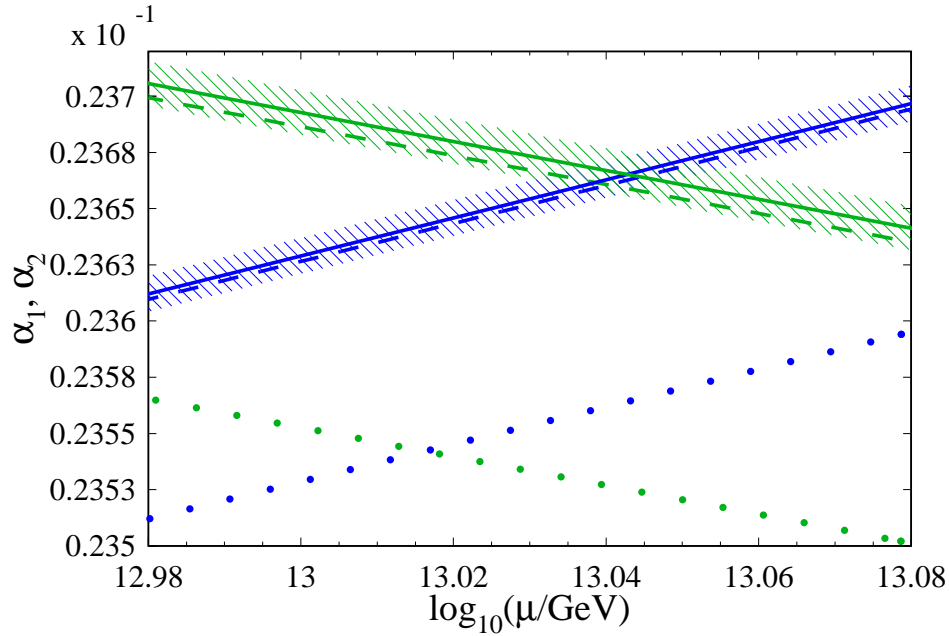}
  \caption{The running of the electroweak gauge couplings in the SM. The lines
    with positive slope correspond to $\alpha_1$, the lines with negative
    slope to $\alpha_2$.  The dotted, dashed and solid lines correspond to
    one-, two- and three-loop precision, respectively.  The bands around the
    three-loop curves visualize the experimental
    uncertainty.}\label{fig::run2}
  \end{center}
\end{figure}

The differences between the loop orders can be seen in Fig.~\ref{fig::run2}
which magnifies the intersection point between $\alpha_1$ and $\alpha_2$.
There is a clear jump between the one- (dotted) and two-loop (dashed)
prediction. The difference between two and three loops (solid curves) is
significantly smaller which implies that perturbation theory converges
very well. 

The experimental uncertainties for $\alpha_1(M_Z)$ and $\alpha_2(M_Z)$ as
given in Eq.~(\ref{eq::values}) are reflected by the bands around the
three-loop results. Defining the difference between the two- and three-loop
result as theoretical uncertainty one observes that it is smaller than the
experimental one, however, of the same order of magnitude.
Without the new three-loop calculation performed in this paper the theory
uncertainty is much larger than the experimental one.

\begin{figure}
  \begin{center}
  \includegraphics[width=0.8\linewidth]{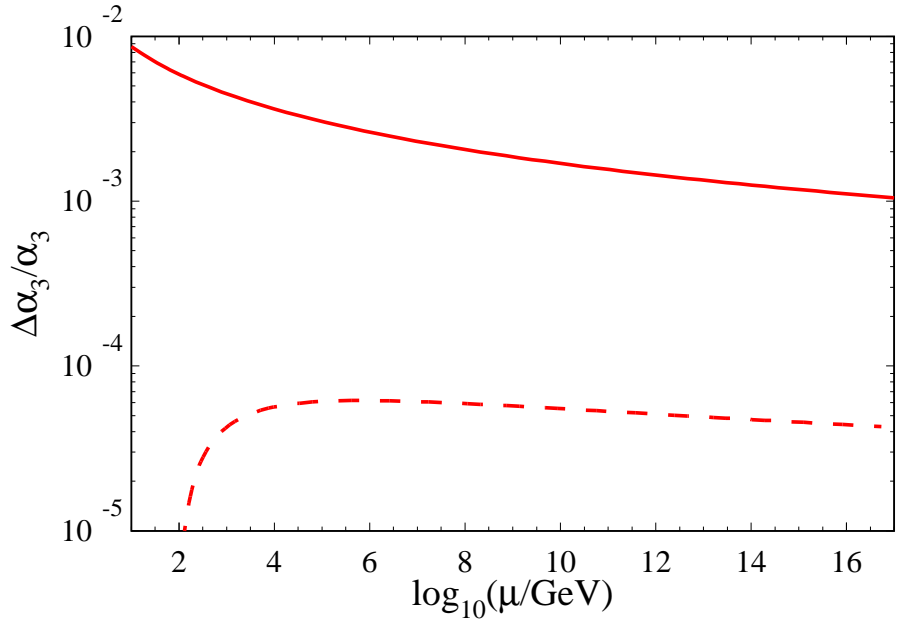}
  \caption{Comparison of the relative experimental and theoretical
    uncertainty of $\alpha_3$. The theoretical uncertainty is given by the
    dashed line, the solid curve corresponds to the experimental
    one.}\label{fig::alpha3}
  \end{center}
\end{figure}

Also in the case of $\alpha_3$ perturbation theory seems to converge
well. However, in contrast to $\alpha_1$ and $\alpha_2$ the experimental
uncertainty turns out to be much larger than 
the theoretical uncertainty.  This is
not surprising as the relative experimental uncertainty of $\alpha_3$ at the
electroweak scale is quite large compared to its electroweak counterparts.
The relative experimental and theoretical uncertainty of $\alpha_3$ is plotted
as a function of the renormalization scale in Fig.~\ref{fig::alpha3}.
Note that by construction we have that $\Delta\alpha_3/\alpha_3|_{\rm theory}$
approaches zero for $\mu\to M_Z$.

Let us finally identify the numerically most important contributions.  
This is done by running from $\mu=M_Z$ to $\mu=10^{16}$~GeV and by comparing
the contribution of each individual term to the total difference between
the two- and three-loop prediction. Similar
results are also obtained for lower scales.

About
90\% of the three-loop corrections to the running of the gauge couplings
arises from only a few terms. In the case of $\alpha_1$ there is only one term
which dominates, namely the one of order $\alpha_1^2\alpha_3^2$. For
$\alpha_2$ one has a contribution of $+56\%$ from the
$\mathcal{O}(\alpha_2^2\alpha_3^2)$ term, $+13\%$ from order
$\alpha_2^3\alpha_3$ and $+37\%$ from $\mathcal{O}(\alpha_2^4)$. All other
terms contribute at most 5\% and partly also cancel each other. Except for the
term of $\mathcal{O}(\alpha_2^4)$
all these terms are presented in this paper for the first
time.  

The beta function $\beta_3$ is dominated by the strong corrections,
however, large cancellations between the $\alpha_3^4$ ($+137\%$),
$\alpha_3^3\alpha_2$ ($+45\%$), $\alpha_3^2\alpha_t^2$ ($+28\%$)
and $\alpha_3^2\alpha_2^2$ ($+17\%$) terms on
the one hand and the $\alpha_3^3\alpha_t$ ($-112\%$) and
$\alpha_3^2\alpha_2\alpha_t$ ($-16\%$) on the other hand are observed.
It is worth noting that the four-loop term of order $\alpha_s^5$ amounts
to $-58\%$ of the difference between the two- and three-loop predictions.
This number has been obtained by adding the four-loop QCD
term~\cite{vanRitbergen:1997va,Czakon:2004bu} to $\beta_3$.


\section{\label{sec::concl}Conclusions}

In this paper we present three-loop results for renormalization constants that
are used to compute the three SM gauge coupling beta functions,
taking into account all contributions.
We have checked that our expressions agree with all partial results present in
the literature. Furthermore the two-loop corrections to the Yukawa
couplings have been computed. We have performed the calculation using both
Lorenz gauge within the unbroken phase of the SM and 
BFG in the broken phase. Our final result is valid for a generic flavour
structure with an arbitrary CKM matrix. It is furthermore sufficiently general
to consider a fourth generation of quarks and leptons.

In order to perform the calculation in an automated way we have written an
interface, {\tt FeynArtsToQ2E}, between the package {\tt FeynArts} and our chain of
programs ({\tt QGRAF}, {\tt q2e}, {\tt exp}, {\tt MATAD}, {\tt MINCER})
allowing to handle the ${\cal O}(10^6)$ Feynman diagrams, which have to be
considered in the course of the calculation, in an effective way.
Thus, we could
perform several checks involving various different Green's functions.
{\tt FeynArtsToQ2E} is not limited to the SM but can easily be used for
extensions like supersymmetric models.


\begin{acknowledgments}
  We thank David Kunz for tests concerning \verb|FeynArtsToQ2E| and 
  Miko{\l}aj Misiak for suggesting to generalize the Yukawa structure of the
  final result. 
  We thank Esben M\o{}lgaard for communications concerning Eq.~(\ref{eq::trans}).
  This work is
  supported by DFG through SFB/TR 9 ``Computational Particle Physics''.
\end{acknowledgments}


\appendix


\section{\label{app::renconst}Renormalization constants}

In this Appendix we present analytical results for the renormalization constants
$Z_{\alpha_1}$,
$Z_{\alpha_2}$, 
$Z_{\alpha_3}$, 
$Z_B$, 
$Z_W$, 
$Z_G$, 
$Z_H$, 
$Z_{c_W}$, 
$Z_{c_G}$, 
$Z_{cCW}$, 
$Z_{cCG}$, 
$Z_{WWW}$, 
$Z_{GGG}$, 
$Z_{HHW}$, 
where the
definition of $Z_{\alpha_i}$ is given in Eq.~(\ref{eq::alpha_bare}) and 
the field renormalization is  defined through
\begin{align}
  && B^{\rm bare} = Z_B B\,, && W^{\rm bare} = Z_W W\,, 
  && G^{\rm bare} = Z_G G\,,\nonumber\\
  && c_W^{\rm bare} = Z_{c_W} c_W\,, && c_G^{\rm bare} = Z_{c_G} c_G\,,
  && H^{\rm bare} = Z_H H\,.
\end{align}
$B$, $W$, $G$, $c_W$ and $c_G$ denote the gauge boson and
ghost fields. The scalar field $H$ is defined in Eq.~(\ref{Higgsdublett}).
The renormalization constants for
the three-particle vertices are also defined in a multiplicative way.

Some of the results listed below contain the gauge parameters $\xi_B$, 
$\xi_W$ and
$\xi_G$. They are conveniently defined via the corresponding gauge boson
propagator which is given by
\begin{eqnarray}
  D_X^{\mu\nu}(q) &=& 
  i \frac{-g^{\mu\nu} + \left(1-\xi_X\right) \frac{q^\mu q^\nu}{q^2}}{q^2}
  \,,
\end{eqnarray}
with $X=B,\,W,\,G$. Note that $\xi_X=1$ corresponds to Feynman and $\xi_X=0$
to Landau gauge.
Our analytical results read
\begin{align}
Z_{\alpha_1} &=
  1 + \frac{\alpha_1}{4\pi} \frac{1}{\epsilon} \bigg\{ \frac{1}{10} + \frac{4 n_G}{3} \bigg\} + \frac{\alpha_1}{\left(4\pi\right)^2} \bigg\{ \frac{1}{\epsilon^2} \bigg[ \frac{\alpha_1}{100} + \frac{4 n_G \alpha_1}{15} + \frac{16 n_G^2 \alpha_1}{9} \bigg] \notag \\
 &  + \frac{1}{\epsilon} \bigg[ \frac{9 \alpha_1}{100} + \frac{9 \alpha_2}{20} - \frac{17 \text{tr}\hat{T}}{20} - \frac{\text{tr}\hat{B}}{4} - \frac{3 \text{tr}\hat{L}}{4} + n_G \bigg( \frac{19 \alpha_1}{30} + \frac{3 \alpha_2}{10} + \frac{22 \alpha_3}{15} \bigg) \bigg] \bigg\} \notag \\
 &  + \frac{\alpha_1}{\left(4\pi\right)^3} \bigg\{ \frac{1}{\epsilon^3} \bigg[ \frac{\alpha_1^2}{1000} + \frac{n_G \alpha_1^2}{25} + \frac{8 n_G^2 \alpha_1^2}{15} + \frac{64 n_G^3 \alpha_1^2}{27} \bigg] \notag \\
 &  + \frac{1}{\epsilon^2} \bigg[ \frac{21 \alpha_1^2}{1000} + \frac{9 \alpha_1 \alpha_2}{100} - \frac{43 \alpha_2^2}{40} + \frac{17 \alpha_1 \text{tr}\hat{T}}{240} + \frac{51 \alpha_2 \text{tr}\hat{T}}{80} + \frac{34 \alpha_3 \text{tr}\hat{T}}{15} - \frac{7 \alpha_1 \text{tr}\hat{B}}{240} \notag \\
 &  + \frac{3 \alpha_2 \text{tr}\hat{B}}{16} + \frac{2 \alpha_3 \text{tr}\hat{B}}{3} + \frac{33 \alpha_1 \text{tr}\hat{L}}{80} + \frac{9 \alpha_2 \text{tr}\hat{L}}{16} - \frac{\text{tr}\hat{B}^2}{8} - \frac{(\text{tr}\hat{B})^2}{4} - \frac{5 \text{tr}\hat{B}\text{tr}\hat{L}}{6}\notag \\
 &   - \frac{3 \text{tr}\hat{L}^2}{8} - \frac{(\text{tr}\hat{L})^2}{4} + \frac{11 \text{tr}\hat{T}\hat{B}}{20} - \frac{17 \text{tr}\hat{T}^2}{40} - \frac{11 \text{tr}\hat{T}\text{tr}\hat{B}}{10} - \frac{31 \text{tr}\hat{T}\text{tr}\hat{L}}{30} - \frac{17 (\text{tr}\hat{T})^2}{20} \notag \\
 &  + n_G \bigg( \frac{77 \alpha_1^2}{180} + \frac{63 \alpha_1 \alpha_2}{50} - \frac{31 \alpha_2^2}{60} + \frac{22 \alpha_1 \alpha_3}{75} - \frac{242 \alpha_3^2}{45} - \frac{34 \alpha_1 \text{tr}\hat{T}}{15} - \frac{2 \alpha_1 \text{tr}\hat{B}}{3} - 2 \alpha_1 \text{tr}\hat{L} \bigg) \notag \\
 &  + n_G^2 \bigg( \frac{266 \alpha_1^2}{135} + \frac{4 \alpha_1 \alpha_2}{5} + \frac{2 \alpha_2^2}{15} + \frac{176 \alpha_1 \alpha_3}{45} + \frac{88 \alpha_3^2}{135} \bigg) \bigg] \notag \\
 &  + \frac{1}{\epsilon} \bigg[ \frac{163 \alpha_1^2}{8000} + \frac{261 \alpha_1 \alpha_2}{800} + \frac{3401 \alpha_2^2}{960} + \frac{9 \alpha_1 \hat{\lambda}}{50} + \frac{3 \alpha_2 \hat{\lambda}}{10} - \frac{3 \hat{\lambda}^2}{5} - \frac{2827 \alpha_1 \text{tr}\hat{T}}{2400} - \frac{157 \alpha_2 \text{tr}\hat{T}}{32} \notag \\
 &  - \frac{29 \alpha_3 \text{tr}\hat{T}}{15} - \frac{1267 \alpha_1 \text{tr}\hat{B}}{2400} - \frac{437 \alpha_2 \text{tr}\hat{B}}{160} - \frac{17 \alpha_3 \text{tr}\hat{B}}{15} - \frac{843 \alpha_1 \text{tr}\hat{L}}{800} - \frac{543 \alpha_2 \text{tr}\hat{L}}{160} \notag \\
 &  + \frac{61 \text{tr}\hat{B}^2}{80} + \frac{17 (\text{tr}\hat{B})^2}{40} + \frac{157 \text{tr}\hat{B}\text{tr}\hat{L}}{60} + \frac{87 \text{tr}\hat{L}^2}{80} + \frac{33 (\text{tr}\hat{L})^2}{40} + \frac{\text{tr}\hat{T}\hat{B}}{8} + \frac{113 \text{tr}\hat{T}^2}{80} \notag \\
 &  + \frac{59 \text{tr}\hat{T}\text{tr}\hat{B}}{20} + \frac{199 \text{tr}\hat{T}\text{tr}\hat{L}}{60} + \frac{101 (\text{tr}\hat{T})^2}{40} \notag \\
 &  + n_G \bigg( - \frac{58 \alpha_1^2}{225} - \frac{7 \alpha_1 \alpha_2}{300} + \frac{83 \alpha_2^2}{90} - \frac{137 \alpha_1 \alpha_3}{675} - \frac{\alpha_2 \alpha_3}{15} + \frac{275 \alpha_3^2}{27} \bigg) \notag \\
 &  + n_G^2 \bigg( - \frac{209 \alpha_1^2}{405} - \frac{11 \alpha_2^2}{45} - \frac{484 \alpha_3^2}{405} \bigg) \bigg] \bigg\}
\,,
\end{align}
\begin{align}
Z_{\alpha_2} &=
  1 + \frac{\alpha_2}{4\pi} \frac{1}{\epsilon} \bigg\{ - \frac{43}{6} + \frac{4 n_G}{3} \bigg\} + \frac{\alpha_2}{\lp4\pi\rp^2} \bigg\{ \frac{1}{\epsilon^2} \bigg[ \frac{1849 \alpha_2}{36} - \frac{172 n_G \alpha_2}{9} + \frac{16 n_G^2 \alpha_2}{9} \bigg] \notag \\
 &  + \frac{1}{\epsilon} \bigg[ \frac{3 \alpha_1}{20} - \frac{259 \alpha_2}{12} - \frac{3 \text{tr}\hat{T}}{4} - \frac{3 \text{tr}\hat{B}}{4} - \frac{\text{tr}\hat{L}}{4} + n_G \bigg( \frac{\alpha_1}{10} + \frac{49 \alpha_2}{6} + 2 \alpha_3 \bigg) \bigg] \bigg\} \notag \\
 &  + \frac{\alpha_2}{\lp4\pi\rp^3} \bigg\{ \frac{1}{\epsilon^3} \bigg[ - \frac{79507 \alpha_2^2}{216} + \frac{1849 n_G \alpha_2^2}{9} - \frac{344 n_G^2 \alpha_2^2}{9} + \frac{64 n_G^3 \alpha_2^2}{27} \bigg] \notag \\
 &  + \frac{1}{\epsilon^2} \bigg[ \frac{\alpha_1^2}{200} - \frac{43 \alpha_1 \alpha_2}{20} + \frac{77959 \alpha_2^2}{216}  + \frac{17 \alpha_1 \text{tr}\hat{T}}{80} + \frac{181 \alpha_2 \text{tr}\hat{T}}{16} + 2 \alpha_3 \text{tr}\hat{T} + \frac{\alpha_1 \text{tr}\hat{B}}{16} \notag \\
 &  + \frac{181 \alpha_2 \text{tr}\hat{B}}{16} + 2 \alpha_3 \text{tr}\hat{B} + \frac{3 \alpha_1 \text{tr}\hat{L}}{16} + \frac{181 \alpha_2 \text{tr}\hat{L}}{48} - \frac{3 \text{tr}\hat{B}^2}{8} - \frac{3 (\text{tr}\hat{B})^2}{4} - \frac{\text{tr}\hat{B}\text{tr}\hat{L}}{2} \notag \\
 &  - \frac{(\text{tr}\hat{L})^2}{12} + \frac{3 \text{tr}\hat{T}\hat{B}}{4} - \frac{3 \text{tr}\hat{T}^2}{8} - \frac{3 \text{tr}\hat{T}\text{tr}\hat{B}}{2} - \frac{\text{tr}\hat{T}\text{tr}\hat{L}}{2} - \frac{3 (\text{tr}\hat{T})^2}{4} - \frac{\text{tr}\hat{L}^2}{8} \notag \\
 &  + n_G \bigg( \frac{7 \alpha_1^2}{100} - \frac{31 \alpha_1 \alpha_2}{30} - \frac{22001 \alpha_2^2}{108} - \frac{86 \alpha_2 \alpha_3}{3} - \frac{22 \alpha_3^2}{3} - 2 \alpha_2 \text{tr}\hat{T} - 2 \alpha_2 \text{tr}\hat{B} - \frac{2 \alpha_2 \text{tr}\hat{L}}{3} \bigg) \notag \\
 &  + n_G^2 \bigg( \frac{2 \alpha_1^2}{45} + \frac{4 \alpha_1 \alpha_2}{15} + \frac{686 \alpha_2^2}{27} + \frac{16 \alpha_2 \alpha_3}{3} + \frac{8 \alpha_3^2}{9} \bigg) \bigg] \notag \\
 &  + \frac{1}{\epsilon} \bigg[ \frac{163 \alpha_1^2}{4800} + \frac{187 \alpha_1 \alpha_2}{160} - \frac{667111 \alpha_2^2}{5184} + \frac{\alpha_1 \hat{\lambda}}{10} + \frac{\alpha_2 \hat{\lambda}}{2} - \hat{\lambda}^2 - \frac{593 \alpha_1 \text{tr}\hat{T}}{480} - \frac{243 \alpha_2 \text{tr}\hat{T}}{32} \notag \\
 &  - \frac{7 \alpha_3 \text{tr}\hat{T}}{3} - \frac{533 \alpha_1 \text{tr}\hat{B}}{480} - \frac{243 \alpha_2 \text{tr}\hat{B}}{32} - \frac{7 \alpha_3 \text{tr}\hat{B}}{3} - \frac{17 \alpha_1 \text{tr}\hat{L}}{32} - \frac{81 \alpha_2 \text{tr}\hat{L}}{32} + \frac{19 \text{tr}\hat{B}^2}{16} \notag \\
 &  + \frac{15 (\text{tr}\hat{B})^2}{8} + \frac{5 \text{tr}\hat{B}\text{tr}\hat{L}}{4} + \frac{19 \text{tr}\hat{L}^2}{48} + \frac{5 (\text{tr}\hat{L})^2}{24} + \frac{9 \text{tr}\hat{T}\hat{B}}{8} + \frac{19 \text{tr}\hat{T}^2}{16} + \frac{15 \text{tr}\hat{T}\text{tr}\hat{B}}{4} \notag \\
 &  + \frac{5 \text{tr}\hat{T}\text{tr}\hat{L}}{4} + \frac{15 (\text{tr}\hat{T})^2}{8} + n_G \bigg( - \frac{7 \alpha_1^2}{45} + \frac{13 \alpha_1 \alpha_2}{60} + \frac{6412 \alpha_2^2}{81} - \frac{\alpha_1 \alpha_3}{45} + \frac{13 \alpha_2 \alpha_3}{3} + \frac{125 \alpha_3^2}{9} \bigg) \notag \\
 &  + n_G^2 \bigg( - \frac{11 \alpha_1^2}{135} - \frac{415 \alpha_2^2}{81} -
 \frac{44 \alpha_3^2}{27} \bigg) \bigg] \bigg\}
\,,
\end{align}
\begin{align}
Z_{\alpha_3} &=
  1 + \frac{\alpha_3}{4\pi} \frac{1}{\epsilon} \bigg\{ - 11 + \frac{4 n_G}{3} \bigg\} + \frac{\alpha_3}{\lp4\pi\rp^2} \bigg\{ \frac{1}{\epsilon^2} \bigg[ 121 \alpha_3 - \frac{88 n_G \alpha_3}{3} + \frac{16 n_G^2 \alpha_3}{9} \bigg] \notag \\
 &  + \frac{1}{\epsilon} \bigg[ - 51 \alpha_3 - \text{tr}\hat{T} - \text{tr}\hat{B} + n_G \bigg( \frac{11 \alpha_1}{60} + \frac{3 \alpha_2}{4} + \frac{38 \alpha_3}{3} \bigg) \bigg] \bigg\} \notag \\
 &  + \frac{\alpha_3}{\lp4\pi\rp^3} \bigg\{ \frac{1}{\epsilon^3} \bigg[ - 1331 \alpha_3^2 + 484 n_G \alpha_3^2 - \frac{176 n_G^2 \alpha_3^2}{3} + \frac{64 n_G^3 \alpha_3^2}{27} \bigg] \notag \\
 &  + \frac{1}{\epsilon^2} \bigg[ 1309 \alpha_3^2 + \frac{17 \alpha_1 \text{tr}\hat{T}}{60} + \frac{3 \alpha_2 \text{tr}\hat{T}}{4} + \frac{74 \alpha_3 \text{tr}\hat{T}}{3} + \frac{\alpha_1 \text{tr}\hat{B}}{12} + \frac{3 \alpha_2 \text{tr}\hat{B}}{4} + \frac{74 \alpha_3 \text{tr}\hat{B}}{3} \notag \\
 &  - \frac{\text{tr}\hat{B}^2}{2} - (\text{tr}\hat{B})^2 - \frac{\text{tr}\hat{B}\text{tr}\hat{L}}{3} + \text{tr}\hat{T}\hat{B} - \frac{\text{tr}\hat{T}^2}{2} - 2 \text{tr}\hat{T}\text{tr}\hat{B} - \frac{\text{tr}\hat{T}\text{tr}\hat{L}}{3} - (\text{tr}\hat{T})^2 \notag \\
 &  + n_G \bigg( \frac{11 \alpha_1^2}{1800} - \frac{43 \alpha_2^2}{24} - \frac{121 \alpha_1 \alpha_3}{30} - \frac{33 \alpha_2 \alpha_3}{2} - \frac{4354 \alpha_3^2}{9} - \frac{8 \alpha_3 \text{tr}\hat{T}}{3} - \frac{8 \alpha_3 \text{tr}\hat{B}}{3} \bigg)  \notag \\
 &  + n_G^2 \bigg( \frac{11 \alpha_1^2}{135} + \frac{\alpha_2^2}{3} + \frac{22 \alpha_1 \alpha_3}{45} + 2 \alpha_2 \alpha_3 + \frac{1064 \alpha_3^2}{27} \bigg) \bigg] \notag \\
 &  + \frac{1}{\epsilon} \bigg[ - \frac{2857 \alpha_3^2}{6} - \frac{101 \alpha_1 \text{tr}\hat{T}}{120} - \frac{31 \alpha_2 \text{tr}\hat{T}}{8} - \frac{40 \alpha_3 \text{tr}\hat{T}}{3} - \frac{89 \alpha_1 \text{tr}\hat{B}}{120} - \frac{31 \alpha_2 \text{tr}\hat{B}}{8} - \frac{40 \alpha_3 \text{tr}\hat{B}}{3} \notag \\
 &  + \frac{3 \text{tr}\hat{B}^2}{2} + \frac{7 (\text{tr}\hat{B})^2}{2} + \frac{7 \text{tr}\hat{B}\text{tr}\hat{L}}{6} - \text{tr}\hat{T}\hat{B} + \frac{3 \text{tr}\hat{T}^2}{2} + 7 \text{tr}\hat{T}\text{tr}\hat{B} + \frac{7 \text{tr}\hat{T}\text{tr}\hat{L}}{6} + \frac{7 (\text{tr}\hat{T})^2}{2} \notag \\ 
 &  + n_G \bigg( - \frac{13 \alpha_1^2}{360} - \frac{\alpha_1 \alpha_2}{120} + \frac{241 \alpha_2^2}{72} + \frac{77 \alpha_1 \alpha_3}{135} + \frac{7 \alpha_2 \alpha_3}{3} + \frac{5033 \alpha_3^2}{27} \bigg) \notag \\
 &  + n_G^2 \bigg( - \frac{121 \alpha_1^2}{810} - \frac{11 \alpha_2^2}{18} - \frac{650 \alpha_3^2}{81} \bigg) \bigg] \bigg\}
\,,
\end{align}
\begin{align}
Z_B &=
  1 + \frac{\alpha_1}{4\pi} \frac{1}{\epsilon} \bigg\{ - \frac{1}{10} - \frac{4 n_G}{3} \bigg\} + \frac{\alpha_1}{\lp4\pi\rp^2} \frac{1}{\epsilon} \bigg\{ - \frac{9 \alpha_1}{100} - \frac{9 \alpha_2}{20} + \frac{17 \text{tr}\hat{T}}{20} + \frac{\text{tr}\hat{B}}{4} + \frac{3 \text{tr}\hat{L}}{4} \notag \\
 &  + n_G \bigg[ - \frac{19 \alpha_1}{30} - \frac{3 \alpha_2}{10} - \frac{22 \alpha_3}{15} \bigg] \bigg\} \notag \\
 &  + \frac{\alpha_1}{\lp4\pi\rp^3} \bigg\{ \frac{1}{\epsilon^2} \bigg[ - \frac{3 \alpha_1^2}{1000} + \frac{43 \alpha_2^2}{40} - \frac{289 \alpha_1 \text{tr}\hat{T}}{1200} - \frac{51 \alpha_2 \text{tr}\hat{T}}{80} - \frac{34 \alpha_3 \text{tr}\hat{T}}{15} - \frac{\alpha_1 \text{tr}\hat{B}}{48} \notag \\
 &  - \frac{3 \alpha_2 \text{tr}\hat{B}}{16} - \frac{2 \alpha_3 \text{tr}\hat{B}}{3} - \frac{9 \alpha_1 \text{tr}\hat{L}}{16} - \frac{9 \alpha_2 \text{tr}\hat{L}}{16} + \frac{\text{tr}\hat{B}^2}{8} + \frac{(\text{tr}\hat{B})^2}{4} + \frac{5 \text{tr}\hat{B}\text{tr}\hat{L}}{6} + \frac{3 \text{tr}\hat{L}^2}{8} \notag \\
 &  + \frac{(\text{tr}\hat{L})^2}{4} - \frac{11 \text{tr}\hat{T}\hat{B}}{20} + \frac{17 \text{tr}\hat{T}^2}{40} + \frac{11 \text{tr}\hat{T}\text{tr}\hat{B}}{10} + \frac{31 \text{tr}\hat{T}\text{tr}\hat{L}}{30} + \frac{17 (\text{tr}\hat{T})^2}{20} \notag \\
 &  + n_G \bigg( - \frac{11 \alpha_1^2}{180} + \frac{31 \alpha_2^2}{60} + \frac{242 \alpha_3^2}{45} \bigg) + n_G^2 \bigg( - \frac{38 \alpha_1^2}{135} - \frac{2 \alpha_2^2}{15} - \frac{88 \alpha_3^2}{135} \bigg) \bigg] \notag \\
 &  + \frac{1}{\epsilon} \bigg[ - \frac{163 \alpha_1^2}{8000} - \frac{261 \alpha_1 \alpha_2}{800} - \frac{3401 \alpha_2^2}{960} - \frac{9 \alpha_1 \hat{\lambda}}{50} - \frac{3 \alpha_2 \hat{\lambda}}{10} + \frac{3 \hat{\lambda}^2}{5}  \notag \\
 &  + \frac{2827 \alpha_1 \text{tr}\hat{T}}{2400} + \frac{157 \alpha_2 \text{tr}\hat{T}}{32} + \frac{29 \alpha_3 \text{tr}\hat{T}}{15} + \frac{1267 \alpha_1 \text{tr}\hat{B}}{2400} + \frac{437 \alpha_2 \text{tr}\hat{B}}{160} + \frac{17 \alpha_3 \text{tr}\hat{B}}{15} \notag \\
 &  + \frac{843 \alpha_1 \text{tr}\hat{L}}{800} + \frac{543 \alpha_2 \text{tr}\hat{L}}{160} - \frac{61 \text{tr}\hat{B}^2}{80} - \frac{17 (\text{tr}\hat{B})^2}{40} - \frac{157 \text{tr}\hat{B}\text{tr}\hat{L}}{60} - \frac{87 \text{tr}\hat{L}^2}{80} \notag \\
 &  - \frac{33 (\text{tr}\hat{L})^2}{40} - \frac{\text{tr}\hat{T}\hat{B}}{8} - \frac{113 \text{tr}\hat{T}^2}{80} - \frac{59 \text{tr}\hat{T}\text{tr}\hat{B}}{20} - \frac{199 \text{tr}\hat{T}\text{tr}\hat{L}}{60} - \frac{101 (\text{tr}\hat{T})^2}{40} \notag \\
 &  + n_G \bigg( \frac{58 \alpha_1^2}{225} + \frac{7 \alpha_1 \alpha_2}{300} - \frac{83 \alpha_2^2}{90} + \frac{137 \alpha_1 \alpha_3}{675} + \frac{\alpha_2 \alpha_3}{15} - \frac{275 \alpha_3^2}{27} \bigg) \notag \\
 &  + n_G^2 \bigg( \frac{209 \alpha_1^2}{405} + \frac{11 \alpha_2^2}{45} + \frac{484 \alpha_3^2}{405} \bigg) \bigg] \bigg\}
\,,
\end{align}
\begin{align}
Z_W &=
  1 + \frac{\alpha_2}{4\pi} \frac{1}{\epsilon} \bigg\{ \frac{25}{6} - \frac{4 n_G}{3} - \xi_W \bigg\} \notag \\
 &  + \frac{\alpha_2}{\lp4\pi\rp^2} \bigg\{ \frac{1}{\epsilon^2} \bigg[ - \frac{25 \alpha_2}{4} - \frac{8 \xi_W \alpha_2}{3} + \xi_W^2 \alpha_2 + n_G \bigg( 2 \alpha_2 + \frac{4 \xi_W \alpha_2}{3} \bigg) \bigg] \notag \\
 &  + \frac{1}{\epsilon} \bigg[ - \frac{3 \alpha_1}{20} + \frac{113 \alpha_2}{8} + \frac{3 \text{tr}\hat{T}}{4} + \frac{3 \text{tr}\hat{B}}{4} + \frac{\text{tr}\hat{L}}{4} - \frac{11 \xi_W \alpha_2}{4} - \frac{\xi_W^2 \alpha_2}{2} \notag \\
 &  + n_G \bigg( - \frac{\alpha_1}{10} - \frac{13 \alpha_2}{2} - 2 \alpha_3 \bigg) \bigg] \bigg\} \notag \\
 &  + \frac{\alpha_2}{\lp4\pi\rp^3} \bigg\{ \frac{1}{\epsilon^3} \bigg[ \frac{1525 \alpha_2^2}{72} + \frac{8 n_G^2 \alpha_2^2}{9} + \frac{89 \xi_W \alpha_2^2}{12} + \frac{7 \xi_W^2 \alpha_2^2}{6} - \xi_W^3 \alpha_2^2 \notag \\
 &  + n_G \bigg( - \frac{86 \alpha_2^2}{9} - \frac{10 \xi_W \alpha_2^2}{3} - \frac{4 \xi_W^2 \alpha_2^2}{3} \bigg) \bigg] \notag \\
 &  + \frac{1}{\epsilon^2} \bigg[ - \frac{\alpha_1^2}{200} + \frac{3 \alpha_1 \alpha_2}{20} - \frac{29629 \alpha_2^2}{432} - \frac{17 \alpha_1 \text{tr}\hat{T}}{80} - \frac{21 \alpha_2 \text{tr}\hat{T}}{16} - 2 \alpha_3 \text{tr}\hat{T}\notag \\
 &   - \frac{\alpha_1 \text{tr}\hat{B}}{16} - \frac{21 \alpha_2 \text{tr}\hat{B}}{16} - 2 \alpha_3 \text{tr}\hat{B} - \frac{3 \alpha_1 \text{tr}\hat{L}}{16} - \frac{7 \alpha_2 \text{tr}\hat{L}}{16} + \frac{3 \text{tr}\hat{B}^2}{8} + \frac{3 (\text{tr}\hat{B})^2}{4}\notag \\
 &  + \frac{\text{tr}\hat{B}\text{tr}\hat{L}}{2} + \frac{\text{tr}\hat{L}^2}{8} + \frac{(\text{tr}\hat{L})^2}{12} - \frac{3 \text{tr}\hat{T}\hat{B}}{4} + \frac{3 \text{tr}\hat{T}^2}{8} + \frac{3 \text{tr}\hat{T}\text{tr}\hat{B}}{2} + \frac{\text{tr}\hat{T}\text{tr}\hat{L}}{2} + \frac{3 (\text{tr}\hat{T})^2}{4} \notag \\
 &  + \xi_W \bigg( \frac{3 \alpha_1 \alpha_2}{20} - \frac{271 \alpha_2^2}{24} - \frac{3 \alpha_2 \text{tr}\hat{T}}{4} - \frac{3 \alpha_2 \text{tr}\hat{B}}{4} - \frac{\alpha_2 \text{tr}\hat{L}}{4} \bigg) + \frac{53 \xi_W^2 \alpha_2^2}{12} + \frac{7 \xi_W^3 \alpha_2^2}{6}\notag \\
 &  + n_G \bigg( - \frac{7 \alpha_1^2}{100} + \frac{\alpha_1 \alpha_2}{10} + \frac{4273 \alpha_2^2}{108} + 2 \alpha_2 \alpha_3 + \frac{22 \alpha_3^2}{3} \notag \\
 &  + \xi_W \bigg( \frac{\alpha_1 \alpha_2}{10} + \frac{47 \alpha_2^2}{6} + 2 \alpha_2 \alpha_3 \bigg) + \frac{2 \xi_W^2 \alpha_2^2}{3} \bigg) + n_G^2 \bigg( - \frac{2 \alpha_1^2}{45} - \frac{118 \alpha_2^2}{27} - \frac{8 \alpha_3^2}{9} \bigg) \bigg] \notag \\
 &  + \frac{1}{\epsilon} \bigg[ - \frac{163 \alpha_1^2}{4800} - \frac{11 \alpha_1 \alpha_2}{32} - \frac{3 \zeta_3 \alpha_1 \alpha_2}{10} + \frac{143537 \alpha_2^2}{1728} + \frac{\zeta_3 \alpha_2^2}{2} - \frac{\alpha_1 \hat{\lambda}}{10} - \frac{\alpha_2 \hat{\lambda}}{2} + \hat{\lambda}^2 \notag \\
 &  + \frac{593 \alpha_1 \text{tr}\hat{T}}{480} + \frac{79 \alpha_2 \text{tr}\hat{T}}{32} + \frac{7 \alpha_3 \text{tr}\hat{T}}{3} + \frac{533 \alpha_1 \text{tr}\hat{B}}{480} + \frac{79 \alpha_2 \text{tr}\hat{B}}{32} + \frac{7 \alpha_3 \text{tr}\hat{B}}{3} + \frac{17 \alpha_1 \text{tr}\hat{L}}{32} \notag \\
 &  + \frac{79 \alpha_2 \text{tr}\hat{L}}{96} - \frac{19 \text{tr}\hat{B}^2}{16} - \frac{15 (\text{tr}\hat{B})^2}{8} - \frac{5 \text{tr}\hat{B}\text{tr}\hat{L}}{4} - \frac{19 \text{tr}\hat{L}^2}{48} - \frac{5 (\text{tr}\hat{L})^2}{24} - \frac{9 \text{tr}\hat{T}\hat{B}}{8} \notag \\
 &  - \frac{19 \text{tr}\hat{T}^2}{16} - \frac{15 \text{tr}\hat{T}\text{tr}\hat{B}}{4} - \frac{5 \text{tr}\hat{T}\text{tr}\hat{L}}{4} - \frac{15 (\text{tr}\hat{T})^2}{8} \notag \\
 &  + \xi_W \bigg( - \frac{105 \alpha_2^2}{8} - 2 \zeta_3 \alpha_2^2 \bigg) + \xi_W^2 \bigg( - \frac{11 \alpha_2^2}{4} - \frac{\zeta_3 \alpha_2^2}{2} \bigg) - \frac{7 \xi_W^3 \alpha_2^2}{12} \notag \\
 &  + n_G \bigg( \frac{7 \alpha_1^2}{45} + \frac{8 \alpha_1 \alpha_2}{15} - \frac{4 \zeta_3 \alpha_1 \alpha_2}{5} - \frac{7025 \alpha_2^2}{108} + 12 \zeta_3 \alpha_2^2 + \frac{\alpha_1 \alpha_3}{45} + \frac{32 \alpha_2 \alpha_3}{3} \notag \\
 &  - 16 \zeta_3 \alpha_2 \alpha_3 - \frac{125 \alpha_3^2}{9} + \frac{8 \xi_W \alpha_2^2}{3} \bigg) + n_G^2 \bigg( \frac{11 \alpha_1^2}{135} + \frac{185 \alpha_2^2}{27} + \frac{44 \alpha_3^2}{27} \bigg) \bigg] \bigg\}
\,,
\end{align}
\begin{align}
Z_G &=
  1 + \frac{\alpha_3}{4\pi} \frac{1}{\epsilon} \bigg\{ \frac{13}{2} - \frac{4 n_G}{3} - \frac{3 \xi_G}{2} \bigg\} \notag \\
 &  + \frac{\alpha_3}{\lp4\pi\rp^2} \bigg\{ \frac{1}{\epsilon^2} \bigg[ - \frac{117 \alpha_3}{8} - \frac{51 \xi_G \alpha_3}{8} + \frac{9 \xi_G^2 \alpha_3}{4} + n_G \bigg( 3 \alpha_3 + 2 \xi_G \alpha_3 \bigg) \bigg] \notag \\
 &  + \frac{1}{\epsilon} \bigg[ + \frac{531 \alpha_3}{16}  + \text{tr}\hat{T} + \text{tr}\hat{B} - \frac{99 \xi_G \alpha_3}{16} - \frac{9 \xi_G^2 \alpha_3}{8} + n_G \bigg( - \frac{11 \alpha_1}{60} - \frac{3 \alpha_2}{4} - \frac{61 \alpha_3}{6} \bigg) \bigg] \bigg\} \notag \\
 &  + \frac{\alpha_3}{\lp4\pi\rp^3} \bigg\{ \frac{1}{\epsilon^3} \bigg[ \frac{1209 \alpha_3^2}{16} + \frac{4 n_G^2 \alpha_3^2}{3} + \frac{423 \xi_G \alpha_3^2}{16} + \frac{9 \xi_G^2 \alpha_3^2}{2} - \frac{27 \xi_G^3 \alpha_3^2}{8} \notag \\
 &  + n_G \bigg( - 22 \alpha_3^2 - \frac{15 \xi_G \alpha_3^2}{2} - 3 \xi_G^2 \alpha_3^2 \bigg) \bigg] \notag \\
 &  + \frac{1}{\epsilon^2} \bigg[ - \frac{7957 \alpha_3^2}{32} - \frac{17 \alpha_1 \text{tr}\hat{T}}{60} - \frac{3 \alpha_2 \text{tr}\hat{T}}{4} - \frac{25 \alpha_3 \text{tr}\hat{T}}{6} - \frac{\alpha_1 \text{tr}\hat{B}}{12} - \frac{3 \alpha_2 \text{tr}\hat{B}}{4} - \frac{25 \alpha_3 \text{tr}\hat{B}}{6}\notag \\
 &  + \frac{\text{tr}\hat{B}^2}{2} + (\text{tr}\hat{B})^2 + \frac{\text{tr}\hat{B}\text{tr}\hat{L}}{3} - \text{tr}\hat{T}\hat{B} + \frac{\text{tr}\hat{T}^2}{2} + 2 \text{tr}\hat{T}\text{tr}\hat{B} + \frac{\text{tr}\hat{T}\text{tr}\hat{L}}{3} + (\text{tr}\hat{T})^2 \notag \\
 &  + \xi_G \bigg( - \frac{1287 \alpha_3^2}{32} - \frac{3 \alpha_3 \text{tr}\hat{T}}{2} - \frac{3 \alpha_3 \text{tr}\hat{B}}{2} \bigg)+ \frac{117 \xi_G^2 \alpha_3^2}{8} + \frac{63 \xi_G^3 \alpha_3^2}{16} \notag \\
 &  + n_G \bigg( - \frac{11 \alpha_1^2}{1800} + \frac{43 \alpha_2^2}{24} + \frac{11 \alpha_1 \alpha_3}{40} + \frac{9 \alpha_2 \alpha_3}{8} + \frac{1691 \alpha_3^2}{18} \notag \\
 &  + \xi_G \bigg( \frac{11 \alpha_1 \alpha_3}{40} + \frac{9 \alpha_2 \alpha_3}{8} + \frac{73 \alpha_3^2}{4} \bigg) + \frac{3 \xi_G^2 \alpha_3^2}{2} \bigg) + n_G^2 \bigg( - \frac{11 \alpha_1^2}{135} - \frac{\alpha_2^2}{3} - \frac{182 \alpha_3^2}{27} \bigg) \bigg] \notag \\
 &  + \frac{1}{\epsilon} \bigg[ \frac{9965 \alpha_3^2}{32} - \frac{81 \zeta_3 \alpha_3^2}{16} + \frac{101 \alpha_1 \text{tr}\hat{T}}{120} + \frac{31 \alpha_2 \text{tr}\hat{T}}{8} + \frac{91 \alpha_3 \text{tr}\hat{T}}{12} + \frac{89 \alpha_1 \text{tr}\hat{B}}{120} + \frac{31 \alpha_2 \text{tr}\hat{B}}{8} \notag \\
 &  + \frac{91 \alpha_3 \text{tr}\hat{B}}{12} - \frac{3 \text{tr}\hat{B}^2}{2} - \frac{7 (\text{tr}\hat{B})^2}{2} - \frac{7 \text{tr}\hat{B}\text{tr}\hat{L}}{6} + \text{tr}\hat{T}\hat{B} - \frac{3 \text{tr}\hat{T}^2}{2} - 7 \text{tr}\hat{T}\text{tr}\hat{B} - \frac{7 \text{tr}\hat{T}\text{tr}\hat{L}}{6} \notag \\
 &  - \frac{7 (\text{tr}\hat{T})^2}{2} + \xi_G \bigg( - \frac{1503 \alpha_3^2}{32} - \frac{27 \zeta_3 \alpha_3^2}{4} \bigg) + \xi_G^2 \bigg( - \frac{297 \alpha_3^2}{32} - \frac{27 \zeta_3 \alpha_3^2}{16} \bigg) - \frac{63 \xi_G^3 \alpha_3^2}{32} \notag \\
 &  + n_G \bigg( \frac{13 \alpha_1^2}{360} + \frac{\alpha_1 \alpha_2}{120} - \frac{241 \alpha_2^2}{72} + \frac{3223 \alpha_1 \alpha_3}{2160} - \frac{11 \zeta_3 \alpha_1 \alpha_3}{5} + \frac{293 \alpha_2 \alpha_3}{48} - 9 \zeta_3 \alpha_2 \alpha_3 \notag \\
 &  - \frac{8155 \alpha_3^2}{54} + 22 \zeta_3 \alpha_3^2 + 6 \xi_G \alpha_3^2 \bigg) + n_G^2 \bigg( \frac{121 \alpha_1^2}{810} + \frac{11 \alpha_2^2}{18} + \frac{860 \alpha_3^2}{81} \bigg) \bigg] \bigg\} 
\,,
\end{align}
\begin{align}
Z_{c_W} &=
  1 + \frac{\alpha_2}{4\pi} \frac{1}{\epsilon} \bigg\{ \frac{3}{2} - \frac{\xi_W}{2} \bigg\} \notag \\
 &  + \frac{\alpha_2}{\lp4\pi\rp^2} \bigg\{ \frac{1}{\epsilon^2} \bigg[ - \frac{17 \alpha_2}{4} + n_G \alpha_2 + \frac{3 \xi_W^2 \alpha_2}{8} \bigg] + \frac{1}{\epsilon} \bigg[ \frac{179 \alpha_2}{48} - \frac{5 n_G \alpha_2}{6} + \frac{\xi_W \alpha_2}{8} \bigg] \bigg\} \notag \\
 &  + \frac{\alpha_2}{\lp4\pi\rp^3} \bigg\{ \frac{1}{\epsilon^3} \bigg[ \frac{1309 \alpha_2^2}{72} + \frac{8 n_G^2 \alpha_2^2}{9} + \frac{17 \xi_W \alpha_2^2}{24} - \frac{9 \xi_W^2 \alpha_2^2}{16} - \frac{5 \xi_W^3 \alpha_2^2}{16} \notag \\
 &  + n_G \bigg( - \frac{145 \alpha_2^2}{18} - \frac{\xi_W \alpha_2^2}{6} \bigg) \bigg] + \frac{1}{\epsilon^2} \bigg[ \frac{3 \alpha_1 \alpha_2}{20} - \frac{29209 \alpha_2^2}{864} - \frac{20 n_G^2 \alpha_2^2}{27} - \frac{3 \alpha_2 \text{tr}\hat{T}}{4} - \frac{3 \alpha_2 \text{tr}\hat{B}}{4} \notag \\
 &  - \frac{\alpha_2 \text{tr}\hat{L}}{4} + \frac{37 \xi_W \alpha_2^2}{96} + \frac{13 \xi_W^2 \alpha_2^2}{16} + \frac{\xi_W^3 \alpha_2^2}{6} + n_G \bigg( \frac{\alpha_1 \alpha_2}{10} + \frac{1535 \alpha_2^2}{108} + 2 \alpha_2 \alpha_3 - \frac{\xi_W \alpha_2^2}{12} \bigg) \bigg] \notag \\
 &  + \frac{1}{\epsilon} \bigg[ - \frac{33 \alpha_1 \alpha_2}{80} + \frac{3 \zeta_3 \alpha_1 \alpha_2}{20} + \frac{59125 \alpha_2^2}{2592} - \frac{70 n_G^2 \alpha_2^2}{81} - \frac{\zeta_3 \alpha_2^2}{4} + \frac{41 \alpha_2 \text{tr}\hat{T}}{16} + \frac{41 \alpha_2 \text{tr}\hat{B}}{16} \notag \\
 &  + \frac{41 \alpha_2 \text{tr}\hat{L}}{48} + \xi_W \bigg( - \frac{29 \alpha_2^2}{24} + \zeta_3 \alpha_2^2 \bigg) + \xi_W^2 \bigg( - \frac{\alpha_2^2}{4} + \frac{\zeta_3 \alpha_2^2}{4} \bigg) - \frac{\xi_W^3 \alpha_2^2}{8} \notag \\
 &  + n_G \bigg( - \frac{3 \alpha_1 \alpha_2}{8} + \frac{2 \zeta_3 \alpha_1 \alpha_2}{5} - \frac{4573 \alpha_2^2}{648} - 6 \zeta_3 \alpha_2^2 - \frac{15 \alpha_2 \alpha_3}{2} + 8 \zeta_3 \alpha_2 \alpha_3 + \frac{7 \xi_W \alpha_2^2}{6} \bigg) \bigg] \bigg\} 
\,,
\end{align}
\begin{align}
Z_{c_G} &=
  1 + \frac{\alpha_3}{4\pi} \frac{1}{\epsilon} \bigg\{ \frac{9}{4} - \frac{3 \xi_G}{4} \bigg\} \notag \\
 &  + \frac{\alpha_3}{\lp4\pi\rp^2} \bigg\{ \frac{1}{\epsilon^2} \bigg[ - \frac{315 \alpha_3}{32} + \frac{3 n_G \alpha_3}{2} + \frac{27 \xi_G^2 \alpha_3}{32} \bigg] + \frac{1}{\epsilon} \bigg[ \frac{285 \alpha_3}{32} - \frac{5 n_G \alpha_3}{4} + \frac{9 \xi_G \alpha_3}{32} \bigg] \bigg\} \notag \\
 &  + \frac{\alpha_3}{\lp4\pi\rp^3} \bigg\{ \frac{1}{\epsilon^3} \bigg[ \frac{8295 \alpha_3^2}{128} + \frac{4 n_G^2 \alpha_3^2}{3} + \frac{315 \xi_G \alpha_3^2}{128} - \frac{243 \xi_G^2 \alpha_3^2}{128} - \frac{135 \xi_G^3 \alpha_3^2}{128} \notag \\
 &  + n_G \bigg( - \frac{149 \alpha_3^2}{8} - \frac{3 \xi_G \alpha_3^2}{8} \bigg) \bigg] + \frac{1}{\epsilon^2} \bigg[ - \frac{15587 \alpha_3^2}{128} - \frac{10 n_G^2 \alpha_3^2}{9} - \frac{3 \alpha_3 \text{tr}\hat{T}}{2} - \frac{3 \alpha_3 \text{tr}\hat{B}}{2} \notag \\
 &  + \frac{45 \xi_G \alpha_3^2}{32} + \frac{351 \xi_G^2 \alpha_3^2}{128} + \frac{9 \xi_G^3 \alpha_3^2}{16} + n_G \bigg( \frac{11 \alpha_1 \alpha_3}{40} + \frac{9 \alpha_2 \alpha_3}{8} + \frac{1597 \alpha_3^2}{48} - \frac{3 \xi_G \alpha_3^2}{16} \bigg) \bigg] \notag \\
 &  + \frac{1}{\epsilon} \bigg[ \frac{15817 \alpha_3^2}{192} - \frac{35 n_G^2 \alpha_3^2}{27} + \frac{81 \zeta_3 \alpha_3^2}{32} + \frac{23 \alpha_3 \text{tr}\hat{T}}{8} + \frac{23 \alpha_3 \text{tr}\hat{B}}{8} \notag \\
 &  + \xi_G \bigg( - \frac{153 \alpha_3^2}{32} + \frac{27 \zeta_3 \alpha_3^2}{8} \bigg) + \xi_G^2 \bigg( - \frac{27 \alpha_3^2}{32} + \frac{27 \zeta_3 \alpha_3^2}{32} \bigg) - \frac{27 \xi_G^3 \alpha_3^2}{64} \notag \\
 &  + n_G \bigg( - \frac{33 \alpha_1 \alpha_3}{32} + \frac{11 \zeta_3 \alpha_1 \alpha_3}{10} - \frac{135 \alpha_2 \alpha_3}{32} + \frac{9 \zeta_3 \alpha_2 \alpha_3}{2} - \frac{637 \alpha_3^2}{36} + \frac{21 \xi_G \alpha_3^2}{8} - 11 \zeta_3 \alpha_3^2 \bigg) \bigg] \bigg\} 
\,,
\end{align}
\begin{align}
Z_H &=
  1 + \frac{1}{4\pi} \frac{1}{\epsilon} \bigg\{ \frac{9 \alpha_1}{20} + \frac{9 \alpha_2}{4} - 3 \text{tr}\hat{T} - 3 \text{tr}\hat{B} - \text{tr}\hat{L} - \frac{3 \xi_B \alpha_1}{20} - \frac{3 \xi_W \alpha_2}{4} \bigg\} \notag \\
 &  + \frac{1}{\lp4\pi\rp^2} \bigg\{ \frac{1}{\epsilon^2} \bigg[ \frac{99 \alpha_1^2}{800} + \frac{81 \alpha_1 \alpha_2}{80} - \frac{177 \alpha_2^2}{32} - \frac{3 \alpha_1 \text{tr}\hat{T}}{40} - \frac{27 \alpha_2 \text{tr}\hat{T}}{8} + 12 \alpha_3 \text{tr}\hat{T} - \frac{39 \alpha_1 \text{tr}\hat{B}}{40} \notag \\
 &  - \frac{27 \alpha_2 \text{tr}\hat{B}}{8} + 12 \alpha_3 \text{tr}\hat{B} + \frac{27 \alpha_1 \text{tr}\hat{L}}{40} - \frac{9 \alpha_2 \text{tr}\hat{L}}{8} - \frac{9 \text{tr}\hat{B}^2}{4} - \frac{3 \text{tr}\hat{L}^2}{4} + \frac{9 \text{tr}\hat{T}\hat{B}}{2} - \frac{9 \text{tr}\hat{T}^2}{4} \notag \\
 &  + \xi_B \bigg( - \frac{27 \alpha_1^2}{400} - \frac{27 \alpha_1 \alpha_2}{80} + \frac{9 \xi_W \alpha_1 \alpha_2}{80} + \frac{9 \alpha_1 \text{tr}\hat{T}}{20} + \frac{9 \alpha_1 \text{tr}\hat{B}}{20} + \frac{3 \alpha_1 \text{tr}\hat{L}}{20} \bigg) + \frac{9 \xi_B^2 \alpha_1^2}{800} \notag \\
 &  + \xi_W \bigg( - \frac{27 \alpha_1 \alpha_2}{80} - \frac{9 \alpha_2^2}{16} + \frac{9 \alpha_2 \text{tr}\hat{T}}{4} + \frac{9 \alpha_2 \text{tr}\hat{B}}{4} + \frac{3 \alpha_2 \text{tr}\hat{L}}{4} \bigg) + \frac{21 \xi_W^2 \alpha_2^2}{32} \notag \\
 &  + n_G \bigg( \frac{3 \alpha_1^2}{10} + \frac{3 \alpha_2^2}{2} \bigg) \bigg] \notag \\
 &  + \frac{1}{\epsilon} \bigg[ - \frac{93 \alpha_1^2}{1600} - \frac{27 \alpha_1 \alpha_2}{160} + \frac{511 \alpha_2^2}{64} - 3 \hat{\lambda}^2 - \frac{17 \alpha_1 \text{tr}\hat{T}}{16} - \frac{45 \alpha_2 \text{tr}\hat{T}}{16} - 10 \alpha_3 \text{tr}\hat{T} - \frac{5 \alpha_1 \text{tr}\hat{B}}{16} \notag \\
 &  - \frac{45 \alpha_2 \text{tr}\hat{B}}{16} - 10 \alpha_3 \text{tr}\hat{B} - \frac{15 \alpha_1 \text{tr}\hat{L}}{16} - \frac{15 \alpha_2 \text{tr}\hat{L}}{16} + \frac{27 \text{tr}\hat{B}^2}{8} + \frac{9 \text{tr}\hat{L}^2}{8} - \frac{3 \text{tr}\hat{T}\hat{B}}{4} + \frac{27 \text{tr}\hat{T}^2}{8} \notag \\
 &  - \frac{3 \xi_W \alpha_2^2}{2} - \frac{3 \xi_W^2 \alpha_2^2}{16} + n_G \bigg( - \frac{\alpha_1^2}{4} - \frac{5 \alpha_2^2}{4} \bigg) \bigg] \bigg\} \notag \\
 &  + \frac{1}{\lp4\pi\rp^3} \bigg\{ \frac{1}{\epsilon^3} \bigg[ \frac{429 \alpha_1^3}{16000} + \frac{891 \alpha_1^2 \alpha_2}{3200} - \frac{1593 \alpha_1 \alpha_2^2}{640} + \frac{8555 \alpha_2^3}{384} - \frac{93 \alpha_1^2 \text{tr}\hat{T}}{800} - \frac{9 \alpha_1 \alpha_2 \text{tr}\hat{T}}{16} \notag \\
 &  + \frac{435 \alpha_2^2 \text{tr}\hat{T}}{32} - \frac{7 \alpha_1 \alpha_3 \text{tr}\hat{T}}{5} + 9 \alpha_2 \alpha_3 \text{tr}\hat{T} - 76 \alpha_3^2 \text{tr}\hat{T} - \frac{177 \alpha_1^2 \text{tr}\hat{B}}{800} - \frac{99 \alpha_1 \alpha_2 \text{tr}\hat{B}}{80} \notag \\
 &  + \frac{435 \alpha_2^2 \text{tr}\hat{B}}{32} + \frac{17 \alpha_1 \alpha_3 \text{tr}\hat{B}}{5} + 9 \alpha_2 \alpha_3 \text{tr}\hat{B} - 76 \alpha_3^2 \text{tr}\hat{B} - \frac{339 \alpha_1^2 \text{tr}\hat{L}}{800} + \frac{27 \alpha_1 \alpha_2 \text{tr}\hat{L}}{80} \notag \\
 &  + \frac{145 \alpha_2^2 \text{tr}\hat{L}}{32} - \frac{9 \text{tr}\hat{B}^3}{4} - \frac{9 \alpha_1 \text{tr}\hat{B}^2}{20} + 18 \alpha_3 \text{tr}\hat{B}^2 - \frac{9 \text{tr}\hat{T} \text{tr}\hat{B}^2}{4} - \frac{9 \text{tr}\hat{B} \text{tr}\hat{B}^2}{4} - \frac{3 \text{tr}\hat{L} \text{tr}\hat{B}^2}{4} \notag \\
 &  - \frac{3 \text{tr}\hat{L}^3}{4} + \frac{27 \alpha_1 \text{tr}\hat{L}^2}{20} - \frac{3 \text{tr}\hat{T} \text{tr}\hat{L}^2}{4} - \frac{3 \text{tr}\hat{B} \text{tr}\hat{L}^2}{4} - \frac{\text{tr}\hat{L} \text{tr}\hat{L}^2}{4} + \frac{9 \text{tr}\hat{T}^2\hat{B}}{4} - \frac{9 \text{tr}\hat{T}^3}{4} \notag \\
 &  - \frac{9 \alpha_1 \text{tr}\hat{T}\hat{B}}{20} - 36 \alpha_3 \text{tr}\hat{T}\hat{B} + \frac{9 \text{tr}\hat{T} \text{tr}\hat{T}\hat{B}}{2} + \frac{9 \text{tr}\hat{B} \text{tr}\hat{T}\hat{B}}{2} + \frac{3 \text{tr}\hat{L} \text{tr}\hat{T}\hat{B}}{2} + \frac{9 \text{tr}\hat{T}\hat{B}^2}{4} \notag \\
 &  + \frac{9 \alpha_1 \text{tr}\hat{T}^2}{10} + 18 \alpha_3 \text{tr}\hat{T}^2 - \frac{9 \text{tr}\hat{T} \text{tr}\hat{T}^2}{4} - \frac{9 \text{tr}\hat{B} \text{tr}\hat{T}^2}{4} - \frac{3 \text{tr}\hat{L} \text{tr}\hat{T}^2}{4} \notag \\
 &  + \xi_B \bigg( - \frac{297 \alpha_1^3}{16000} - \frac{243 \alpha_1^2 \alpha_2}{1600} + \frac{531 \alpha_1 \alpha_2^2}{640} - \frac{63 \xi_W^2 \alpha_1 \alpha_2^2}{640} + \frac{9 \alpha_1^2 \text{tr}\hat{T}}{800} + \frac{81 \alpha_1 \alpha_2 \text{tr}\hat{T}}{160} \notag \\
 &  - \frac{9 \alpha_1 \alpha_3 \text{tr}\hat{T}}{5} + \frac{117 \alpha_1^2 \text{tr}\hat{B}}{800} + \frac{81 \alpha_1 \alpha_2 \text{tr}\hat{B}}{160} - \frac{9 \alpha_1 \alpha_3 \text{tr}\hat{B}}{5} - \frac{81 \alpha_1^2 \text{tr}\hat{L}}{800} + \frac{27 \alpha_1 \alpha_2 \text{tr}\hat{L}}{160} \notag \\
 &  + \frac{27 \alpha_1 \text{tr}\hat{B}^2}{80} + \frac{9 \alpha_1 \text{tr}\hat{L}^2}{80} - \frac{27 \alpha_1 \text{tr}\hat{T}\hat{B}}{40} + \frac{27 \alpha_1 \text{tr}\hat{T}^2}{80} \notag \\
 &  + \xi_W \bigg( \frac{81 \alpha_1^2 \alpha_2}{1600} + \frac{27 \alpha_1 \alpha_2^2}{320} - \frac{27 \alpha_1 \alpha_2 \text{tr}\hat{T}}{80} - \frac{27 \alpha_1 \alpha_2 \text{tr}\hat{B}}{80} - \frac{9 \alpha_1 \alpha_2 \text{tr}\hat{L}}{80} \bigg) \bigg) \notag \\
 &  + \xi_W \bigg( - \frac{297 \alpha_1^2 \alpha_2}{3200} - \frac{81 \alpha_1 \alpha_2^2}{320} + \frac{367 \alpha_2^3}{128} + \frac{9 \alpha_1 \alpha_2 \text{tr}\hat{T}}{160} - \frac{27 \alpha_2^2 \text{tr}\hat{T}}{32} - 9 \alpha_2 \alpha_3 \text{tr}\hat{T} \notag \\
 &  + \frac{117 \alpha_1 \alpha_2 \text{tr}\hat{B}}{160} - \frac{27 \alpha_2^2 \text{tr}\hat{B}}{32} - 9 \alpha_2 \alpha_3 \text{tr}\hat{B} - \frac{81 \alpha_1 \alpha_2 \text{tr}\hat{L}}{160} - \frac{9 \alpha_2^2 \text{tr}\hat{L}}{32} + \frac{27 \alpha_2 \text{tr}\hat{B}^2}{16} \notag \\
 &  + \frac{9 \alpha_2 \text{tr}\hat{L}^2}{16} - \frac{27 \alpha_2 \text{tr}\hat{T}\hat{B}}{8} + \frac{27 \alpha_2 \text{tr}\hat{T}^2}{16} \bigg) \notag \\
 &  + \xi_B^2 \bigg( \frac{81 \alpha_1^3}{16000} + \frac{81 \alpha_1^2 \alpha_2}{3200} - \frac{27 \xi_W \alpha_1^2 \alpha_2}{3200} - \frac{27 \alpha_1^2 \text{tr}\hat{T}}{800} - \frac{27 \alpha_1^2 \text{tr}\hat{B}}{800} - \frac{9 \alpha_1^2 \text{tr}\hat{L}}{800} \bigg) \notag \\
 &  + \xi_W^2 \bigg( \frac{189 \alpha_1 \alpha_2^2}{640} - \frac{63 \alpha_2^3}{128} - \frac{63 \alpha_2^2 \text{tr}\hat{T}}{32} - \frac{63 \alpha_2^2 \text{tr}\hat{B}}{32} - \frac{21 \alpha_2^2 \text{tr}\hat{L}}{32} \bigg) - \frac{9 \xi_B^3 \alpha_1^3}{16000} - \frac{77 \xi_W^3 \alpha_2^3}{128} \notag \\
 &  + n_G \bigg( \frac{7 \alpha_1^3}{40} + \frac{27 \alpha_1^2 \alpha_2}{40} + \frac{27 \alpha_1 \alpha_2^2}{40} - \frac{263 \alpha_2^3}{24} - \frac{\alpha_1^2 \text{tr}\hat{T}}{3} - 3 \alpha_2^2 \text{tr}\hat{T} + \frac{16 \alpha_3^2 \text{tr}\hat{T}}{3} - \frac{11 \alpha_1^2 \text{tr}\hat{B}}{15} \notag \\
 &  - 3 \alpha_2^2 \text{tr}\hat{B} + \frac{16 \alpha_3^2 \text{tr}\hat{B}}{3} + \frac{\alpha_1^2 \text{tr}\hat{L}}{5} - \alpha_2^2 \text{tr}\hat{L} + \xi_B \bigg( - \frac{9 \alpha_1^3}{200} - \frac{9 \alpha_1 \alpha_2^2}{40} \bigg) \notag \\
 &  + \xi_W \bigg( - \frac{9 \alpha_1^2 \alpha_2}{40} - \frac{5 \alpha_2^3}{8} \bigg) \bigg) + n_G^2 \bigg( \frac{4 \alpha_1^3}{15} + \frac{4 \alpha_2^3}{3} \bigg) \bigg] \notag \\
 &  + \frac{1}{\epsilon^2} \bigg[ - \frac{97 \alpha_1^3}{32000} - \frac{99 \alpha_1^2 \alpha_2}{1280} + \frac{4917 \alpha_1 \alpha_2^2}{1280} - \frac{121093 \alpha_2^3}{2304} - \frac{27 \alpha_1^2 \hat{\lambda}}{200} - \frac{9 \alpha_1 \alpha_2 \hat{\lambda}}{20} - \frac{9 \alpha_2^2 \hat{\lambda}}{8} \notag \\
 &  + \frac{9 \alpha_1 \hat{\lambda}^2}{20} + \frac{9 \alpha_2 \hat{\lambda}^2}{4} - 24 \hat{\lambda}^3 - \frac{541 \alpha_1^2 \text{tr}\hat{T}}{1600} - \frac{177 \alpha_1 \alpha_2 \text{tr}\hat{T}}{160} - \frac{885 \alpha_2^2 \text{tr}\hat{T}}{64} - \frac{\alpha_1 \alpha_3 \text{tr}\hat{T}}{10} \notag \\
 &  - \frac{33 \alpha_2 \alpha_3 \text{tr}\hat{T}}{2} + 198 \alpha_3^2 \text{tr}\hat{T} - 9 \hat{\lambda}^2 \text{tr}\hat{T} + \frac{191 \alpha_1^2 \text{tr}\hat{B}}{1600} + \frac{57 \alpha_1 \alpha_2 \text{tr}\hat{B}}{160} - \frac{885 \alpha_2^2 \text{tr}\hat{B}}{64} \notag \\
 &  - \frac{49 \alpha_1 \alpha_3 \text{tr}\hat{B}}{10} - \frac{33 \alpha_2 \alpha_3 \text{tr}\hat{B}}{2} + 198 \alpha_3^2 \text{tr}\hat{B} - 9 \hat{\lambda}^2 \text{tr}\hat{B} - \frac{3 \alpha_1^2 \text{tr}\hat{L}}{1600} - \frac{45 \alpha_1 \alpha_2 \text{tr}\hat{L}}{32} \notag \\
 &  - \frac{295 \alpha_2^2 \text{tr}\hat{L}}{64} - 3 \hat{\lambda}^2 \text{tr}\hat{L} + \frac{15 \text{tr}\hat{B}^3}{8} - \frac{123 \alpha_1 \text{tr}\hat{B}^2}{80} - \frac{117 \alpha_2 \text{tr}\hat{B}^2}{16} - 39 \alpha_3 \text{tr}\hat{B}^2 + 18 \hat{\lambda} \text{tr}\hat{B}^2 \notag \\
 &  
 + \frac{81 \text{tr}\hat{B} \text{tr}\hat{B}^2}{8} + \frac{27 \text{tr}\hat{L} \text{tr}\hat{B}^2}{8} + \frac{5 \text{tr}\hat{L}^3}{8} - \frac{261 \alpha_1 \text{tr}\hat{L}^2}{80} - \frac{39 \alpha_2 \text{tr}\hat{L}^2}{16} + 6 \hat{\lambda} \text{tr}\hat{L}^2 \notag \\
 &  + \frac{27 \text{tr}\hat{T} \text{tr}\hat{L}^2}{8} + \frac{27
   \text{tr}\hat{B} \text{tr}\hat{L}^2}{8} + \frac{9 \text{tr}\hat{L} \text{tr}\hat{L}^2}{8} 
+ \frac{15 \text{tr}\hat{T}^3}{8} + \frac{43 \alpha_1
  \text{tr}\hat{T}\hat{B}}{20} \notag \\
 &  + \frac{9 \alpha_2 \text{tr}\hat{T}\hat{B}}{8} 
   + 46 \alpha_3 \text{tr}\hat{T}\hat{B} 
- \frac{11 \text{tr}\hat{L}
   \text{tr}\hat{T}\hat{B}}{4} 
- \frac{297 \alpha_1 \text{tr}\hat{T}^2}{80} \notag \\
 &  - \frac{117 \alpha_2 \text{tr}\hat{T}^2}{16} - 39 \alpha_3
 \text{tr}\hat{T}^2 + 18 \hat{\lambda} \text{tr}\hat{T}^2 + \frac{81
   \text{tr}\hat{T} \text{tr}\hat{T}^2}{8} 
+ \frac{27 \text{tr}\hat{L} \text{tr}\hat{T}^2}{8} \notag \\
 &  + \xi_B \bigg( \frac{279 \alpha_1^3}{32000} + \frac{81 \alpha_1^2 \alpha_2}{3200} - \frac{1533 \alpha_1 \alpha_2^2}{1280} + \frac{9 \xi_W \alpha_1 \alpha_2^2}{40} + \frac{9 \xi_W^2 \alpha_1 \alpha_2^2}{320} + \frac{9 \alpha_1 \hat{\lambda}^2}{20} + \frac{51 \alpha_1^2 \text{tr}\hat{T}}{320} \notag \\
 &  + \frac{27 \alpha_1 \alpha_2 \text{tr}\hat{T}}{64} + \frac{3 \alpha_1 \alpha_3 \text{tr}\hat{T}}{2} + \frac{3 \alpha_1^2 \text{tr}\hat{B}}{64} + \frac{27 \alpha_1 \alpha_2 \text{tr}\hat{B}}{64} + \frac{3 \alpha_1 \alpha_3 \text{tr}\hat{B}}{2} + \frac{9 \alpha_1^2 \text{tr}\hat{L}}{64} \notag \\
 &  + \frac{9 \alpha_1 \alpha_2 \text{tr}\hat{L}}{64} - \frac{81 \alpha_1 \text{tr}\hat{B}^2}{160} - \frac{27 \alpha_1 \text{tr}\hat{L}^2}{160} + \frac{9 \alpha_1 \text{tr}\hat{T}\hat{B}}{80} - \frac{81 \alpha_1 \text{tr}\hat{T}^2}{160} \bigg) \notag \\
 &  + \xi_W \bigg( \frac{279 \alpha_1^2 \alpha_2}{6400} - \frac{351 \alpha_1 \alpha_2^2}{640} - \frac{141 \alpha_2^3}{256} + \frac{9 \alpha_2 \hat{\lambda}^2}{4} + \frac{51 \alpha_1 \alpha_2 \text{tr}\hat{T}}{64} + \frac{423 \alpha_2^2 \text{tr}\hat{T}}{64} \notag \\
 &  + \frac{15 \alpha_2 \alpha_3 \text{tr}\hat{T}}{2} + \frac{15 \alpha_1 \alpha_2 \text{tr}\hat{B}}{64} + \frac{423 \alpha_2^2 \text{tr}\hat{B}}{64} + \frac{15 \alpha_2 \alpha_3 \text{tr}\hat{B}}{2} + \frac{45 \alpha_1 \alpha_2 \text{tr}\hat{L}}{64} + \frac{141 \alpha_2^2 \text{tr}\hat{L}}{64} \notag \\
 &  - \frac{81 \alpha_2 \text{tr}\hat{B}^2}{32} - \frac{27 \alpha_2 \text{tr}\hat{L}^2}{32} + \frac{9 \alpha_2 \text{tr}\hat{T}\hat{B}}{16} - \frac{81 \alpha_2 \text{tr}\hat{T}^2}{32} \bigg) \notag \\
 &  + \xi_W^2 \bigg( - \frac{27 \alpha_1 \alpha_2^2}{320} + \frac{189 \alpha_2^3}{64} + \frac{9 \alpha_2^2 \text{tr}\hat{T}}{16} + \frac{9 \alpha_2^2 \text{tr}\hat{B}}{16} + \frac{3 \alpha_2^2 \text{tr}\hat{L}}{16} \bigg) + \frac{33 \xi_W^3 \alpha_2^3}{64} \notag \\
 &  + n_G \bigg( \frac{11 \alpha_1^3}{1200} - \frac{219 \alpha_1^2 \alpha_2}{400} - \frac{39 \alpha_1 \alpha_2^2}{80} + \frac{3241 \alpha_2^3}{144} + \frac{11 \alpha_1^2 \alpha_3}{25} + 3 \alpha_2^2 \alpha_3 - \frac{11 \alpha_1^2 \text{tr}\hat{T}}{30} + \frac{3 \alpha_2^2 \text{tr}\hat{T}}{2} \notag \\
 &  - \frac{40 \alpha_3^2 \text{tr}\hat{T}}{3} + \frac{19 \alpha_1^2 \text{tr}\hat{B}}{30} + \frac{3 \alpha_2^2 \text{tr}\hat{B}}{2} - \frac{40 \alpha_3^2 \text{tr}\hat{B}}{3} - \frac{9 \alpha_1^2 \text{tr}\hat{L}}{10} + \frac{\alpha_2^2 \text{tr}\hat{L}}{2} 
   + \xi_B \bigg( \frac{3 \alpha_1^3}{80} + \frac{3 \alpha_1 \alpha_2^2}{16}
 \bigg) 
 \notag \\
 &
 + \xi_W \bigg( \frac{3 \alpha_1^2 \alpha_2}{16} - \frac{9
   \alpha_2^3}{16} \bigg) \bigg) + n_G^2 \bigg( - \frac{2 \alpha_1^3}{9} -
 \frac{10 \alpha_2^3}{9} \bigg) 
 -3\alpha_b\alpha_t^2 -3\alpha_b^2\alpha_t
 \bigg] \notag \\
 &  + \frac{1}{\epsilon} \bigg[ - \frac{413 \alpha_1^3}{6000} + \frac{27 \zeta_3 \alpha_1^3}{2000} - \frac{279 \alpha_1^2 \alpha_2}{800} - \frac{27 \zeta_3 \alpha_1^2 \alpha_2}{400} - \frac{51 \alpha_1 \alpha_2^2}{64} + \frac{9 \zeta_3 \alpha_1 \alpha_2^2}{80} + \frac{70519 \alpha_2^3}{1728} \notag \\
 &  + \frac{69 \zeta_3 \alpha_2^3}{16} - \frac{117 \alpha_1^2 \hat{\lambda}}{400} + \frac{27 \zeta_3 \alpha_1^2 \hat{\lambda}}{50} - \frac{39 \alpha_1 \alpha_2 \hat{\lambda}}{40} + \frac{9 \zeta_3 \alpha_1 \alpha_2 \hat{\lambda}}{5} - \frac{39 \alpha_2^2 \hat{\lambda}}{16} + \frac{9 \zeta_3 \alpha_2^2 \hat{\lambda}}{2} \notag \\
 &  - 3 \alpha_1 \hat{\lambda}^2 - 15 \alpha_2 \hat{\lambda}^2 + 12 \hat{\lambda}^3 + \frac{52831 \alpha_1^2 \text{tr}\hat{T}}{28800} - \frac{\zeta_3 \alpha_1^2 \text{tr}\hat{T}}{100} - \frac{371 \alpha_1 \alpha_2 \text{tr}\hat{T}}{320} \notag \\
 &  - \frac{27 \zeta_3 \alpha_1 \alpha_2 \text{tr}\hat{T}}{10} - \frac{2433 \alpha_2^2 \text{tr}\hat{T}}{128} + \frac{63 \zeta_3 \alpha_2^2 \text{tr}\hat{T}}{4} + \frac{2419 \alpha_1 \alpha_3 \text{tr}\hat{T}}{180} - \frac{68 \zeta_3 \alpha_1 \alpha_3 \text{tr}\hat{T}}{5} \notag \\
 &  + \frac{163 \alpha_2 \alpha_3 \text{tr}\hat{T}}{4} - 36 \zeta_3 \alpha_2 \alpha_3 \text{tr}\hat{T} - \frac{910 \alpha_3^2 \text{tr}\hat{T}}{9} + 8 \zeta_3 \alpha_3^2 \text{tr}\hat{T} + \frac{45 \hat{\lambda}^2 \text{tr}\hat{T}}{2} + \frac{27 \alpha_1 (\text{tr}\hat{T})^2}{20} \notag \\
 &  + \frac{27 \alpha_2 (\text{tr}\hat{T})^2}{4} + \frac{5479 \alpha_1^2 \text{tr}\hat{B}}{28800} + \frac{29 \zeta_3 \alpha_1^2 \text{tr}\hat{B}}{100} - \frac{671 \alpha_1 \alpha_2 \text{tr}\hat{B}}{320} + \frac{9 \zeta_3 \alpha_1 \alpha_2 \text{tr}\hat{B}}{5} \notag \\
 &  - \frac{2433 \alpha_2^2 \text{tr}\hat{B}}{128} + \frac{63 \zeta_3 \alpha_2^2 \text{tr}\hat{B}}{4} + \frac{991 \alpha_1 \alpha_3 \text{tr}\hat{B}}{180} - 4 \zeta_3 \alpha_1 \alpha_3 \text{tr}\hat{B} + \frac{163 \alpha_2 \alpha_3 \text{tr}\hat{B}}{4} \notag \\
 &  - 36 \zeta_3 \alpha_2 \alpha_3 \text{tr}\hat{B} - \frac{910 \alpha_3^2 \text{tr}\hat{B}}{9} + 8 \zeta_3 \alpha_3^2 \text{tr}\hat{B} + \frac{45 \hat{\lambda}^2 \text{tr}\hat{B}}{2} + \frac{27 \alpha_1 \text{tr}\hat{T} \text{tr}\hat{B}}{10} + \frac{27 \alpha_2 \text{tr}\hat{T} \text{tr}\hat{B}}{2} \notag \\
 &  + \frac{27 \alpha_1 (\text{tr}\hat{B})^2}{20} + \frac{27 \alpha_2 (\text{tr}\hat{B})^2}{4} + \frac{8517 \alpha_1^2 \text{tr}\hat{L}}{3200} - \frac{117 \zeta_3 \alpha_1^2 \text{tr}\hat{L}}{100} + \frac{411 \alpha_1 \alpha_2 \text{tr}\hat{L}}{320} - \frac{18 \zeta_3 \alpha_1 \alpha_2 \text{tr}\hat{L}}{5} \notag \\
 &  - \frac{811 \alpha_2^2 \text{tr}\hat{L}}{128} + \frac{21 \zeta_3 \alpha_2^2 \text{tr}\hat{L}}{4} + \frac{15 \hat{\lambda}^2 \text{tr}\hat{L}}{2} + \frac{9 \alpha_1 \text{tr}\hat{T} \text{tr}\hat{L}}{10} + \frac{9 \alpha_2 \text{tr}\hat{T} \text{tr}\hat{L}}{2} + \frac{9 \alpha_1 \text{tr}\hat{B} \text{tr}\hat{L}}{10} \notag \\
 &  + \frac{9 \alpha_2 \text{tr}\hat{B} \text{tr}\hat{L}}{2} + \frac{3 \alpha_1 (\text{tr}\hat{L})^2}{20} + \frac{3 \alpha_2 (\text{tr}\hat{L})^2}{4} + \frac{25 \text{tr}\hat{B}^3}{16} - 3 \zeta_3 \text{tr}\hat{B}^3 + \frac{303 \alpha_1 \text{tr}\hat{B}^2}{80} - \frac{9 \zeta_3 \alpha_1 \text{tr}\hat{B}^2}{5} \notag \\
 &  + \frac{279 \alpha_2 \text{tr}\hat{B}^2}{16} - 9 \zeta_3 \alpha_2
 \text{tr}\hat{B}^2 - \frac{5 \alpha_3 \text{tr}\hat{B}^2}{2} + 24 \zeta_3
 \alpha_3 \text{tr}\hat{B}^2 - 15 \hat{\lambda} \text{tr}\hat{B}^2  
 \notag \\
 &  - 18 \text{tr}\hat{B} \text{tr}\hat{B}^2 - 6 \text{tr}\hat{L} \text{tr}\hat{B}^2 + \frac{25 \text{tr}\hat{L}^3}{48} - \zeta_3 \text{tr}\hat{L}^3 + \frac{33 \alpha_1 \text{tr}\hat{L}^2}{80} + \frac{9 \zeta_3 \alpha_1 \text{tr}\hat{L}^2}{5} + \frac{93 \alpha_2 \text{tr}\hat{L}^2}{16} \notag \\
 &  - 3 \zeta_3 \alpha_2 \text{tr}\hat{L}^2 - 5 \hat{\lambda}
 \text{tr}\hat{L}^2 - 6 \text{tr}\hat{T} \text{tr}\hat{L}^2 - 6
 \text{tr}\hat{B} \text{tr}\hat{L}^2 - 2 \text{tr}\hat{L} \text{tr}\hat{L}^2 
+ \frac{25 \text{tr}\hat{T}^3}{16} \notag \\
 &  - 3 \zeta_3 \text{tr}\hat{T}^3 + \frac{31 \alpha_1 \text{tr}\hat{T}\hat{B}}{40} - \frac{8 \zeta_3 \alpha_1 \text{tr}\hat{T}\hat{B}}{5} + \frac{21 \alpha_2 \text{tr}\hat{T}\hat{B}}{8} - 19 \alpha_3 \text{tr}\hat{T}\hat{B} + 16 \zeta_3 \alpha_3 \text{tr}\hat{T}\hat{B} \notag \\
 &  
+ \frac{\text{tr}\hat{L} \text{tr}\hat{T}\hat{B}}{2} 
+ \frac{211 \alpha_1 \text{tr}\hat{T}^2}{80} + \frac{3 \zeta_3 \alpha_1 \text{tr}\hat{T}^2}{5} \notag \\
 &  + \frac{279 \alpha_2 \text{tr}\hat{T}^2}{16} - 9 \zeta_3 \alpha_2 \text{tr}\hat{T}^2 - \frac{5 \alpha_3 \text{tr}\hat{T}^2}{2} + 24 \zeta_3 \alpha_3 \text{tr}\hat{T}^2 - 15 \hat{\lambda} \text{tr}\hat{T}^2 - 18 \text{tr}\hat{T} \text{tr}\hat{T}^2 \notag \\
 &  
- 6 \text{tr}\hat{L} \text{tr}\hat{T}^2 - \frac{507 \xi_W \alpha_2^3}{64} + \xi_W^2 \bigg( - \frac{39 \alpha_2^3}{32} - \frac{3 \zeta_3 \alpha_2^3}{8} \bigg) - \frac{5 \xi_W^3 \alpha_2^3}{16} \notag \\
 &  + n_G \bigg( - \frac{158 \alpha_1^3}{225} + \frac{19 \zeta_3 \alpha_1^3}{25} - \frac{3 \alpha_1^2 \alpha_2}{40} + \frac{9 \zeta_3 \alpha_1^2 \alpha_2}{25} - \frac{3 \alpha_1 \alpha_2^2}{10} + \frac{3 \zeta_3 \alpha_1 \alpha_2^2}{5} - \frac{2381 \alpha_2^3}{216} \notag \\
 &  - 15 \zeta_3 \alpha_2^3 - \frac{33 \alpha_1^2 \alpha_3}{20} + \frac{44 \zeta_3 \alpha_1^2 \alpha_3}{25} - \frac{45 \alpha_2^2 \alpha_3}{4} + 12 \zeta_3 \alpha_2^2 \alpha_3 + \frac{127 \alpha_1^2 \text{tr}\hat{T}}{120}  + \frac{21 \alpha_2^2 \text{tr}\hat{T}}{8} \notag \\
 &  + \frac{32 \alpha_3^2 \text{tr}\hat{T}}{3} + \frac{31 \alpha_1^2 \text{tr}\hat{B}}{120} + \frac{21 \alpha_2^2 \text{tr}\hat{B}}{8} + \frac{32 \alpha_3^2 \text{tr}\hat{B}}{3} + \frac{39 \alpha_1^2 \text{tr}\hat{L}}{40} + \frac{7 \alpha_2^2 \text{tr}\hat{L}}{8} + \frac{17 \xi_W \alpha_2^3}{8} \bigg) \notag \\
 &  + n_G^2 \bigg( - \frac{7 \alpha_1^3}{27} - \frac{35 \alpha_2^3}{27} \bigg) 
 -\frac{277\alpha_t\alpha_b^2}{16}  -\frac{277\alpha_b\alpha_t^2}{16}
\bigg] \bigg\} 
\,,
\end{align}
\begin{align}
Z_{cCW} &=
  1 - \frac{\xi_W \alpha_2}{4\pi} \frac{1}{\epsilon}+ \frac{\xi_W \alpha_2}{\lp4\pi\rp^2} \bigg\{ \frac{1}{\epsilon} \bigg[ - \frac{5 \alpha_2}{4} - \frac{\xi_W \alpha_2}{4} \bigg] + \frac{1}{\epsilon^2} \bigg[ \frac{3 \alpha_2}{2} + \xi_W \alpha_2 \bigg] \bigg\} \notag \\
 &  + \frac{\xi_W \alpha_2}{\lp4\pi\rp^3} \bigg\{ \frac{1}{\epsilon^3} \bigg[ - \frac{61 \alpha_2^2}{12} + \frac{2 n_G \alpha_2^2}{3} - 3 \xi_W \alpha_2^2 - \xi_W^2 \alpha_2^2 \bigg] \notag \\
 &  + \frac{1}{\epsilon^2} \bigg[ \frac{221 \alpha_2^2}{24} - \frac{5 n_G \alpha_2^2}{3} + 4 \xi_W \alpha_2^2 + \frac{3 \xi_W^2 \alpha_2^2}{4} \bigg] \notag \\
 &  + \frac{1}{\epsilon} \bigg[ - \frac{373 \alpha_2^2}{48} + \frac{5 n_G \alpha_2^2}{2} - \frac{13 \xi_W \alpha_2^2}{8} - \frac{5 \xi_W^2 \alpha_2^2}{12} \bigg] \bigg\} 
\,,
\end{align}
\begin{align}
Z_{cCG} &=
   1 - \frac{3}{2} \frac{\xi_G \alpha_3}{4\pi} \frac{1}{\epsilon} + \frac{\xi_G \alpha_3}{\lp4\pi\rp^2} \bigg\{ \frac{1}{\epsilon^2} \bigg[ \frac{27 \alpha_3}{8} + \frac{9 \xi_G \alpha_3}{4} \bigg] + \frac{1}{\epsilon} \bigg[ - \frac{45 \alpha_3}{16} - \frac{9 \xi_G \alpha_3}{16} \bigg] \bigg\} \notag \\
 &  + \frac{\xi_G \alpha_3}{\lp4\pi\rp^3} \bigg\{ \frac{1}{\epsilon^3} \bigg[ - \frac{279 \alpha_3^2}{16} + \frac{3 n_G \alpha_3^2}{2} - \frac{81 \xi_G \alpha_3^2}{8} - \frac{27 \xi_G^2 \alpha_3^2}{8} \bigg] \notag \\
 &  + \frac{1}{\epsilon^2} \bigg[ \frac{1035 \alpha_3^2}{32} - \frac{15 n_G \alpha_3^2}{4} + \frac{27 \xi_G \alpha_3^2}{2} + \frac{81 \xi_G^2 \alpha_3^2}{32} \bigg] \notag \\
 &  + \frac{1}{\epsilon} \bigg[ - \frac{1809 \alpha_3^2}{64} + \frac{45 n_G \alpha_3^2}{8} - \frac{351 \xi_G \alpha_3^2}{64} - \frac{45 \xi_G^2 \alpha_3^2}{32} \bigg] \bigg\} 
\,,
\end{align}
\begin{align}
Z_{WWW} &=
  1 + \frac{1}{\epsilon} \frac{\alpha_2}{4\pi} \bigg\{ \frac{8}{3} - \frac{4 n_G}{3} - \frac{3 \xi_W}{2} \bigg\} \notag \\
 &  + \frac{\alpha_2}{\lp4\pi\rp^2} \bigg\{ \frac{1}{\epsilon^2} \bigg[ - 6 \alpha_2 - \frac{7 \xi_W \alpha_2}{4} + \frac{15 \xi_W^2 \alpha_2}{8} + n_G \bigg( 3 \alpha_2 + 2 \xi_W \alpha_2 \bigg) \bigg] \notag \\
 &  + \frac{1}{\epsilon} \bigg[ - \frac{3 \alpha_1}{20} + \frac{499 \alpha_2}{48} + \frac{3 \text{tr}\hat{T}}{4} + \frac{3 \text{tr}\hat{B}}{4} + \frac{\text{tr}\hat{L}}{4} - \frac{33 \xi_W \alpha_2}{8} - \frac{3 \xi_W^2 \alpha_2}{4} \notag \\
 &  + n_G \bigg( - \frac{\alpha_1}{10} - \frac{17 \alpha_2}{3} - 2 \alpha_3 \bigg) \bigg] \bigg\} \notag \\
 &  + \frac{\alpha_2}{\lp4\pi\rp^3} \bigg\{ \frac{1}{\epsilon^3} \bigg[ \frac{70 \alpha_2^2}{3} + \frac{4 n_G^2 \alpha_2^2}{3} + \frac{59 \xi_W \alpha_2^2}{8} - \frac{5 \xi_W^2 \alpha_2^2}{8} - \frac{35 \xi_W^3 \alpha_2^2}{16} \notag \\
 &  + n_G \bigg( - \frac{43 \alpha_2^2}{3} - \frac{13 \xi_W \alpha_2^2}{2} - \frac{5 \xi_W^2 \alpha_2^2}{2} \bigg) \bigg] \notag \\
 &  + \frac{1}{\epsilon^2} \bigg[ - \frac{\alpha_1^2}{200} + \frac{9 \alpha_1 \alpha_2}{40} - \frac{26057 \alpha_2^2}{432} - \frac{17 \alpha_1 \text{tr}\hat{T}}{80} - \frac{27 \alpha_2 \text{tr}\hat{T}}{16} - 2 \alpha_3 \text{tr}\hat{T} - \frac{\alpha_1 \text{tr}\hat{B}}{16}\notag \\
 &  - \frac{27 \alpha_2 \text{tr}\hat{B}}{16} - 2 \alpha_3 \text{tr}\hat{B} - \frac{3 \alpha_1 \text{tr}\hat{L}}{16} - \frac{9 \alpha_2 \text{tr}\hat{L}}{16} + \frac{3 \text{tr}\hat{B}^2}{8} + \frac{3 (\text{tr}\hat{B})^2}{4} + \frac{\text{tr}\hat{B}\text{tr}\hat{L}}{2} + \frac{\text{tr}\hat{L}^2}{8} \notag \\
 &  + \frac{(\text{tr}\hat{L})^2}{12} - \frac{3 \text{tr}\hat{T}\hat{B}}{4} + \frac{3 \text{tr}\hat{T}^2}{8} + \frac{3 \text{tr}\hat{T}\text{tr}\hat{B}}{2} + \frac{\text{tr}\hat{T}\text{tr}\hat{L}}{2} + \frac{3 (\text{tr}\hat{T})^2}{4} \notag \\
 &  + \xi_W \bigg( \frac{9 \alpha_1 \alpha_2}{40} - \frac{165 \alpha_2^2}{32} - \frac{9 \alpha_2 \text{tr}\hat{T}}{8} - \frac{9 \alpha_2 \text{tr}\hat{B}}{8} - \frac{3 \alpha_2 \text{tr}\hat{L}}{8} \bigg) + \frac{157 \xi_W^2 \alpha_2^2}{16} + \frac{17 \xi_W^3 \alpha_2^2}{8} \notag \\
 &  + n_G \bigg( - \frac{7 \alpha_1^2}{100} + \frac{3 \alpha_1 \alpha_2}{20} + \frac{4433 \alpha_2^2}{108} + 3 \alpha_2 \alpha_3 + \frac{22 \alpha_3^2}{3} \notag \\
 &  + \xi_W \bigg( \frac{3 \alpha_1 \alpha_2}{20} + \frac{21 \alpha_2^2}{2} + 3 \alpha_2 \alpha_3 \bigg) + \xi_W^2 \alpha_2^2 \bigg) + n_G^2 \bigg( - \frac{2 \alpha_1^2}{45} - \frac{128 \alpha_2^2}{27} - \frac{8 \alpha_3^2}{9} \bigg) \bigg] \notag \\
 &  + \frac{1}{\epsilon} \bigg[ - \frac{163 \alpha_1^2}{4800} + \frac{11 \alpha_1 \alpha_2}{160} - \frac{9 \zeta_3 \alpha_1 \alpha_2}{20} + \frac{312361 \alpha_2^2}{5184}  + \frac{3 \zeta_3 \alpha_2^2}{4} - \frac{\alpha_1 \hat{\lambda}}{10} - \frac{\alpha_2 \hat{\lambda}}{2} + \hat{\lambda}^2 \notag \\
 &  + \frac{593 \alpha_1 \text{tr}\hat{T}}{480} - \frac{3 \alpha_2 \text{tr}\hat{T}}{32} + \frac{7 \alpha_3 \text{tr}\hat{T}}{3} + \frac{533 \alpha_1 \text{tr}\hat{B}}{480} - \frac{3 \alpha_2 \text{tr}\hat{B}}{32} + \frac{7 \alpha_3 \text{tr}\hat{B}}{3} + \frac{17 \alpha_1 \text{tr}\hat{L}}{32} \notag \\
 &  - \frac{\alpha_2 \text{tr}\hat{L}}{32} - \frac{19 \text{tr}\hat{B}^2}{16} - \frac{15 (\text{tr}\hat{B})^2}{8} - \frac{5 \text{tr}\hat{B}\text{tr}\hat{L}}{4} - \frac{19 \text{tr}\hat{L}^2}{48} - \frac{5 (\text{tr}\hat{L})^2}{24} - \frac{9 \text{tr}\hat{T}\hat{B}}{8} \notag \\
 &  - \frac{19 \text{tr}\hat{T}^2}{16} - \frac{15 \text{tr}\hat{T}\text{tr}\hat{B}}{4} - \frac{5 \text{tr}\hat{T}\text{tr}\hat{L}}{4} - \frac{15 (\text{tr}\hat{T})^2}{8} + \xi_W \bigg( - \frac{315 \alpha_2^2}{16} - 3 \zeta_3 \alpha_2^2 \bigg)\notag \\
 &  + \xi_W^2 \bigg( - \frac{33 \alpha_2^2}{8} - \frac{3 \zeta_3 \alpha_2^2}{4} \bigg) - \frac{7 \xi_W^3 \alpha_2^2}{8} + n_G \bigg( \frac{7 \alpha_1^2}{45} + \frac{109 \alpha_1 \alpha_2}{120} - \frac{6 \zeta_3 \alpha_1 \alpha_2}{5} \notag \\
 &  - \frac{37577 \alpha_2^2}{648} + 18 \zeta_3 \alpha_2^2 + \frac{\alpha_1 \alpha_3}{45} + \frac{109 \alpha_2 \alpha_3}{6} - 24 \zeta_3 \alpha_2 \alpha_3 - \frac{125 \alpha_3^2}{9} + 4 \xi_W \alpha_2^2 \bigg) \notag \\
 &  + n_G^2 \bigg( \frac{11 \alpha_1^2}{135} + \frac{625 \alpha_2^2}{81} + \frac{44 \alpha_3^2}{27} \bigg) \bigg] \bigg\}
\,,
\end{align}
\begin{align}
Z_{GGG} &=
  1 + \frac{1}{\epsilon} \frac{\alpha_3}{4\pi} \bigg\{ \frac{17}{4} - \frac{4 n_G}{3} - \frac{9 \xi_G}{4} \bigg\} \notag \\
 &  + \frac{\alpha_3}{\lp4\pi\rp^2} \bigg\{ \frac{1}{\epsilon^2} \bigg[ - \frac{459 \alpha_3}{32} - \frac{9 \xi_G \alpha_3}{2} + \frac{135 \xi_G^2 \alpha_3}{32} + n_G \bigg( \frac{9 \alpha_3}{2} + 3 \xi_G \alpha_3 \bigg) \bigg] \notag \\
 &  + \frac{1}{\epsilon} \bigg[\frac{777 \alpha_3}{32} + \text{tr}\hat{T} + \text{tr}\hat{B} - \frac{297 \xi_G \alpha_3}{32} - \frac{27 \xi_G^2 \alpha_3}{16} + n_G \bigg( - \frac{11 \alpha_1}{60} - \frac{3 \alpha_2}{4} - \frac{107 \alpha_3}{12} \bigg) \bigg] \bigg\} \notag \\
 &  + \frac{\alpha_3}{\lp4\pi\rp^3} \bigg\{ \frac{1}{\epsilon^3} \bigg[ \frac{10863 \alpha_3^2}{128} + 2 n_G^2 \alpha_3^2 + \frac{3537 \xi_G \alpha_3^2}{128} - \frac{135 \xi_G^2 \alpha_3^2}{128} - \frac{945 \xi_G^3 \alpha_3^2}{128} \notag \\
 &  + n_G \bigg( - 33 \alpha_3^2 - \frac{117 \xi_G \alpha_3^2}{8} - \frac{45 \xi_G^2 \alpha_3^2}{8} \bigg) \bigg] \notag \\
 &  + \frac{1}{\epsilon^2} \bigg[ - \frac{28079 \alpha_3^2}{128}  - \frac{17 \alpha_1 \text{tr}\hat{T}}{60} - \frac{3 \alpha_2 \text{tr}\hat{T}}{4} - \frac{59 \alpha_3 \text{tr}\hat{T}}{12} - \frac{\alpha_1 \text{tr}\hat{B}}{12} - \frac{3 \alpha_2 \text{tr}\hat{B}}{4} - \frac{59 \alpha_3 \text{tr}\hat{B}}{12} \notag \\
 &  + \frac{\text{tr}\hat{B}^2}{2} + (\text{tr}\hat{B})^2 + \frac{\text{tr}\hat{B}\text{tr}\hat{L}}{3} - \text{tr}\hat{T}\hat{B} + \frac{\text{tr}\hat{T}^2}{2} + 2 \text{tr}\hat{T}\text{tr}\hat{B} + \frac{\text{tr}\hat{T}\text{tr}\hat{L}}{3} + (\text{tr}\hat{T})^2 \notag \\
 &  + \xi_G \bigg( - \frac{621 \alpha_3^2}{32} - \frac{9 \alpha_3 \text{tr}\hat{T}}{4} - \frac{9 \alpha_3 \text{tr}\hat{B}}{4} \bigg) + \frac{4185 \xi_G^2 \alpha_3^2}{128} + \frac{459 \xi_G^3 \alpha_3^2}{64}\notag \\
 &  + n_G \bigg( - \frac{11 \alpha_1^2}{1800} + \frac{43 \alpha_2^2}{24} + \frac{33 \alpha_1 \alpha_3}{80} + \frac{27 \alpha_2 \alpha_3}{16} + \frac{14101 \alpha_3^2}{144} + \frac{9 \xi_G^2 \alpha_3^2}{4}\notag \\
 &  + \xi_G \bigg( \frac{33 \alpha_1 \alpha_3}{80} + \frac{27 \alpha_2 \alpha_3}{16} + \frac{393 \alpha_3^2}{16} \bigg) \bigg) + n_G^2 \bigg( - \frac{11 \alpha_1^2}{135} - \frac{\alpha_2^2}{3} - \frac{197 \alpha_3^2}{27} \bigg) \bigg] \notag \\
 &  + \frac{1}{\epsilon} \bigg[ \frac{43973 \alpha_3^2}{192} - \frac{243 \zeta_3 \alpha_3^2}{32} + \frac{101 \alpha_1 \text{tr}\hat{T}}{120} + \frac{31 \alpha_2 \text{tr}\hat{T}}{8} + \frac{113 \alpha_3 \text{tr}\hat{T}}{24} + \frac{89 \alpha_1 \text{tr}\hat{B}}{120} + \frac{31 \alpha_2 \text{tr}\hat{B}}{8} \notag \\
 &  + \frac{113 \alpha_3 \text{tr}\hat{B}}{24} - \frac{3 \text{tr}\hat{B}^2}{2} - \frac{7 (\text{tr}\hat{B})^2}{2} - \frac{7 \text{tr}\hat{B}\text{tr}\hat{L}}{6} + \text{tr}\hat{T}\hat{B} - \frac{3 \text{tr}\hat{T}^2}{2} - 7 \text{tr}\hat{T}\text{tr}\hat{B} - \frac{7 \text{tr}\hat{T}\text{tr}\hat{L}}{6} \notag \\
 &  - \frac{7 (\text{tr}\hat{T})^2}{2} + \xi_G \bigg( - \frac{4509 \alpha_3^2}{64} - \frac{81 \zeta_3 \alpha_3^2}{8} \bigg) + \xi_G^2 \bigg( - \frac{891 \alpha_3^2}{64} - \frac{81 \zeta_3 \alpha_3^2}{32} \bigg) - \frac{189 \xi_G^3 \alpha_3^2}{64} \notag \\
 &  + n_G \bigg( \frac{13 \alpha_1^2}{360} + \frac{\alpha_1 \alpha_2}{120} - \frac{241 \alpha_2^2}{72} + \frac{10901 \alpha_1 \alpha_3}{4320} - \frac{33 \zeta_3 \alpha_1 \alpha_3}{10} + \frac{991 \alpha_2 \alpha_3}{96} - \frac{27 \zeta_3 \alpha_2 \alpha_3}{2} \notag \\
 &  - \frac{14399 \alpha_3^2}{108} + 9 \xi_G \alpha_3^2 + 33 \zeta_3 \alpha_3^2 \bigg) + n_G^2 \bigg( \frac{121 \alpha_1^2}{810} + \frac{11 \alpha_2^2}{18} + \frac{965 \alpha_3^2}{81} \bigg) \bigg] \bigg\}
\,,
\end{align}
\begin{align}
Z_{HHW} =
& 1 + \frac{1}{4\pi}  \frac{1}{\epsilon} \bigg\{ \frac{9 \alpha_1}{20} + \frac{3 \alpha_2}{4} - 3 \text{tr}\hat{T} - 3 \text{tr}\hat{B} - \text{tr}\hat{L} - \frac{3 \xi_B \alpha_1}{20} - \frac{5 \xi_W \alpha_2}{4} \bigg\} \notag \\
 &  +\frac{1}{\lp4\pi\rp^2} \bigg\{ \frac{1}{\epsilon^2} \bigg[ \frac{99 \alpha_1^2}{800} + \frac{27 \alpha_1 \alpha_2}{80} - \frac{77 \alpha_2^2}{32} - \frac{3 \alpha_1 \text{tr}\hat{T}}{40} + \frac{9 \alpha_2 \text{tr}\hat{T}}{8} + 12 \alpha_3 \text{tr}\hat{T} - \frac{39 \alpha_1 \text{tr}\hat{B}}{40} \notag \\
 &  + \frac{9 \alpha_2 \text{tr}\hat{B}}{8} + 12 \alpha_3 \text{tr}\hat{B} + \frac{27 \alpha_1 \text{tr}\hat{L}}{40} + \frac{3 \alpha_2 \text{tr}\hat{L}}{8} - \frac{9 \text{tr}\hat{B}^2}{4} - \frac{3 \text{tr}\hat{L}^2}{4} + \frac{9 \text{tr}\hat{T}\hat{B}}{2} - \frac{9 \text{tr}\hat{T}^2}{4} \notag \\
 & + \xi_B \bigg( - \frac{27 \alpha_1^2}{400} - \frac{9 \alpha_1 \alpha_2}{80} + \frac{9 \alpha_1 \text{tr}\hat{T}}{20} + \frac{9 \alpha_1 \text{tr}\hat{B}}{20} + \frac{3 \alpha_1 \text{tr}\hat{L}}{20} + \frac{3 \xi_W \alpha_1 \alpha_2}{16} \bigg) \notag \\
 &  + \xi_W \bigg( - \frac{9 \alpha_1 \alpha_2}{16} + \frac{15 \alpha_2^2}{16} + \frac{15 \alpha_2 \text{tr}\hat{T}}{4} + \frac{15 \alpha_2 \text{tr}\hat{B}}{4} + \frac{5 \alpha_2 \text{tr}\hat{L}}{4} \bigg) + \frac{9 \xi_B^2 \alpha_1^2}{800} + \frac{45 \xi_W^2 \alpha_2^2}{32} \notag \\
 &  + n_G \bigg( \frac{3 \alpha_1^2}{10} + \frac{\alpha_2^2}{2} \bigg) \bigg] \notag \\
 &  + \frac{1}{\epsilon} \bigg[ - \frac{93 \alpha_1^2}{1600} - \frac{27 \alpha_1 \alpha_2}{160} + \frac{817 \alpha_2^2}{192} - 3 \hat{\lambda}^2 - \frac{17 \alpha_1 \text{tr}\hat{T}}{16} - \frac{45 \alpha_2 \text{tr}\hat{T}}{16} - 10 \alpha_3 \text{tr}\hat{T} \notag \\
 &  - \frac{5 \alpha_1 \text{tr}\hat{B}}{16} - \frac{45 \alpha_2 \text{tr}\hat{B}}{16} - 10 \alpha_3 \text{tr}\hat{B} - \frac{15 \alpha_1 \text{tr}\hat{L}}{16} - \frac{15 \alpha_2 \text{tr}\hat{L}}{16} + \frac{27 \text{tr}\hat{B}^2}{8} + \frac{9 \text{tr}\hat{L}^2}{8} \notag \\
 &  - \frac{3 \text{tr}\hat{T}\hat{B}}{4} + \frac{27 \text{tr}\hat{T}^2}{8} - \frac{23 \xi_W \alpha_2^2}{8} - \frac{7 \xi_W^2 \alpha_2^2}{16} + n_G \bigg( - \frac{\alpha_1^2}{4} - \frac{5 \alpha_2^2}{12} \bigg) \bigg] \bigg\} \notag \\
 &  + \frac{1}{\lp4\pi\rp^3} \bigg\{ \frac{1}{\epsilon^3} \bigg[ \frac{429 \alpha_1^3}{16000} + \frac{297 \alpha_1^2 \alpha_2}{3200} - \frac{693 \alpha_1 \alpha_2^2}{640} + \frac{12551 \alpha_2^3}{1152} - \frac{93 \alpha_1^2 \text{tr}\hat{T}}{800} - \frac{9 \alpha_1 \alpha_2 \text{tr}\hat{T}}{20} \notag \\
 &  - \frac{27 \alpha_2^2 \text{tr}\hat{T}}{32} - \frac{7 \alpha_1 \alpha_3 \text{tr}\hat{T}}{5} - 9 \alpha_2 \alpha_3 \text{tr}\hat{T} - 76 \alpha_3^2 \text{tr}\hat{T} - \frac{177 \alpha_1^2 \text{tr}\hat{B}}{800} + \frac{9 \alpha_1 \alpha_2 \text{tr}\hat{B}}{40} \notag \\
 &  - \frac{27 \alpha_2^2 \text{tr}\hat{B}}{32} + \frac{17 \alpha_1 \alpha_3 \text{tr}\hat{B}}{5} - 9 \alpha_2 \alpha_3 \text{tr}\hat{B} - 76 \alpha_3^2 \text{tr}\hat{B} - \frac{339 \alpha_1^2 \text{tr}\hat{L}}{800} - \frac{27 \alpha_1 \alpha_2 \text{tr}\hat{L}}{40} \notag \\
 &  - \frac{9 \alpha_2^2 \text{tr}\hat{L}}{32} - \frac{9 \text{tr}\hat{B}^3}{4} - \frac{9 \alpha_1 \text{tr}\hat{B}^2}{20} + \frac{27 \alpha_2 \text{tr}\hat{B}^2}{8} + 18 \alpha_3 \text{tr}\hat{B}^2 - \frac{9 \text{tr}\hat{T} \text{tr}\hat{B}^2}{4} \notag \\
 &  - \frac{9 \text{tr}\hat{B} \text{tr}\hat{B}^2}{4} - \frac{3 \text{tr}\hat{L} \text{tr}\hat{B}^2}{4} - \frac{3 \text{tr}\hat{L}^3}{4} + \frac{27 \alpha_1 \text{tr}\hat{L}^2}{20} + \frac{9 \alpha_2 \text{tr}\hat{L}^2}{8} - \frac{3 \text{tr}\hat{T} \text{tr}\hat{L}^2}{4} \notag \\
 &  - \frac{3 \text{tr}\hat{B} \text{tr}\hat{L}^2}{4} - \frac{\text{tr}\hat{L} \text{tr}\hat{L}^2}{4} + \frac{9 \text{tr}\hat{T}^2\hat{B}}{4} - \frac{9 \text{tr}\hat{T}^3}{4} - \frac{9 \alpha_1 \text{tr}\hat{T}\hat{B}}{20} - \frac{27 \alpha_2 \text{tr}\hat{T}\hat{B}}{4} \notag \\
 &  - 36 \alpha_3 \text{tr}\hat{T}\hat{B} + \frac{9 \text{tr}\hat{T} \text{tr}\hat{T}\hat{B}}{2} + \frac{9 \text{tr}\hat{B} \text{tr}\hat{T}\hat{B}}{2} + \frac{3 \text{tr}\hat{L} \text{tr}\hat{T}\hat{B}}{2} + \frac{9 \text{tr}\hat{T}\hat{B}^2}{4} + \frac{9 \alpha_1 \text{tr}\hat{T}^2}{10} \notag \\
 &  + \frac{27 \alpha_2 \text{tr}\hat{T}^2}{8} + 18 \alpha_3 \text{tr}\hat{T}^2 - \frac{9 \text{tr}\hat{T} \text{tr}\hat{T}^2}{4} - \frac{9 \text{tr}\hat{B} \text{tr}\hat{T}^2}{4} - \frac{3 \text{tr}\hat{L} \text{tr}\hat{T}^2}{4} \notag \\
 &  + \xi_B \bigg( - \frac{297 \alpha_1^3}{16000} - \frac{81 \alpha_1^2 \alpha_2}{1600} + \frac{231 \alpha_1 \alpha_2^2}{640} + \frac{9 \alpha_1^2 \text{tr}\hat{T}}{800} - \frac{27 \alpha_1 \alpha_2 \text{tr}\hat{T}}{160} \notag \\
 &  - \frac{9 \alpha_1 \alpha_3 \text{tr}\hat{T}}{5} + \frac{117 \alpha_1^2 \text{tr}\hat{B}}{800} - \frac{27 \alpha_1 \alpha_2 \text{tr}\hat{B}}{160} - \frac{9 \alpha_1 \alpha_3 \text{tr}\hat{B}}{5} - \frac{81 \alpha_1^2 \text{tr}\hat{L}}{800} \notag \\
 &  - \frac{9 \alpha_1 \alpha_2 \text{tr}\hat{L}}{160} + \frac{27 \alpha_1 \text{tr}\hat{B}^2}{80} + \frac{9 \alpha_1 \text{tr}\hat{L}^2}{80} - \frac{27 \alpha_1 \text{tr}\hat{T}\hat{B}}{40} + \frac{27 \alpha_1 \text{tr}\hat{T}^2}{80} \notag \\
 &  + \xi_W \bigg( \frac{27 \alpha_1^2 \alpha_2}{320} - \frac{9 \alpha_1 \alpha_2^2}{64} - \frac{9 \alpha_1 \alpha_2 \text{tr}\hat{T}}{16} - \frac{9 \alpha_1 \alpha_2 \text{tr}\hat{B}}{16} - \frac{3 \alpha_1 \alpha_2 \text{tr}\hat{L}}{16} \bigg) - \frac{27 \xi_W^2 \alpha_1 \alpha_2^2}{128} \bigg) \notag \\
 &  + \xi_W \bigg( - \frac{99 \alpha_1^2 \alpha_2}{640} + \frac{27 \alpha_1 \alpha_2^2}{64} - \frac{745 \alpha_2^3}{384} + \frac{3 \alpha_1 \alpha_2 \text{tr}\hat{T}}{32} - \frac{225 \alpha_2^2 \text{tr}\hat{T}}{32} - 15 \alpha_2 \alpha_3 \text{tr}\hat{T} \notag \\
 &  + \frac{39 \alpha_1 \alpha_2 \text{tr}\hat{B}}{32} - \frac{225 \alpha_2^2 \text{tr}\hat{B}}{32} - 15 \alpha_2 \alpha_3 \text{tr}\hat{B} - \frac{27 \alpha_1 \alpha_2 \text{tr}\hat{L}}{32} - \frac{75 \alpha_2^2 \text{tr}\hat{L}}{32} + \frac{45 \alpha_2 \text{tr}\hat{B}^2}{16} \notag \\
 &  + \frac{15 \alpha_2 \text{tr}\hat{L}^2}{16} - \frac{45 \alpha_2 \text{tr}\hat{T}\hat{B}}{8} + \frac{45 \alpha_2 \text{tr}\hat{T}^2}{16} \bigg) + \xi_B^2 \bigg( \frac{81 \alpha_1^3}{16000} + \frac{27 \alpha_1^2 \alpha_2}{3200} \notag \\
 &  - \frac{27 \alpha_1^2 \text{tr}\hat{T}}{800} - \frac{27 \alpha_1^2 \text{tr}\hat{B}}{800} - \frac{9 \alpha_1^2 \text{tr}\hat{L}}{800} - \frac{9 \xi_W \alpha_1^2 \alpha_2}{640} \bigg) \notag \\
 &  + \xi_W^2 \bigg( \frac{81 \alpha_1 \alpha_2^2}{128} - \frac{405 \alpha_2^3}{128} - \frac{135 \alpha_2^2 \text{tr}\hat{T}}{32} - \frac{135 \alpha_2^2 \text{tr}\hat{B}}{32} - \frac{45 \alpha_2^2 \text{tr}\hat{L}}{32} \bigg) \notag \\
 &  - \frac{9 \xi_B^3 \alpha_1^3}{16000} - \frac{195 \xi_W^3 \alpha_2^3}{128} \notag \\
 &  + n_G \bigg( \frac{7 \alpha_1^3}{40} + \frac{9 \alpha_1^2 \alpha_2}{40} + \frac{9 \alpha_1 \alpha_2^2}{40} - \frac{317 \alpha_2^3}{72} - \frac{\alpha_1^2 \text{tr}\hat{T}}{3} + \frac{16 \alpha_3^2 \text{tr}\hat{T}}{3} - \frac{11 \alpha_1^2 \text{tr}\hat{B}}{15} \notag \\
 &  + \frac{16 \alpha_3^2 \text{tr}\hat{B}}{3} + \frac{\alpha_1^2 \text{tr}\hat{L}}{5} + \xi_B \bigg( - \frac{9 \alpha_1^3}{200} - \frac{3 \alpha_1 \alpha_2^2}{40} \bigg) + \xi_W \bigg( - \frac{3 \alpha_1^2 \alpha_2}{8} + \frac{5 \alpha_2^3}{24} \bigg) \bigg) \notag \\
 &  + n_G^2 \bigg( \frac{4 \alpha_1^3}{15} + \frac{4 \alpha_2^3}{9} \bigg) \bigg] \notag \\
 &  + \frac{1}{\epsilon^2} \bigg[ - \frac{97 \alpha_1^3}{32000} + \frac{63 \alpha_1^2 \alpha_2}{6400} + \frac{2901 \alpha_1 \alpha_2^2}{1280} - \frac{193057 \alpha_2^3}{6912} - \frac{27 \alpha_1^2 \hat{\lambda}}{200} - \frac{9 \alpha_1 \alpha_2 \hat{\lambda}}{20} \notag \\
 &  - \frac{9 \alpha_2^2 \hat{\lambda}}{8} + \frac{9 \alpha_1 \hat{\lambda}^2}{20} + \frac{27 \alpha_2 \hat{\lambda}^2}{4} - 24 \hat{\lambda}^3 - \frac{541 \alpha_1^2 \text{tr}\hat{T}}{1600} + \frac{39 \alpha_1 \alpha_2 \text{tr}\hat{T}}{80} + \frac{149 \alpha_2^2 \text{tr}\hat{T}}{64} \notag \\
 &  - \frac{\alpha_1 \alpha_3 \text{tr}\hat{T}}{10} - \frac{3 \alpha_2 \alpha_3 \text{tr}\hat{T}}{2} + 198 \alpha_3^2 \text{tr}\hat{T} - 9 \hat{\lambda}^2 \text{tr}\hat{T} + \frac{191 \alpha_1^2 \text{tr}\hat{B}}{1600} + \frac{33 \alpha_1 \alpha_2 \text{tr}\hat{B}}{40} \notag \\
 &  + \frac{149 \alpha_2^2 \text{tr}\hat{B}}{64} - \frac{49 \alpha_1 \alpha_3 \text{tr}\hat{B}}{10} - \frac{3 \alpha_2 \alpha_3 \text{tr}\hat{B}}{2} + 198 \alpha_3^2 \text{tr}\hat{B} - 9 \hat{\lambda}^2 \text{tr}\hat{B} - \frac{3 \alpha_1^2 \text{tr}\hat{L}}{1600} \notag \\
 &  + \frac{149 \alpha_2^2 \text{tr}\hat{L}}{192} - 3 \hat{\lambda}^2 \text{tr}\hat{L} + \frac{15 \text{tr}\hat{B}^3}{8} - \frac{123 \alpha_1 \text{tr}\hat{B}^2}{80} - \frac{99 \alpha_2 \text{tr}\hat{B}^2}{8} - 39 \alpha_3 \text{tr}\hat{B}^2 \notag \\
 &  + 18 \hat{\lambda} \text{tr}\hat{B}^2 
+ \frac{81 \text{tr}\hat{B} \text{tr}\hat{B}^2}{8} + \frac{27 \text{tr}\hat{L} \text{tr}\hat{B}^2}{8} + \frac{5 \text{tr}\hat{L}^3}{8} - \frac{261 \alpha_1 \text{tr}\hat{L}^2}{80} \notag \\
 &  - \frac{33 \alpha_2 \text{tr}\hat{L}^2}{8} + 6 \hat{\lambda}
 \text{tr}\hat{L}^2 + \frac{27 \text{tr}\hat{T} \text{tr}\hat{L}^2}{8} +
 \frac{27 \text{tr}\hat{B} \text{tr}\hat{L}^2}{8} + \frac{9 \text{tr}\hat{L}
   \text{tr}\hat{L}^2}{8} 
\notag \\
 &  + \frac{15 \text{tr}\hat{T}^3}{8} + \frac{43 \alpha_1
   \text{tr}\hat{T}\hat{B}}{20} + \frac{9 \alpha_2 \text{tr}\hat{T}\hat{B}}{4}
 + 46 \alpha_3 \text{tr}\hat{T}\hat{B} 
+ \frac{81 \text{tr}\hat{T} \text{tr}\hat{T}^2}{8} + \frac{27 \text{tr}\hat{L} \text{tr}\hat{T}^2}{8} 
 \notag \\
 &  - \frac{11 \text{tr}\hat{L} \text{tr}\hat{T}\hat{B}}{4} 
- \frac{297 \alpha_1 \text{tr}\hat{T}^2}{80} - \frac{99 \alpha_2 \text{tr}\hat{T}^2}{8} - 39 \alpha_3 \text{tr}\hat{T}^2 + 18 \hat{\lambda} \text{tr}\hat{T}^2 \notag \\
 &  + \xi_B \bigg( \frac{279 \alpha_1^3}{32000} + \frac{81 \alpha_1^2 \alpha_2}{3200} - \frac{817 \alpha_1 \alpha_2^2}{1280} + \frac{9 \alpha_1 \hat{\lambda}^2}{20} + \frac{51 \alpha_1^2 \text{tr}\hat{T}}{320} + \frac{27 \alpha_1 \alpha_2 \text{tr}\hat{T}}{64} \notag \\
 &  + \frac{3 \alpha_1 \alpha_3 \text{tr}\hat{T}}{2} + \frac{3 \alpha_1^2 \text{tr}\hat{B}}{64} + \frac{27 \alpha_1 \alpha_2 \text{tr}\hat{B}}{64} + \frac{3 \alpha_1 \alpha_3 \text{tr}\hat{B}}{2} + \frac{9 \alpha_1^2 \text{tr}\hat{L}}{64} + \frac{9 \alpha_1 \alpha_2 \text{tr}\hat{L}}{64} \notag \\
 &  - \frac{81 \alpha_1 \text{tr}\hat{B}^2}{160} - \frac{27 \alpha_1 \text{tr}\hat{L}^2}{160} + \frac{9 \alpha_1 \text{tr}\hat{T}\hat{B}}{80} - \frac{81 \alpha_1 \text{tr}\hat{T}^2}{160} + \frac{69 \xi_W \alpha_1 \alpha_2^2}{160} + \frac{21 \xi_W^2 \alpha_1 \alpha_2^2}{320} \bigg) \notag \\
 &  + \xi_W \bigg( \frac{93 \alpha_1^2 \alpha_2}{1280} - \frac{693 \alpha_1 \alpha_2^2}{640} + \frac{6515 \alpha_2^3}{768} + \frac{15 \alpha_2 \hat{\lambda}^2}{4} + \frac{85 \alpha_1 \alpha_2 \text{tr}\hat{T}}{64} + \frac{777 \alpha_2^2 \text{tr}\hat{T}}{64} \notag \\
 &  + \frac{25 \alpha_2 \alpha_3 \text{tr}\hat{T}}{2} + \frac{25 \alpha_1 \alpha_2 \text{tr}\hat{B}}{64} + \frac{777 \alpha_2^2 \text{tr}\hat{B}}{64} + \frac{25 \alpha_2 \alpha_3 \text{tr}\hat{B}}{2} + \frac{75 \alpha_1 \alpha_2 \text{tr}\hat{L}}{64} \notag \\
 &  + \frac{259 \alpha_2^2 \text{tr}\hat{L}}{64} - \frac{135 \alpha_2 \text{tr}\hat{B}^2}{32} - \frac{45 \alpha_2 \text{tr}\hat{L}^2}{32} + \frac{15 \alpha_2 \text{tr}\hat{T}\hat{B}}{16} - \frac{135 \alpha_2 \text{tr}\hat{T}^2}{32} \bigg) \notag \\
 &  + \xi_W^2 \bigg( - \frac{63 \alpha_1 \alpha_2^2}{320} + \frac{473 \alpha_2^3}{64} + \frac{21 \alpha_2^2 \text{tr}\hat{T}}{16} + \frac{21 \alpha_2^2 \text{tr}\hat{B}}{16} + \frac{7 \alpha_2^2 \text{tr}\hat{L}}{16} \bigg) + \frac{241 \xi_W^3 \alpha_2^3}{192} \notag \\
 &  + n_G \bigg( \frac{11 \alpha_1^3}{1200} - \frac{69 \alpha_1^2 \alpha_2}{400} - \frac{17 \alpha_1 \alpha_2^2}{80} + \frac{4123 \alpha_2^3}{432} + \frac{11 \alpha_1^2 \alpha_3}{25} + \alpha_2^2 \alpha_3 - \frac{11 \alpha_1^2 \text{tr}\hat{T}}{30} \notag \\
 &  - \alpha_2^2 \text{tr}\hat{T} - \frac{40 \alpha_3^2 \text{tr}\hat{T}}{3} + \frac{19 \alpha_1^2 \text{tr}\hat{B}}{30} - \alpha_2^2 \text{tr}\hat{B} - \frac{40 \alpha_3^2 \text{tr}\hat{B}}{3} - \frac{9 \alpha_1^2 \text{tr}\hat{L}}{10} - \frac{\alpha_2^2 \text{tr}\hat{L}}{3} \notag \\
 &  + \xi_B \bigg( \frac{3 \alpha_1^3}{80} + \frac{\alpha_1 \alpha_2^2}{16}
 \bigg) + \xi_W \bigg( \frac{5 \alpha_1^2 \alpha_2}{16} - \frac{103
   \alpha_2^3}{48} \bigg) \bigg) + n_G^2 \bigg( - \frac{2 \alpha_1^3}{9} -
 \frac{10 \alpha_2^3}{27} \bigg) 
 \notag \\
 & -3 \alpha_t\alpha_b^2 -3 \alpha_t^2\alpha_b 
 \bigg] \notag \\
 &  + \frac{1}{\epsilon} \bigg[ - \frac{413 \alpha_1^3}{6000} + \frac{27 \zeta_3 \alpha_1^3}{2000} - \frac{279 \alpha_1^2 \alpha_2}{800} - \frac{27 \zeta_3 \alpha_1^2 \alpha_2}{400} - \frac{123 \alpha_1 \alpha_2^2}{320} - \frac{3 \zeta_3 \alpha_1 \alpha_2^2}{80} \notag \\
 &  + \frac{93307 \alpha_2^3}{5184} + \frac{73 \zeta_3 \alpha_2^3}{16} - \frac{117 \alpha_1^2 \hat{\lambda}}{400} + \frac{27 \zeta_3 \alpha_1^2 \hat{\lambda}}{50} - \frac{39 \alpha_1 \alpha_2 \hat{\lambda}}{40} + \frac{9 \zeta_3 \alpha_1 \alpha_2 \hat{\lambda}}{5} - \frac{39 \alpha_2^2 \hat{\lambda}}{16} \notag \\
 &  + \frac{9 \zeta_3 \alpha_2^2 \hat{\lambda}}{2} - 3 \alpha_1 \hat{\lambda}^2 - 15 \alpha_2 \hat{\lambda}^2 + 12 \hat{\lambda}^3 + \frac{52831 \alpha_1^2 \text{tr}\hat{T}}{28800} - \frac{\zeta_3 \alpha_1^2 \text{tr}\hat{T}}{100} - \frac{371 \alpha_1 \alpha_2 \text{tr}\hat{T}}{320} \notag \\
 &  - \frac{27 \zeta_3 \alpha_1 \alpha_2 \text{tr}\hat{T}}{10} - \frac{2761 \alpha_2^2 \text{tr}\hat{T}}{128} + \frac{63 \zeta_3 \alpha_2^2 \text{tr}\hat{T}}{4} + \frac{2419 \alpha_1 \alpha_3 \text{tr}\hat{T}}{180} - \frac{68 \zeta_3 \alpha_1 \alpha_3 \text{tr}\hat{T}}{5} \notag \\
 &  + \frac{163 \alpha_2 \alpha_3 \text{tr}\hat{T}}{4} - 36 \zeta_3 \alpha_2 \alpha_3 \text{tr}\hat{T} - \frac{910 \alpha_3^2 \text{tr}\hat{T}}{9} + 8 \zeta_3 \alpha_3^2 \text{tr}\hat{T} + \frac{45 \hat{\lambda}^2 \text{tr}\hat{T}}{2} \notag \\
 &  + \frac{27 \alpha_1 (\text{tr}\hat{T})^2}{20} + \frac{27 \alpha_2 (\text{tr}\hat{T})^2}{4} + \frac{5479 \alpha_1^2 \text{tr}\hat{B}}{28800} + \frac{29 \zeta_3 \alpha_1^2 \text{tr}\hat{B}}{100} - \frac{671 \alpha_1 \alpha_2 \text{tr}\hat{B}}{320} \notag \\
 &  + \frac{9 \zeta_3 \alpha_1 \alpha_2 \text{tr}\hat{B}}{5} - \frac{2761 \alpha_2^2 \text{tr}\hat{B}}{128} + \frac{63 \zeta_3 \alpha_2^2 \text{tr}\hat{B}}{4} + \frac{991 \alpha_1 \alpha_3 \text{tr}\hat{B}}{180} - 4 \zeta_3 \alpha_1 \alpha_3 \text{tr}\hat{B} \notag \\
 &  + \frac{163 \alpha_2 \alpha_3 \text{tr}\hat{B}}{4} - 36 \zeta_3 \alpha_2 \alpha_3 \text{tr}\hat{B} - \frac{910 \alpha_3^2 \text{tr}\hat{B}}{9} + 8 \zeta_3 \alpha_3^2 \text{tr}\hat{B} + \frac{45 \hat{\lambda}^2 \text{tr}\hat{B}}{2} \notag \\
 &  + \frac{27 \alpha_1 \text{tr}\hat{T} \text{tr}\hat{B}}{10} + \frac{27 \alpha_2 \text{tr}\hat{T} \text{tr}\hat{B}}{2} + \frac{27 \alpha_1 (\text{tr}\hat{B})^2}{20} + \frac{27 \alpha_2 (\text{tr}\hat{B})^2}{4} + \frac{8517 \alpha_1^2 \text{tr}\hat{L}}{3200} \notag \\
 &  - \frac{117 \zeta_3 \alpha_1^2 \text{tr}\hat{L}}{100} + \frac{411 \alpha_1 \alpha_2 \text{tr}\hat{L}}{320} - \frac{18 \zeta_3 \alpha_1 \alpha_2 \text{tr}\hat{L}}{5} - \frac{2761 \alpha_2^2 \text{tr}\hat{L}}{384} + \frac{21 \zeta_3 \alpha_2^2 \text{tr}\hat{L}}{4} \notag \\
 &  + \frac{15 \hat{\lambda}^2 \text{tr}\hat{L}}{2} + \frac{9 \alpha_1 \text{tr}\hat{T} \text{tr}\hat{L}}{10} + \frac{9 \alpha_2 \text{tr}\hat{T} \text{tr}\hat{L}}{2} + \frac{9 \alpha_1 \text{tr}\hat{B} \text{tr}\hat{L}}{10} + \frac{9 \alpha_2 \text{tr}\hat{B} \text{tr}\hat{L}}{2} \notag \\
 &  + \frac{3 \alpha_1 (\text{tr}\hat{L})^2}{20} + \frac{3 \alpha_2 (\text{tr}\hat{L})^2}{4} + \frac{25 \text{tr}\hat{B}^3}{16} - 3 \zeta_3 \text{tr}\hat{B}^3 + \frac{303 \alpha_1 \text{tr}\hat{B}^2}{80} - \frac{9 \zeta_3 \alpha_1 \text{tr}\hat{B}^2}{5} \notag \\
 &  + \frac{279 \alpha_2 \text{tr}\hat{B}^2}{16} - 9 \zeta_3 \alpha_2 \text{tr}\hat{B}^2 - \frac{5 \alpha_3 \text{tr}\hat{B}^2}{2} + 24 \zeta_3 \alpha_3 \text{tr}\hat{B}^2 - 15 \hat{\lambda} \text{tr}\hat{B}^2 \notag \\
 &  
- 18 \text{tr}\hat{B} \text{tr}\hat{B}^2 - 6 \text{tr}\hat{L} \text{tr}\hat{B}^2 + \frac{25 \text{tr}\hat{L}^3}{48} - \zeta_3 \text{tr}\hat{L}^3 + \frac{33 \alpha_1 \text{tr}\hat{L}^2}{80} \notag \\
 &  + \frac{9 \zeta_3 \alpha_1 \text{tr}\hat{L}^2}{5} + \frac{93 \alpha_2 \text{tr}\hat{L}^2}{16} - 3 \zeta_3 \alpha_2 \text{tr}\hat{L}^2 - 5 \hat{\lambda} \text{tr}\hat{L}^2 - 6 \text{tr}\hat{T} \text{tr}\hat{L}^2 - 6 \text{tr}\hat{B} \text{tr}\hat{L}^2 \notag \\
 &  - 2 \text{tr}\hat{L} \text{tr}\hat{L}^2 
+ \frac{25 \text{tr}\hat{T}^3}{16} - 3 \zeta_3 \text{tr}\hat{T}^3 + \frac{31 \alpha_1 \text{tr}\hat{T}\hat{B}}{40} - \frac{8 \zeta_3 \alpha_1 \text{tr}\hat{T}\hat{B}}{5} \notag \\
 &  + \frac{21 \alpha_2 \text{tr}\hat{T}\hat{B}}{8} - 19 \alpha_3
 \text{tr}\hat{T}\hat{B} + 16 \zeta_3 \alpha_3 \text{tr}\hat{T}\hat{B} 
\notag \\
 &  + \frac{\text{tr}\hat{L} \text{tr}\hat{T}\hat{B}}{2} 
+ \frac{211 \alpha_1 \text{tr}\hat{T}^2}{80} + \frac{3 \zeta_3 \alpha_1 \text{tr}\hat{T}^2}{5} + \frac{279 \alpha_2 \text{tr}\hat{T}^2}{16} - 9 \zeta_3 \alpha_2 \text{tr}\hat{T}^2 \notag \\
 &  - \frac{5 \alpha_3 \text{tr}\hat{T}^2}{2} + 24 \zeta_3 \alpha_3
 \text{tr}\hat{T}^2 - 15 \hat{\lambda} \text{tr}\hat{T}^2 
- 18 \text{tr}\hat{T} \text{tr}\hat{T}^2 
- 6 \text{tr}\hat{L} \text{tr}\hat{T}^2 \notag \\
 &  + \xi_W \bigg( - \frac{927 \alpha_2^3}{64} - \zeta_3 \alpha_2^3 \bigg) + \xi_W^2 \bigg( - \frac{83 \alpha_2^3}{32} - \frac{5 \zeta_3 \alpha_2^3}{8} \bigg) - \frac{29 \xi_W^3 \alpha_2^3}{48} \notag \\
 &  + n_G \bigg( - \frac{158 \alpha_1^3}{225} + \frac{19 \zeta_3 \alpha_1^3}{25} - \frac{3 \alpha_1^2 \alpha_2}{40} + \frac{9 \zeta_3 \alpha_1^2 \alpha_2}{25} + \frac{3 \alpha_1 \alpha_2^2}{40} + \frac{\zeta_3 \alpha_1 \alpha_2^2}{5} - \frac{1285 \alpha_2^3}{324} \notag \\
 &  - 9 \zeta_3 \alpha_2^3 - \frac{33 \alpha_1^2 \alpha_3}{20} + \frac{44 \zeta_3 \alpha_1^2 \alpha_3}{25} - \frac{15 \alpha_2^2 \alpha_3}{4} + 4 \zeta_3 \alpha_2^2 \alpha_3 + \frac{127 \alpha_1^2 \text{tr}\hat{T}}{120} + \frac{21 \alpha_2^2 \text{tr}\hat{T}}{8} \notag \\
 & + \frac{32 \alpha_3^2 \text{tr}\hat{T}}{3} + \frac{31 \alpha_1^2 \text{tr}\hat{B}}{120} + \frac{21 \alpha_2^2 \text{tr}\hat{B}}{8} + \frac{32 \alpha_3^2 \text{tr}\hat{B}}{3} + \frac{39 \alpha_1^2 \text{tr}\hat{L}}{40} + \frac{7 \alpha_2^2 \text{tr}\hat{L}}{8} + \frac{83 \xi_W \alpha_2^3}{24} \bigg) \notag \\
 & + n_G^2 \bigg( - \frac{7 \alpha_1^3}{27} - \frac{35 \alpha_2^3}{81} \bigg) 
 -\frac{277\alpha_t\alpha_b^2}{16}  -\frac{277\alpha_b\alpha_t^2}{16} \bigg] \bigg\}\,.
\end{align}

Note that in the results for $Z_H$ and $Z_{HHW}$ there are terms which
contain explicitly $\alpha_b$ and $\alpha_t$ since we have not been able to
reconstruct the corresponding expressions in terms of $\hat{B}$ and
$\hat{T}$. They drop out in the final result for
$Z_{\alpha_2}$. 


\section{\label{app::renconst2}Two-loop Yukawa coupling renormalization constants}

This Subsection contains the two-loop results for the Yukawa coupling
renormalization constants defined through
\begin{eqnarray}
  \alpha_i^{\rm bare} = Z_{\alpha_i} \alpha_i
  \,,
\end{eqnarray}
with $i=t,b,\tau$. They read
\begin{align}
  Z_{\alpha_t} &=
  1 + \frac{1}{4\pi} \frac{1}{\epsilon} \bigg\{ - \frac{17 \alpha_1}{20} - \frac{9 \alpha_2}{4} - 8 \alpha_3 + \frac{3 \alpha_t}{2} - \frac{3 \alpha_b}{2} + 3 \text{tr}\hat{T} + 3 \text{tr}\hat{B} + \text{tr}\hat{L} \bigg\} \notag \\
 &  + \frac{1}{\lp4\pi\rp^2} \bigg\{ \frac{1}{\epsilon^2} \bigg[ \frac{51 \alpha_1^2}{160} + \frac{153 \alpha_1 \alpha_2}{80} + \frac{339 \alpha_2^2}{32} + \frac{34 \alpha_1 \alpha_3}{5} + 18 \alpha_2 \alpha_3 + 76 \alpha_3^2 - \frac{459 \alpha_1 \alpha_t}{80} \notag \\
 &  - \frac{243 \alpha_2 \alpha_t}{16} - 54 \alpha_3 \alpha_t + \frac{81 \alpha_t^2}{4} - \frac{117 \alpha_1 \alpha_b}{80} - \frac{81 \alpha_2 \alpha_b}{16} - 18 \alpha_3 \alpha_b + \frac{45 \alpha_t \alpha_b}{4} + \frac{9 \alpha_b^2}{2} \notag \\
 &  - \frac{79 \alpha_1 \alpha_{\tau}}{40} - \frac{27 \alpha_2 \alpha_{\tau}}{8} - 8 \alpha_3 \alpha_{\tau} + \frac{33 \alpha_t \alpha_{\tau}}{4} + \frac{15 \alpha_b \alpha_{\tau}}{4} + \frac{7 \alpha_{\tau}^2}{4} \notag \\
 &  + n_G \bigg( - \frac{17 \alpha_1^2}{30} - \frac{3 \alpha_2^2}{2} - \frac{16 \alpha_3^2}{3} \bigg)  \bigg] \notag \\
 &  + \frac{1}{\epsilon} \bigg[ \frac{9 \alpha_1^2}{400} - \frac{9 \alpha_1 \alpha_2}{40} - \frac{35 \alpha_2^2}{8} + \frac{19 \alpha_1 \alpha_3}{30} + \frac{9 \alpha_2 \alpha_3}{2} - \frac{202 \alpha_3^2}{3} + 3 \hat{\lambda}^2 + \frac{393 \alpha_1 \alpha_t}{160} \notag \\
 &  + \frac{225 \alpha_2 \alpha_t}{32} + 18 \alpha_3 \alpha_t - 6 \hat{\lambda} \alpha_t - 6 \alpha_t^2 + \frac{7 \alpha_1 \alpha_b}{160} + \frac{99 \alpha_2 \alpha_b}{32} + 2 \alpha_3 \alpha_b - \frac{11 \alpha_t \alpha_b}{8} - \frac{\alpha_b^2}{8} \notag \\
 &  + \frac{15 \alpha_1 \alpha_{\tau}}{16} + \frac{15 \alpha_2 \alpha_{\tau}}{16} - \frac{9 \alpha_t \alpha_{\tau}}{8} + \frac{5 \alpha_b \alpha_{\tau}}{8} - \frac{9 \alpha_{\tau}^2}{8}  + n_G \bigg( \frac{29 \alpha_1^2}{90} + \frac{\alpha_2^2}{2} + \frac{40 \alpha_3^2}{9} \bigg) \bigg] \bigg\}
\,,
\end{align}
\begin{align}
  Z_{\alpha_b} &=
  1 + \frac{1}{4\pi} \frac{1}{\epsilon} \bigg\{ - \frac{\alpha_1}{4} - \frac{9 \alpha_2}{4} - 8 \alpha_3 - \frac{3 \alpha_t}{2} + \frac{3 \alpha_b}{2} + 3 \text{tr}\hat{T} + 3 \text{tr}\hat{B} + \text{tr}\hat{L} \bigg\} \notag \\
 &  + \frac{1}{\lp4\pi\rp^2} \bigg\{\frac{1}{\epsilon^2} \bigg[ \frac{3 \alpha_1^2}{160} + \frac{9 \alpha_1 \alpha_2}{16} + \frac{339 \alpha_2^2}{32} + 2 \alpha_1 \alpha_3 + 18 \alpha_2 \alpha_3 + 76 \alpha_3^2 - \frac{81 \alpha_1 \alpha_t}{80} - \frac{81 \alpha_2 \alpha_t}{16} \notag \\
 &  - 18 \alpha_3 \alpha_t + \frac{9 \alpha_t^2}{2} - \frac{27 \alpha_1 \alpha_b}{16} - \frac{243 \alpha_2 \alpha_b}{16} - 54 \alpha_3 \alpha_b + \frac{45 \alpha_t \alpha_b}{4} + \frac{81 \alpha_b^2}{4} - \frac{11 \alpha_1 \alpha_{\tau}}{8} \notag \\
 &  - \frac{27 \alpha_2 \alpha_{\tau}}{8} - 8 \alpha_3 \alpha_{\tau} + \frac{15 \alpha_t \alpha_{\tau}}{4} + \frac{33 \alpha_b \alpha_{\tau}}{4} + \frac{7 \alpha_{\tau}^2}{4} + n_G \bigg( - \frac{\alpha_1^2}{6} - \frac{3 \alpha_2^2}{2} - \frac{16 \alpha_3^2}{3} \bigg) \bigg] \notag \\
 &  + \frac{1}{\epsilon} \bigg[ - \frac{29 \alpha_1^2}{400} - \frac{27 \alpha_1 \alpha_2}{40} - \frac{35 \alpha_2^2}{8} + \frac{31 \alpha_1 \alpha_3}{30} + \frac{9 \alpha_2 \alpha_3}{2} - \frac{202 \alpha_3^2}{3} + 3 \hat{\lambda}^2 + \frac{91 \alpha_1 \alpha_t}{160} \notag \\
 &  + \frac{99 \alpha_2 \alpha_t}{32} + 2 \alpha_3 \alpha_t - \frac{\alpha_t^2}{8} + \frac{237 \alpha_1 \alpha_b}{160} + \frac{225 \alpha_2 \alpha_b}{32} + 18 \alpha_3 \alpha_b - 6 \hat{\lambda} \alpha_b - \frac{11 \alpha_t \alpha_b}{8} - 6 \alpha_b^2 \notag \\
 &  + \frac{15 \alpha_1 \alpha_{\tau}}{16} + \frac{15 \alpha_2 \alpha_{\tau}}{16} + \frac{5 \alpha_t \alpha_{\tau}}{8} - \frac{9 \alpha_b \alpha_{\tau}}{8} - \frac{9 \alpha_{\tau}^2}{8} + n_G \bigg( - \frac{\alpha_1^2}{90} + \frac{\alpha_2^2}{2} + \frac{40 \alpha_3^2}{9} \bigg) \bigg] \bigg\}
\,,
\end{align}
\begin{align}
  Z_{\alpha_{\tau}} &=
  1 + \frac{1}{4\pi} \frac{1}{\epsilon} \bigg\{ - \frac{9 \alpha_1}{4} - \frac{9 \alpha_2}{4} + \frac{3 \alpha_{\tau}}{2} + 3 \text{tr}\hat{T} + 3 \text{tr}\hat{B} + \text{tr}\hat{L} \bigg\} \notag \\
 &  + \frac{1}{\lp4\pi\rp^2} \bigg\{ \frac{1}{\epsilon^2} \bigg[ \frac{387 \alpha_1^2}{160} + \frac{81 \alpha_1 \alpha_2}{16} + \frac{339 \alpha_2^2}{32} - \frac{321 \alpha_1 \alpha_t}{40} - \frac{81 \alpha_2 \alpha_t}{8} - 12 \alpha_3 \alpha_t + \frac{45 \alpha_t^2}{4} \notag \\
 &  - \frac{57 \alpha_1 \alpha_b}{8} - \frac{81 \alpha_2 \alpha_b}{8} - 12 \alpha_3 \alpha_b + \frac{27 \alpha_t \alpha_b}{2} + \frac{45 \alpha_b^2}{4} - \frac{135 \alpha_1 \alpha_{\tau}}{16} - \frac{135 \alpha_2 \alpha_{\tau}}{16} \notag \\
 &  + \frac{51 \alpha_t \alpha_{\tau}}{4} + \frac{51 \alpha_b \alpha_{\tau}}{4} + \frac{25 \alpha_{\tau}^2}{4} + n_G \bigg( - \frac{3 \alpha_1^2}{2} - \frac{3 \alpha_2^2}{2} \bigg) \bigg] \notag \\
 &  + \frac{1}{\epsilon} \bigg[ \frac{51 \alpha_1^2}{400} + \frac{27 \alpha_1 \alpha_2}{40} - \frac{35 \alpha_2^2}{8} + 3 \hat{\lambda}^2 + \frac{17 \alpha_1 \alpha_t}{16} + \frac{45 \alpha_2 \alpha_t}{16} + 10 \alpha_3 \alpha_t - \frac{27 \alpha_t^2}{8} \notag \\
 &  + \frac{5 \alpha_1 \alpha_b}{16} + \frac{45 \alpha_2 \alpha_b}{16} + 10 \alpha_3 \alpha_b + \frac{3 \alpha_t \alpha_b}{4} - \frac{27 \alpha_b^2}{8} + \frac{537 \alpha_1 \alpha_{\tau}}{160} + \frac{165 \alpha_2 \alpha_{\tau}}{32} - 6 \hat{\lambda} \alpha_{\tau} \notag \\
 &  - \frac{27 \alpha_t \alpha_{\tau}}{8} - \frac{27 \alpha_b \alpha_{\tau}}{8} - \frac{3 \alpha_{\tau}^2}{2} + n_G \bigg( \frac{11 \alpha_1^2}{10} + \frac{\alpha_2^2}{2} \bigg) \bigg] \bigg\}
\,.
\end{align}


\section{\label{app::betaQED}Beta functions for $\alpha_\text{QED}$ and $\sin^2 \theta_W$}

This Appendix contains explicit results up to three-loop order for the QED
coupling $\alpha_{\rm QED}$ 
and the weak mixing angle defined in the $\overline{\rm MS}$ scheme. 
We refrain from providing expressions for
the renormalization constants but directly list the beta functions. They are
obtained in a straightforward way from Eq.~(\ref{eq::alpha_123}) and are given
by
\begin{align} 
\beta_{\alpha_\text{QED}} &=
      \frac{\alpha_{\text{QED}}^2}{\lp4\pi\rp^2} \bigg\{ - 28 + \frac{128 n_G}{9} \bigg\} + \frac{\alpha_{\text{QED}}^2}{\lp4\pi\rp^3} \bigg\{ - \frac{500 \alpha_{\text{QED}}}{3 \sin^2\theta_W} + \frac{4 \alpha_{\text{QED}}}{\cos^2\theta_W} \notag \\
 &  - \frac{52 \text{tr}\hat{T}}{3} - \frac{28 \text{tr}\hat{B}}{3} - 12 \text{tr}\hat{L} + n_G \bigg[ \frac{208 \alpha_{\text{QED}}}{3 \sin^2\theta_W} + \frac{416 \alpha_{\text{QED}}}{27 \cos^2\theta_W} + \frac{320 \alpha_3}{9} \bigg] \bigg\} \notag \\
 &  + \frac{\alpha_{\text{QED}}^2}{\lp4\pi\rp^4} \bigg\{ \frac{137 \alpha_{\text{QED}}^2}{4 \sin^2\theta_W \cos^2\theta_W} - \frac{318251 \alpha_{\text{QED}}^2}{216 \sin^4\theta_W} + \frac{163 \alpha_{\text{QED}}^2}{72 \cos^4\theta_W} + \frac{12 \alpha_{\text{QED}} \hat{\lambda}}{\sin^2\theta_W} + \frac{8 \alpha_{\text{QED}} \hat{\lambda}}{\cos^2\theta_W} \notag \\
 &  - 24 \hat{\lambda}^2 - \frac{757 \alpha_{\text{QED}} \text{tr}\hat{T}}{4 \sin^2\theta_W} - \frac{2303 \alpha_{\text{QED}} \text{tr}\hat{T}}{36 \cos^2\theta_W} - \frac{200 \alpha_3 \text{tr}\hat{T}}{3} - \frac{583 \alpha_{\text{QED}} \text{tr}\hat{B}}{4 \sin^2\theta_W} - \frac{1433 \alpha_{\text{QED}} \text{tr}\hat{B}}{36 \cos^2\theta_W}  \notag \\
 &  - \frac{152 \alpha_3 \text{tr}\hat{B}}{3} - \frac{393 \alpha_{\text{QED}} \text{tr}\hat{L}}{4 \sin^2\theta_W} - \frac{183 \alpha_{\text{QED}} \text{tr}\hat{L}}{4 \cos^2\theta_W} + \frac{59 \text{tr}\hat{B}^2}{2} + 31 (\text{tr}\hat{B})^2 + \frac{202 \text{tr}\hat{B}\text{tr}\hat{L}}{3} \notag \\
 &  + \frac{53 \text{tr}\hat{L}^2}{2} + 19 (\text{tr}\hat{L})^2 + 16 \text{tr}\hat{T}\hat{B} + \frac{85 \text{tr}\hat{T}^2}{2} + 104 \text{tr}\hat{T}\text{tr}\hat{B} + \frac{244 \text{tr}\hat{T}\text{tr}\hat{L}}{3} + 73 (\text{tr}\hat{T})^2  \notag \\
 &  + n_G \bigg[ \frac{32 \alpha_{\text{QED}}^2}{9 \sin^2\theta_W \cos^2\theta_W} + \frac{26146 \alpha_{\text{QED}}^2}{27 \sin^4\theta_W} - \frac{1580 \alpha_{\text{QED}}^2}{81 \cos^4\theta_W} + \frac{152 \alpha_{\text{QED}} \alpha_3}{3 \sin^2\theta_W} \notag \\
 &  - \frac{584 \alpha_{\text{QED}} \alpha_3}{81 \cos^2\theta_W} + \frac{10000 \alpha_3^2}{27} \bigg] + n_G^2 \bigg[ - \frac{1792 \alpha_{\text{QED}}^2}{27 \sin^4\theta_W} - \frac{22880 \alpha_{\text{QED}}^2}{729 \cos^4\theta_W} - \frac{3520 \alpha_3^2}{81} \bigg] \bigg\}
\,,
\end{align}
\begin{align}
\beta_{\sin^2 \theta_W} &= 
\mu^2 \frac{{\rm d}\sin^2 \theta_W}{{\rm d}\mu^2}\notag \\
&=
      \frac{\alpha_{\text{QED}}}{4\pi} \bigg\{ \frac{\sin^2\theta_W}{6} + \frac{43 \cos^2\theta_W}{6} + n_G \bigg[ \frac{20 \sin^2\theta_W}{9} - \frac{4 \cos^2\theta_W}{3} \bigg] \bigg\} \notag \\
 &  + \frac{\alpha_{\text{QED}}}{\lp4\pi\rp^2} \bigg\{ \alpha_{\text{QED}} + \frac{\alpha_{\text{QED}} \tan^2\theta_W}{2} + \frac{259 \alpha_{\text{QED}} \cot^2\theta_W}{6} - \frac{17 \sin^2\theta_W \text{tr}\hat{T}}{6} \notag \\
 &  + \frac{3 \cos^2\theta_W \text{tr}\hat{T}}{2} - \frac{5 \sin^2\theta_W \text{tr}\hat{B}}{6} + \frac{3 \cos^2\theta_W \text{tr}\hat{B}}{2} - \frac{5 \sin^2\theta_W \text{tr}\hat{L}}{2} + \frac{\cos^2\theta_W \text{tr}\hat{L}}{2} \notag \\
 &  + n_G \bigg[ \frac{2 \alpha_{\text{QED}}}{3} + \frac{95 \alpha_{\text{QED}} \tan^2\theta_W}{27} - \frac{49 \alpha_{\text{QED}} \cot^2\theta_W}{3} + \frac{44 \sin^2\theta_W \alpha_3}{9} - 4 \cos^2\theta_W \alpha_3 \bigg] \bigg\} \notag \\
 &  + \frac{\alpha_{\text{QED}}}{\lp4\pi\rp^3} \bigg\{ \frac{2279 \alpha_{\text{QED}}^2}{192 \sin^2\theta_W} + \frac{1403 \alpha_{\text{QED}}^2}{576 \cos^2\theta_W} + \frac{163 \alpha_{\text{QED}}^2 \tan^2\theta_W}{576 \cos^2\theta_W} + \frac{667111 \alpha_{\text{QED}}^2 \cot^2\theta_W}{1728 \sin^2\theta_W} \notag \\
 &  + \alpha_{\text{QED}} \hat{\lambda} + \frac{3 \alpha_{\text{QED}} \tan^2\theta_W \hat{\lambda}}{2} - \frac{3 \alpha_{\text{QED}} \cot^2\theta_W \hat{\lambda}}{2} - 3 \sin^2\theta_W \hat{\lambda}^2 + 3 \cos^2\theta_W \hat{\lambda}^2 \notag \\
 &  - \frac{881 \alpha_{\text{QED}} \text{tr}\hat{T}}{48} - \frac{2827 \alpha_{\text{QED}} \tan^2\theta_W \text{tr}\hat{T}}{288} + \frac{729 \alpha_{\text{QED}} \cot^2\theta_W \text{tr}\hat{T}}{32} - \frac{29 \sin^2\theta_W \alpha_3 \text{tr}\hat{T}}{3} \notag \\
 &  + 7 \cos^2\theta_W \alpha_3 \text{tr}\hat{T} - \frac{389 \alpha_{\text{QED}} \text{tr}\hat{B}}{48} - \frac{1267 \alpha_{\text{QED}} \tan^2\theta_W \text{tr}\hat{B}}{288} + \frac{729 \alpha_{\text{QED}} \cot^2\theta_W \text{tr}\hat{B}}{32} \notag \\
 &  - \frac{17 \sin^2\theta_W \alpha_3 \text{tr}\hat{B}}{3} + 7 \cos^2\theta_W \alpha_3 \text{tr}\hat{B} - \frac{229 \alpha_{\text{QED}} \text{tr}\hat{L}}{16} - \frac{281 \alpha_{\text{QED}} \tan^2\theta_W \text{tr}\hat{L}}{32} \notag \\
 &  + \frac{243 \alpha_{\text{QED}} \cot^2\theta_W \text{tr}\hat{L}}{32} + \frac{61 \sin^2\theta_W \text{tr}\hat{B}^2}{16} - \frac{57 \cos^2\theta_W \text{tr}\hat{B}^2}{16} + \frac{17 \sin^2\theta_W (\text{tr}\hat{B})^2}{8} \notag \\
 &  - \frac{45 \cos^2\theta_W (\text{tr}\hat{B})^2}{8} + \frac{157 \sin^2\theta_W \text{tr}\hat{B}\text{tr}\hat{L}}{12} - \frac{15 \cos^2\theta_W \text{tr}\hat{B}\text{tr}\hat{L}}{4} + \frac{87 \sin^2\theta_W \text{tr}\hat{L}^2}{16} \notag \\
 &  - \frac{19 \cos^2\theta_W \text{tr}\hat{L}^2}{16} + \frac{33 \sin^2\theta_W (\text{tr}\hat{L})^2}{8} - \frac{5 \cos^2\theta_W (\text{tr}\hat{L})^2}{8} + \frac{5 \sin^2\theta_W \text{tr}\hat{T}\hat{B}}{8} \notag \\
 &  - \frac{27 \cos^2\theta_W \text{tr}\hat{T}\hat{B}}{8} + \frac{113 \sin^2\theta_W \text{tr}\hat{T}^2}{16} - \frac{57 \cos^2\theta_W \text{tr}\hat{T}^2}{16} + \frac{59 \sin^2\theta_W \text{tr}\hat{T}\text{tr}\hat{B}}{4} \notag \\
 &  - \frac{45 \cos^2\theta_W \text{tr}\hat{T}\text{tr}\hat{B}}{4} + \frac{199 \sin^2\theta_W \text{tr}\hat{T}\text{tr}\hat{L}}{12} - \frac{15 \cos^2\theta_W \text{tr}\hat{T}\text{tr}\hat{L}}{4} \notag \\
 &  + \frac{101 \sin^2\theta_W (\text{tr}\hat{T})^2}{8}- \frac{45 \cos^2\theta_W (\text{tr}\hat{T})^2}{8} + n_G \bigg[ \frac{127 \alpha_{\text{QED}}^2}{36 \sin^2\theta_W} + \frac{119 \alpha_{\text{QED}}^2}{108 \cos^2\theta_W} \notag \\
 &  - \frac{290 \alpha_{\text{QED}}^2 \tan^2\theta_W}{81 \cos^2\theta_W} - \frac{6412 \alpha_{\text{QED}}^2 \cot^2\theta_W}{27 \sin^2\theta_W} - \frac{2 \alpha_{\text{QED}} \alpha_3}{9} - \frac{137 \alpha_{\text{QED}} \tan^2\theta_W \alpha_3}{81} \notag \\
 &  - 13 \alpha_{\text{QED}} \cot^2\theta_W \alpha_3 + \frac{1375 \sin^2\theta_W \alpha_3^2}{27} - \frac{125 \cos^2\theta_W \alpha_3^2}{3} \bigg] \notag \\
 &  + n_G^2 \bigg[ - \frac{11 \alpha_{\text{QED}}^2}{9 \sin^2\theta_W} + \frac{55 \alpha_{\text{QED}}^2}{81 \cos^2\theta_W} - \frac{5225 \alpha_{\text{QED}}^2 \tan^2\theta_W}{729 \cos^2\theta_W} + \frac{415 \alpha_{\text{QED}}^2 \cot^2\theta_W}{27 \sin^2\theta_W} \notag \\
 &  - \frac{484 \sin^2\theta_W \alpha_3^2}{81} + \frac{44 \cos^2\theta_W \alpha_3^2}{9} \bigg] \bigg\}
\,.
\end{align}


\section{\label{app::compare}Comparison with Ref.~\cite{Pickering:2001aq}}

In this Appendix we provide the explicit form of the expressions needed
for the comparison with Ref.~\cite{Pickering:2001aq}. In that paper
two-component Weyl spinors were used. To make contact with our convention
based on
four-component Dirac spinors we define
\begin{equation}
\Psi_D=\left(\begin{matrix}
        \chi  \\
         \xi^{\dagger}
      \end{matrix}
    \right)  \,,
\end{equation}
where $\xi$ and $\chi$ are left-handed Weyl spinors and $\Psi_D$ denotes a
Dirac spinor. Thus, the Lagrange density of the SM can be expressed in
terms of 45 Weyl spinors:
\begin{equation}
\chi_{t}\,,\xi_{t}\,,\chi_{b}\,, \xi_{b}\,,\chi_{\tau}\,,\xi_{\tau}\,,
\chi_{\nu_{\tau}}\,,\chi_{c}\,,\xi_{c}\,,\chi_{s},\xi_{s},\chi_{\mu},\xi_{\mu},
\chi_{\nu_{\upmu}},
\chi_{u},\xi_{u},\chi_{d},\xi_{d},\chi_{e},\xi_{e},
\chi_{\nu_{\textrm{e}}}\,.
\end{equation}
For simplicity we have suppressed the $SU(3)$ color indices for all quark
spinors. Of course, each quark spinor has to be understood as a triplet
in color space. In this basis the Yukawa matrices become $45\times45$
dimensional.

In the notation of~\cite{Pickering:2001aq}
the part of the Lagrange density describing the Yukawa couplings is given by
\begin{equation}
  -\frac{1}{2}
  \left(Y^{aij}\phi^a\psi_i\psi_j
    +\bar{Y}^a_{ij}\phi^a\bar{\psi}^i\bar{\psi}^j\right)
  \,,
\end{equation}
where $Y^a$ and $\bar{Y}^a$ are the (complex conjugated) Yukawa matrices,
$\phi^a$ are real scalar fields, 
$\psi^i$ and $\bar{\psi}^i$ are (Hermitian conjugated) spinor fields. There
are four real scalar fields in the SM, which means that 
we have four Yukawa matrices. They are given by
\begin{gather}
  Y^1 = \frac{1}{\sqrt{2}}
	\begin{pmatrix}
	  0_{3\times3} & y_t\mathbbm{1}_{3\times3} & 0_{3\times3} & 0_{3\times3} & 0_{3\times1} & 0_{3\times1} & 0_{3\times1} & \cdots\\
	  y_t\mathbbm{1}_{3\times3} & 0_{3\times3} & 0_{3\times3} & 0_{3\times3} & 0_{3\times1} & 0_{3\times1} & 0_{3\times1} & \cdots\\
	  0_{3\times3} & 0_{3\times3} & 0_{3\times3} & y_b\mathbbm{1}_{3\times3} & 0_{3\times1} & 0_{3\times1} & 0_{3\times1} & \cdots\\
	  0_{3\times3} & 0_{3\times3} & y_b\mathbbm{1}_{3\times3} & 0_{3\times3} & 0_{3\times1} & 0_{3\times1} & 0_{3\times1} & \cdots\\
	  0_{1\times3} & 0_{1\times3} & 0_{1\times3} & 0_{1\times3} & 0 & y_{\tau} & 0 & \cdots\\
	  0_{1\times3} & 0_{1\times3} & 0_{1\times3} & 0_{1\times3} & y_{\tau} & 0 & 0 & \cdots\\
	  0_{1\times3} & 0_{1\times3} & 0_{1\times3} & 0_{1\times3} & 0 & 0 & 0 & \cdots\\
	  \vdots & \vdots & \vdots & \vdots & \vdots & \vdots & \vdots & \ddots
        \end{pmatrix}\,, \notag \\
  Y^2 = \frac{1}{\sqrt{2}}
	\begin{pmatrix}
	  0_{3\times3} & 0_{3\times3} & 0_{3\times3} & y_b\mathbbm{1}_{3\times3} & 0_{3\times1} & 0_{3\times1} & 0_{3\times1} & \cdots\\
	  0_{3\times3} & 0_{3\times3} & -y_t\mathbbm{1}_{3\times3} & 0_{3\times3} & 0_{3\times1} & 0_{3\times1} & 0_{3\times1} & \cdots\\
	  0_{3\times3} & -y_t\mathbbm{1} _{3\times3} & 0_{3\times3} & 0_{3\times3} & 0_{3\times1} & 0_{3\times1} & 0_{3\times1} & \cdots\\
	  y_b\mathbbm{1}_{3\times3} & 0_{3\times3} & 0_{3\times3} & 0_{3\times3} & 0_{3\times1} & 0_{3\times1} & 0_{3\times1} & \cdots\\
	  0_{1\times3} & 0_{1\times3} & 0_{1\times3} & 0_{1\times3} & 0 & 0 & 0 & \cdots\\
	  0_{1\times3} & 0_{1\times3} & 0_{1\times3} & 0_{1\times3} & 0 & 0 & y_{\tau} & \cdots\\
	  0_{1\times3} & 0_{1\times3} & 0_{1\times3} & 0_{1\times3} & 0 & y_{\tau} & 0 & \cdots\\
	  \vdots & \vdots & \vdots & \vdots & \vdots & \vdots & \vdots & \ddots
        \end{pmatrix}\,, \notag \\
Y^3 = -\frac{1}{\sqrt{2}} 
	\begin{pmatrix}
	  0_{3\times3} & -\textrm{i}y_t\mathbbm{1}_{3\times3} & 0_{3\times3} & 0_{3\times3} & 0_{3\times1} & 0_{3\times1} & 0_{3\times1} & \cdots\\
	  -\textrm{i}y_t\mathbbm{1}_{3\times3} & 0_{3\times3} & 0_{3\times3} & 0_{3\times3} & 0_{3\times1} & 0_{3\times1} & 0_{3\times1} & \cdots\\
	  0_{3\times3} & 0_{3\times3} & 0_{3\times3} & \textrm{i}y_b\mathbbm{1}_{3\times3} & 0_{3\times1} & 0_{3\times1} & 0_{3\times1} & \cdots\\
	  0_{3\times3} & 0_{3\times3} & \textrm{i}y_b\mathbbm{1}_{3\times3} & 0_{3\times3} & 0_{3\times1} & 0_{3\times1} & 0_{3\times1} & \cdots\\
	  0_{1\times3} & 0_{1\times3} & 0_{1\times3} & 0_{1\times3} & 0 & \textrm{i}y_{\tau} & 0 & \cdots\\
	  0_{1\times3} & 0_{1\times3} & 0_{1\times3} & 0_{1\times3} & \textrm{i}y_{\tau} & 0 & 0 & \cdots\\
	  0_{1\times3} & 0_{1\times3} & 0_{1\times3} & 0_{1\times3} & 0 & 0 & 0 & \cdots\\
	  \vdots & \vdots & \vdots & \vdots & \vdots & \vdots & \vdots & \ddots
        \end{pmatrix}\,, \notag \\
Y^4 = -\frac{1}{\sqrt{2}} 
	\begin{pmatrix}
	  0_{3\times3} & 0_{3\times3} & 0_{3\times3} & \textrm{i}y_b\mathbbm{1}_{3\times3} & 0_{3\times1} & 0_{3\times1} & 0_{3\times1} & \cdots\\
	  0_{3\times3} & 0_{3\times3} & \textrm{i}y_t\mathbbm{1}_{3\times3} & 0_{3\times3} & 0_{3\times1} & 0_{3\times1} & 0_{3\times1} & \cdots\\
	  0_{3\times3} & \textrm{i}y_t\mathbbm{1} _{3\times3} & 0_{3\times3} & 0_{3\times3} & 0_{3\times1} & 0_{3\times1} & 0_{3\times1} & \cdots\\
	  \textrm{i}y_b\mathbbm{1}_{3\times3} & 0_{3\times3} & 0_{3\times3} & 0_{3\times3} & 0_{3\times1} & 0_{3\times1} & 0_{3\times1} & \cdots\\
	  0_{1\times3} & 0_{1\times3} & 0_{1\times3} & 0_{1\times3} & 0 & 0 & 0 & \cdots\\
	  0_{1\times3} & 0_{1\times3} & 0_{1\times3} & 0_{1\times3} & 0 & 0 & \textrm{i}y_{\tau} & \cdots\\
	  0_{1\times3} & 0_{1\times3} & 0_{1\times3} & 0_{1\times3} & 0 & \textrm{i}y_{\tau} & 0 & \cdots\\
	  \vdots & \vdots & \vdots & \vdots & \vdots & \vdots & \vdots & \ddots
        \end{pmatrix}\,.
\end{gather}
In the above formulas,  all matrix elements not
explicitly given are zero. 
$Y^1$/$Y^3$ is the Yukawa matrix of the real/the imaginary part of the isospin
down component of the SM Higgs doublet. 
$Y^2$/$Y^4$ is the Yukawa matrix of the real/the imaginary part of the isospin
up component of the SM Higgs doublet. 
So we have
\begin{equation}\label{Higgsdublett}
  H = \frac{1}{\sqrt{2}}\begin{pmatrix}
    \phi^2 + \textrm{i} \phi^4\\
    \phi^1 + \textrm{i} \phi^3
  \end{pmatrix}\,.
\end{equation}

The part of the Lagrange density describing the Higgs self-interaction
reads in the notation of Ref.~\cite{Pickering:2001aq}
\begin{equation}
  -\frac{1}{4!}\lambda_{abcd}\phi^a\phi^b\phi^c\phi^d\,. 
\end{equation}
In the SM we have
\begin{align}
  \lambda_{aaaa} &= 6\times(4\pi\hat{\lambda}),\notag\\
  \lambda_{aabb} = \lambda_{abab} = \lambda_{abba} &= 2\times(4\pi\hat{\lambda})\ (a \neq b),\notag\\
  \lambda_{abcd} &= 0\ \ \ (\text{otherwise})\,.
\end{align}

All relations discussed so far are generic for all three gauge groups.
However, there are  some expressions which depend on the specific
gauge group one wants to consider. The remainder of this
Section lists these expressions. For each of the three gauge groups,
we give the expressions for the generators in the representations of the
scalar fields, $S^A$, and of the 
Weyl spinors, $R^A$.  We also give expressions for the invariants
$T(S),C(S),T(R),C(R),C(G),r$, all of which are symbols used in
Ref.~\cite{Pickering:2001aq}.  They are defined as
\begin{align}
  \text{Tr}\left(S^A S^B\right) &= 
  \delta^{AB}T(S), & S^A_{ac} S^A_{cb} &= C(S)_{ab},\notag\\
  \text{Tr}\left(R^A R^B\right) &= 
  \delta^{AB}T(R), & R^{Ak}_i R^{Aj}_k &= C(R)_i^j,\notag\\
  f^{ACD}f^{BCD} &= \delta^{AB}C(G), & \delta^{AA}& = r\,.
\end{align}
$f^{ABC}$ are the structure constants of the respective gauge
group. The indices $A,B,C$ take values in the range $1,\dots,r$,
where $r$ gives the dimension  of the group.

\subsection{$U(1)$}

For the case of $U(1)$ gauge group,  we have
\begin{gather}
  S^1 = \frac{\text{i}}{2}
  \begin{pmatrix}
    0&  0&  1&  0\\
    0&  0&  0&  1\\
    -1&  0&  0&  0\\
    0&  -1&  0& 0
  \end{pmatrix}\,,\label{S1}\\
  R^1 =
   \Diag\left(\frac{1}{6},\frac{1}{6},\frac{1}{6},
    -\frac{2}{3},-\frac{2}{3},-\frac{2}{3},
    \frac{1}{6},\frac{1}{6},\frac{1}{6},
    \frac{1}{3},\frac{1}{3},\frac{1}{3}, 
    -\frac{1}{2},1,-\frac{1}{2},\dots\right)\,,
\end{gather}
where the ellipsis is to be replaced twice by the first 15
entries. The derivation of $S^1$ is given below. The entries in $R^1$
are the hypercharges $Y$ of the respective spinors. They can be
derived by using the relation $Y=Q-I_3$, where $Q$ corresponds to the electric charge
and $I_3$ is the weak isospin.
Furthermore, we have
\begin{align}
  T(S) &= 1\,,  & C(S) &= \frac{1}{4}\mathbbm{1}_{4\times4}\,,\notag\\
  T(R) &= 10\,, & C(R) &= \left(R^1\right)^2\,,\notag\\
  C(G) &= 0\,,  & r &= 1\,.
\end{align}

Let us now show how to derive Eq.~(\ref{S1}).  We want to find the
$U(1)$ representation transforming the four real scalar fields of the
Higgs doublet of the SM (see Eq.~(\ref{Higgsdublett}). The
matrix $S^1$ is the generator of this transformation,
\begin{equation}\label{Definition S1}
  \begin{pmatrix}
  \phi'_1\\
  \phi'_2\\
  \phi'_3\\
  \phi'_4
  \end{pmatrix}
  =
  \textrm{e}^{\textrm{i}\omega S^1}
  \begin{pmatrix}
  \phi_1\\
  \phi_2\\
  \phi_3\\
  \phi_4
  \end{pmatrix}
  =
  \left(\mathbbm{1}+\textrm{i}\omega S^1+\mathcal{O}(\omega^2)\right)
  \begin{pmatrix}
  \phi_1\\
  \phi_2\\
  \phi_3\\
  \phi_4
  \end{pmatrix}\,.
\end{equation}
with $\omega$ being the transformation parameter.  In a next step we
take advantage of the fact that it is known how the SM Higgs doublet
transforms in order to determine $S^1$. As the SM Higgs doublet has
hypercharge $1/2$, we have
\begin{align}\label{Bestimmung S1}
  \frac{1}{\sqrt{2}}
  \begin{pmatrix}
  \phi'^2 + \textrm{i} \phi'^4\\
  \phi'^1 + \textrm{i} \phi'^3
  \end{pmatrix}
  &= H' = \textrm{e}^{\textrm{i}\omega \left(\frac{1}{2}\mathbbm{1}\right)}H =
  \textrm{e}^{\textrm{i}\omega \left(\frac{1}{2}\mathbbm{1}\right)}
  \frac{1}{\sqrt{2}}
  \begin{pmatrix}
  \phi^2 + \textrm{i} \phi^4\\
  \phi^1 + \textrm{i} \phi^3
  \end{pmatrix}\notag\\
  &= \left[\mathbbm{1}+\textrm{i}\omega
    \left(\frac{1}{2}\mathbbm{1}\right)+\mathcal{O}(\omega^2)\right] 
  \frac{1}{\sqrt{2}}
  \begin{pmatrix}
  \phi^2 + \textrm{i} \phi^4\\
  \phi^1 + \textrm{i} \phi^3
  \end{pmatrix}\notag\\
  &= H + \omega \frac{1}{2} \frac{1}{\sqrt{2}}
  \begin{pmatrix}
  -\phi^4 + \textrm{i} \phi^2\\
  -\phi^3 + \textrm{i} \phi^1
  \end{pmatrix}
  +\mathcal{O}(\omega^2)\,.
\end{align}
With the help of the last equation one can determine 
the transformation of the SM Higgs doublet, like, e.g.,
$\phi'_1=\phi_1-\frac{\omega}{2}\phi_3+\mathcal{O}(\omega^2)$. 
It is then straightforward to determine $S^1$ by inserting the
equations found in this way in Eq.~(\ref{Definition S1}).

\subsection{$SU(2)$}

For the $SU(2)$ group, we have
\begin{equation}
  S^1 = -\frac{\text{i}}{2}
  \begin{pmatrix}
  0&  0&  0& -1 \\
  0&  0&  -1& 0 \\
  0&  1&  0&  0\\
  1&  0&  0& 0
  \end{pmatrix},\,
  S^2 = -\frac{\text{i}}{2}
\begin{pmatrix}
  0&  -1&  0& 0 \\
  1&  0&  0& 0 \\
  0&  0&  0& -1 \\
  0&  0&  1& 0
\end{pmatrix},\,
  S^3 = -\frac{\text{i}}{2}
\begin{pmatrix}
  0&  0&  1& 0 \\
  0&  0&  0& -1 \\
  -1&  0&  0& 0 \\
  0&  1&  0& 0
\end{pmatrix},
\end{equation}
and
\begin{equation}
  R^A = 1/2
\begin{pmatrix}
  \sigma^A_{1,1}\mathbbm{1}_{3\times3} & 0_{3\times3} & \sigma^A_{1,2}\mathbbm{1}_{3\times3} & 0_{3\times3} & 0_{3\times1} & 0_{3\times1} & 0_{3\times1} & \cdots\\
  0_{3\times3} & 0_{3\times3} & 0_{3\times3} & 0_{3\times3} & 0_{3\times1} & 0_{3\times1} & 0_{3\times1} & \cdots \\
  \sigma^A_{2,1}\mathbbm{1}_{3\times3} & 0_{3\times3} & \sigma^A_{2,2}\mathbbm{1}_{3\times3} & 0_{3\times3} & 0_{3\times1} & 0_{3\times1} & 0_{3\times1} & \cdots \\
  0_{3\times3} & 0_{3\times3} & 0_{3\times3} & 0_{3\times3} & 0_{3\times1} & 0_{3\times1} & 0_{3\times1} & \cdots \\
  0_{1\times3} & 0_{1\times3} & 0_{1\times3} & 0_{1\times3} & \sigma^A_{2,2} & 0 & \sigma^A_{2,1} & \cdots\\
  0_{1\times3} & 0_{1\times3} & 0_{1\times3} & 0_{1\times3} & 0 & 0 & 0 & \cdots\\
  0_{1\times3} & 0_{1\times3} & 0_{1\times3} & 0_{1\times3} & \sigma^A_{1,2} & 0 & \sigma^A_{1,1} & \cdots\\
  \vdots&\vdots&\vdots&\vdots&\vdots&\vdots&\vdots&\ddots
\end{pmatrix}\,.
\end{equation}
In the last equation $\sigma^A$ are the Pauli matrices. 
Not all of the matrix elements
left out in $R^A$ vanish, but these elements play no role for the
comparison of our results to~\cite{Pickering:2001aq}.  The derivation
of the generators $S^A$ proceeds in analogy to the derivation of $S^1$
explained in the former Subsection. One merely has to substitute the
generator of $U(1)$, $\frac{1}{2}\mathbbm{1}$, by the generators of
SU(2), $\frac{1}{2}\sigma^A$, in Eq.~(\ref{Bestimmung S1}).

We furthermore have
\begin{align}
  T(S) &= 1, & C(S) &= \frac{3}{4}\mathbbm{1}_{4\times4},\notag \\ T(R) &= 6,
  & C(R) &= \frac{3}{4}
  \Diag\left(1,1,1,0,0,0,1,1,1,0,0,0,1,0,1,\dots\right)\,,\notag \\ C(G)
  &= 2, & r &= 3\,.
\end{align}
The ellipsis has to be replaced by the first entries twice.

\subsection{$SU(3)$}

In the SM the matrices $S^A$ vanish for the group $SU(3)$. The matrices
$R^A$ are block-diagonal and read
\begin{equation}
  R^A =
  \frac{1}{2}\BlockDiag
  \left(\Lambda^A,-(\Lambda^A)^T,\Lambda^A,-(\Lambda^A)^T,0,0,0,\dots\right)
  \,.
\end{equation}
The ellipsis has to be replaced twice by the former entries, which contain the
Gell-Mann matrices $\Lambda^A$.

Finally, we have
\begin{align}
  T(S) &= 0,  & C(S) &= 0_{4\times4},\notag\\
  T(R) &= 6, & C(R) &= \frac{4}{3} \BlockDiag \left(\mathbbm{1}_{12\times12},0_{3\times3},\mathbbm{1}_{12\times12},0_{3\times3},\mathbbm{1}_{12\times12},0_{3\times3}\right),\notag\\
  C(G) &= 3,  & r &= 8\,.
\end{align}




\end{document}